\documentclass[prd,twocolumn,showpacs,showkeys,preprintnumbers,floatfix,nofootinbib,superscriptaddress]{revtex4-1}
\usepackage{amsfonts} 
\usepackage{amssymb} 
\usepackage{amsmath} 
\usepackage{graphicx} 
\usepackage{subfigure} 
\usepackage{array} 
\usepackage{dcolumn} 
\usepackage{bm} 
\usepackage{latexsym} 
\usepackage{longtable} 
\usepackage{hyperref} 
\usepackage{bbold}
\usepackage{color}
\usepackage{multirow}
\usepackage{rotating}

\graphicspath{{./figs/}}

\newcommand{\qslash}[1]{\text{$\not \! #1$}}

\newcommand{\tsep}{\mathop{t_{\rm sep}}\nolimits}

\newcommand{\gsim}{\raisebox{-0.7ex}{$\stackrel{\textstyle >}{\sim}$ }}
\newcommand{\lsim}{\raisebox{-0.7ex}{$\stackrel{\textstyle <}{\sim}$ }}

\DeclareMathOperator{\Tr}{Tr}

\definecolor{green}{rgb}{0.1, 0.8, 0.1}

\begin{document}


\title{Controlling Excited-State Contamination in Nucleon Matrix Elements}
\author{Boram Yoon}
\email{boram@lanl.gov}
\affiliation{Los Alamos National Laboratory, Theoretical Division T-2, Los Alamos, NM 87545}

\author{Rajan Gupta}
\email{rajan@lanl.gov}
\affiliation{Los Alamos National Laboratory, Theoretical Division T-2, Los Alamos, NM 87545}

\author{Tanmoy Bhattacharya}
\affiliation{Los Alamos National Laboratory, Theoretical Division T-2, Los Alamos, NM 87545}


\author{Michael Engelhardt}
\affiliation{Department of Physics, New Mexico State University, Las Cruces, NM 88003-8001, USA}

\author{Jeremy Green}
\affiliation{Institut f\"ur Kernphysik, Johannes Gutenberg-Universit\"at Mainz, D-55099 Mainz, Germany}

\author{B\'alint Jo\'o}
\affiliation{Jefferson Lab, 12000 Jefferson Avenue, Newport News, Virginia 23606, USA}
 \author{Huey-Wen Lin}
 \affiliation{Physics Department, University of California, Berkeley, CA 94720}
 \author{John Negele}
 \affiliation{Center for Theoretical Physics, Massachusetts Institute of Technology, Cambridge, Massachusetts 02139, USA}
 \author{Kostas Orginos}
 \affiliation{Department of Physics, College of William and Mary, Williamsburg, Virginia 23187-8795, USA and 
 Jefferson Lab, 12000 Jefferson Avenue, Newport News, Virginia 23606, USA}
 \author{Andrew Pochinsky}
 \affiliation{Center for Theoretical Physics, Massachusetts Institute of Technology, Cambridge, Massachusetts 02139, USA}
 \author{David Richards}
 \affiliation{Jefferson Lab, 12000 Jefferson Avenue, Newport News, Virginia 23606, USA}

\author{Sergey Syritsyn}
\affiliation{Jefferson Lab, 12000 Jefferson Avenue, Newport News, Virginia 23606, USA}
\author{Frank Winter}
\affiliation{Jefferson Lab, 12000 Jefferson Avenue, Newport News, Virginia 23606, USA}
 
\collaboration{Nucleon Matrix Elements (NME) Collaboration}
\preprint{LA-UR-16-20524}
%
\date{\today}
\begin{abstract}
We present a detailed analysis of methods to reduce statistical errors
and excited-state contamination in the calculation of matrix elements
of quark bilinear operators in nucleon states.  All the calculations
were done on a 2+1 flavor ensemble with lattices of size $32^3 \times
64$ generated using the rational hybrid Monte Carlo algorithm at
$a=0.081$~fm and with $M_\pi=312$~MeV. The statistical precision of
the data is improved using the all-mode-averaging method. We compare
two methods for reducing excited-state contamination: a variational
analysis and a 2-state fit to data at multiple values of the
source-sink separation $\tsep$.  We show that both methods can be
tuned to significantly reduce excited-state contamination and discuss
their relative advantages and cost effectiveness. A detailed analysis
of the size of source smearing used in the calculation of quark
propagators and the range of values of $\tsep$ needed to demonstrate
convergence of the isovector charges of the nucleon to the $\tsep \to
\infty $ estimates is presented.
\end{abstract}
\maketitle
%
%
%
%
\section{Introduction}
\label{sec:into}

The ability to obtain precise estimates of matrix elements of bilinear
quark operators within a nucleon state will allow us to probe a number
of phenomenologically interesting quantities.  These include (i) the
isovector and flavor diagonal charges $g_A$, $g_S$ and $g_T$; (ii) the
electric, magnetic and axial vector form factors; (iii) generalized
parton distribution functions (GPDs); (iv) the nucleon sigma term; (v)
strangeness of the nucleon and (vi) the matrix elements of novel CP
violating operators and their contributions to the neutron electric
dipole moment.  Large scale simulations of lattice QCD provide the
best known method for obtaining precise results with control over all
sources of errors.  In this work we investigate the all-mode-averaging
method for improving the statistical precision of the calculations and
compare two methods for mitigating excited-state contamination in the
results.

The methodology for the lattice QCD calculations of the various matrix
elements within the nucleon is well developed for most of these
quantities~\cite{Lin:2012ev,Syritsyn:2014saa,Green:2014vxa,Constantinou:2014tga,Bhattacharya:2015wna}.
Generation of background gauge configurations with (2+1) or (2+1+1)
flavors is now standard. In these, the strange and the charm quark
masses are fixed to their physical values and the two light quark
masses are varied towards their physical
values~\cite{Bazavov:2012xda}.  In this work on nucleon charges, we
use 2+1 flavor configurations generated with the clover-Wilson action.
In general, all zero-momentum observables ${\cal O}(a, M_\pi, M_\pi L)$ are
calculated as functions of the lattice spacing $a$, the light quark mass
characterized by the pion mass $M_\pi$, and the lattice size $L$ expressed in
dimensionless units of $M_\pi L$.  Physical results are then obtained
by taking the continuum limit ($a\rightarrow 0$), the physical pion
mass limit ($M_{\pi^0} = 135$~MeV) and the infinite volume limit
($M_\pi L \rightarrow \infty$).  Since most lattice QCD simulations
are done over a range of values of $\{a, M_\pi, M_\pi L \}$, the above
three limits are best taken simultaneously using a combined fit in the
three variables~\cite{Bhattacharya:2015wna}. \looseness-1

The challenges to obtaining precise results for matrix elements within
the nucleon ground state are the following~\cite{Lin:2012ev,Syritsyn:2014saa,Green:2014vxa,Constantinou:2014tga,Bhattacharya:2015wna}:
\begin{itemize}
\item
Excited-state contamination in nucleon matrix elements: Contributions
of excited states to the matrix elements of operators, for example of
the axial and scalar bilinear quark operators discussed in this study,
can be large at values of the source-sink separation $t_{\rm sep}$
accessible with current computational resources and with nucleon
interpolating operators commonly used.
\item
Statistics: The signal in all nucleon correlation functions degrades
exponentially with the source-sink separation $t_{\rm sep}$. Thus very
high statistics are needed to get a good signal at values of $t_{\rm
  sep}$ at which the excited-state contamination is negligible.
\item
Reliable extrapolation to the continuum limit: This requires simulations at 
at least three values of the lattice spacing covering a sufficiently 
large range, such as $0.05\, \lsim a \, \lsim 0.1$~fm.
\item
Chiral extrapolation: Analytic tools such as heavy baryon chiral
perturbation theory used to derive the behavior of ${\cal O}(a, M_\pi,
M_\pi L)$ versus
$M_\pi$~\cite{Bernard:2006gx,Bernard:2006te,deVries:2010ah} and its
application to extrapolating the lattice data to the physical value
are more complex and not fully resolved.  It is, therefore, necessary
to perform simulations close to the physical point to reduce the
extrapolation uncertainty.
\item
Finite volume corrections: These are large in the matrix elements of
bilinear quark operators in the nucleons. Past calculations show that
one needs $M_\pi L \gsim 4 $ to be in a region in which the volume
dependence is small and can be fit by the leading order correction,
$e^{-M_\pi L}$.  Using larger lattices increases the computational
cost which scales as $L^5$ for lattice generation and $L^4$ for
analysis for fixed $a$ and $M_\pi a$.
\end{itemize}

In this work we focus on the first two sources of errors listed above:
statistical errors and excited-state contamination. We show, by
analyzing 96 low precision (LP) measurements on each of the 443
(2+1)-flavor configurations with lattice size $32^3 \times 64$, that
the all-mode-averaging (AMA) error-reduction
technique~\cite{Blum:2012uh} is an inexpensive way to significantly
improve the statistics (see Sec.~\ref{sec:AMA}).  To understand and
control excited-state contamination, we compare estimates from a
variational analysis~\cite{Dragos:2015ocy} to those from 2-state fits
to data with multiple values of $t_{\rm sep}$.  Since the focus of
this work is on comparing methods, all the data presented are for the
unrenormalized charges and without extrapolation to the physical
point. Results for the renormalized charges will be presented in a
separate study.

This paper is organized as follows. In Sec.~\ref{sec:Methodology}, we
describe the parameters of the gauge ensemble analyzed and the
lattice methodology. A discussion of statistical errors in 2-point and
3-point functions is given in Sec.~\ref{sec:statistics}.  A comparison
of the 2-state fit with multiple $t_{\rm sep}$ and the variational
method for reducing excited-state contamination is given in
Sec.~\ref{sec:excited}. cost effectiveness of the two methods in
reducing excited-state contamination is discussed in
Sec.~\ref{sec:Cost} along with a comparison with results
from~\cite{Dragos:2015ocy}. We end with some final conclusions in
Sec.~\ref{sec:conclusions}.

\section{Lattice Methodology}
\label{sec:Methodology}

We analyze one ensemble of (2+1)-flavor QCD generated using the
Sheikholeslami-Wohlert (clover-Wilson) fermion action with stout-link
smearing \cite{Morningstar:2003gk} of the gauge fields and a
tree-level tadpole-improved Symanzik gauge action.  One iteration of
the four-dimensional stout smearing is used with the weight $\rho=0.125$
for the staples in the rational hybrid Monte Carlo (RHMC) algorithm.
After stout smearing, the tadpole-improved tree-level clover
coefficient is very close to the non-perturbative value.  This was
confirmed using the Schr\"odinger functional method for determining
the clover coefficient non-perturbatively.  The strange quark mass is
tuned to its physical value by requiring the ratio $(2M_{K^+}^2 -
M_{\pi^+}^2)/M_{\Omega^-}$, that is independent of the light quark masses
to lowest order in $\chi$PT, take on its physical value
$=0.1678$~\cite{Lin:2008pr}.  This tuning is done in the 3-flavor
theory, and the resulting value of $m_s$ is then kept fixed as the
light-quark masses in the (2+1)-flavor theory are decreased towards
their physical values.  The lattice spacing is estimated to be
$0.081$~fm from heavy baryon spectroscopy. The two light quark
flavors, $u$ and $d$, are taken to be degenerate with a pion mass of
roughly $312$~MeV.  The lattice parameters of the ensemble studied,
$a081m312$, are summarized in Table~\ref{tab:ens}.  Further details
involving the generation of these gauge configurations will be
presented in a separate publication~\cite{JLAB:2016}.
\begin{table}
\begin{center}
\renewcommand{\arraystretch}{1.2} 
\begin{ruledtabular}
\begin{tabular}{l|ccc|cc}
Ensemble    & $a$ (fm)        & $M_\pi$ (MeV) &  $C_{SW}$  & $L^3\times T$   & $M_\pi L$         \\
\hline
a081m312 &  0.081   & 312        &  1.2053658 & $32^3\times 64$ & 4.08      \\
\end{tabular}
\end{ruledtabular}
\caption{Parameters of the (2+1) flavor clover lattices generated by
  the JLab/W\&M Collaboration~\protect\cite{JLAB:2016}. The number of configurations analyzed are 443.
 }
\label{tab:ens}
\end{center}
\end{table}

The 2- and 3-point correlation functions defined in
Eqs.~\eqref{eq:corr_funs2} and ~\eqref{eq:corr_funs3} are constructed using quark propagators
obtained by inverting the clover Dirac matrix with the same parameters
as used in lattice generation. The inversion uses gauge-invariant
Gaussian smeared sources constructed by applying the three-dimensional
Laplacian operator $\nabla^2$ a fixed number of times $N_{\rm GS}$
to a unit point source, $i.e.$, $(1 - \sigma^2\nabla^2/(4N_{\rm
  GS}))^{N_{\rm GS}}$.  The smearing parameters $\{\sigma, N_{\rm
  GS}\}$ for each measurement are given in Table~\ref{tab:4runs}.

\begin{figure*}[tb]
  \subfigure{
     \includegraphics[width=0.495\linewidth]{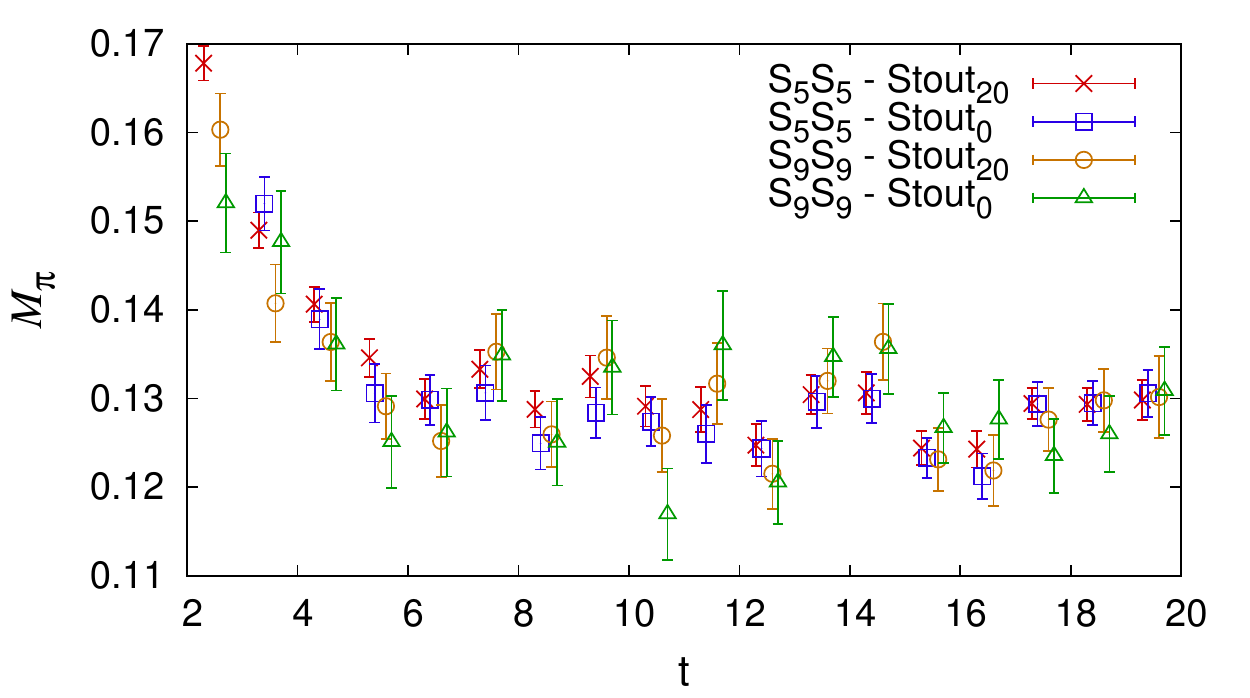}
     \includegraphics[width=0.495\linewidth]{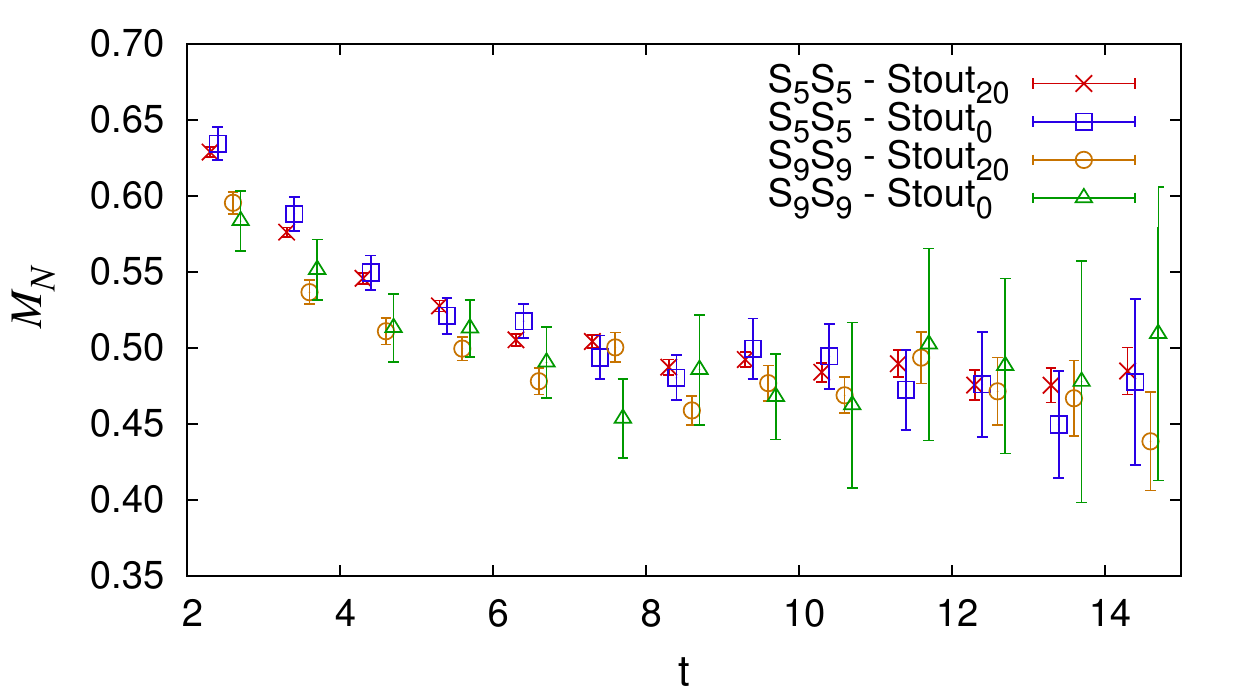}
  }
 \caption{A comparison of the errors in the pion and the proton
   effective mass data with zero versus twenty stout smearing of links
   prior to Gaussian smearing. The data were obtained using 100
   configurations and the AMA setup for both the $S_5S_5$ and $S_9S_9$
   2-point functions.  }
\label{fig:stout}
\end{figure*}

Before constructing the Gaussian smeared sources, we smoothen all the
gauge links by 20 hits of stout smearing with weight $\rho=0.08$. This
is done to reduce the noise in the correlation functions due to
fluctuations in the source. In a related calculation described in
Ref.~\cite{Syritsyn:2009mx}, it was shown that the variance in the rms
radius of the smeared source is significantly reduced with both APE
and Wuppertal smoothening of the links. This reduction in variance
displayed a very steep fall off with the number of smoothening steps
and most of the improvement was achieved at the end of 10--15 hits.  A
similar improvement is expected with the stout smearing. Because stout
smearing is a tiny overhead in our calculation, we conservatively
choose a larger number, 20 hits, to achieve close to the asymptotic
benefit.  In Fig.~\ref{fig:stout}, we show the result of our test
using 100 configurations and the AMA setup (for notation and details
see below): the reduction in errors with 20 stout hits is almost a
factor of three in the nucleon effective mass data and about 50\% in
the pion effective mass data for both $S_5S_5$ and $S_9S_9$ 2-point
functions. A related demonstration of the improvement in the nucleon
effective mass data due to smoothening the links and using smeared
quark sources has previously been discussed in
Ref.~\cite{Bulava:2009jb}. These test calculations have not been
extended to 3-point functions, nevertheless, one expects a similar
level of improvement.

Throughout this paper, the notation $S_i S_j$ will be used to denote a
calculation with source smearing $\sigma=i$ and sink smearing
$\sigma=j$.  Varying the parameter $N_{\rm GS}$ over the values shown 
in Table~\ref{tab:4runs} did not impact any of
the results, so it is dropped from further discussions.  The notation 
V357 implies a $3 \times 3$ variational analysis with $\sigma=3,\ 5, \ 7$.

In this paper we present a detailed analysis with two goals: First, 
to demonstrate that high precision estimates for the
charges and the form-factors can be achieved cost effectively using
the all-mode-averaging (AMA) method~\cite{Blum:2012uh}.  The second
goal is to compare the 2-state fit to data at multiple $\tsep$ and variational
methods~\cite{Dragos:2015ocy} to determine the best strategy for
controlling excited-state contamination in the matrix elements. 

\subsection{Lattice Parameters of the 4 Calculations}
\label{sec:tuning}

We analyze 4 high statistics simulations (labeled runs
R1--R4) carried out on the $a081m312$ lattices. We explore the
efficacy of using quark propagators with different smearing parameters
to reduce the excited-state contamination and obtain estimates in the
$t_{\rm sep} \to \infty$ limit.  We compare three strategies for
reducing excited-state contamination: optimizing the smearing
parameters to reduce excited-state contamination in correlation
functions; using the 2-state fit to correlation functions with data at
multiple values of the source-sink separation $t_{\rm sep}$ (see
Sec.~\ref{sec:2state}); and the variational method using a matrix of
correlation functions constructed using up to three smearings as
discussed in Sec.~\ref{sec:var}. The lattice parameters used in these
four runs are summarized in Table~\ref{tab:4runs}.

These four runs allow us to make four comparisons to understand,
calculate and mitigate the excited-state contributions: (R1) a 2-state
fit to data with smearing $\sigma=5$ and $t_{\rm sep} = 10, 12, 14,
16, 18$; (R2) a variational calculation
with a $3\times 3$ matrix of correlation functions constructed using
$\sigma=3,5,7$ and $t_{\rm sep} = 12 \approx 1$~fm; (R3) a variational
calculation with a $3\times 3$ matrix of correlation functions
constructed using $\sigma=5,7,9$ and $t_{\rm sep} = 12$; and 
(R4) a 2-state fit to data with smearing $\sigma=9$ and
$t_{\rm sep} = 10, 12, 14, 16, 18$. In the analysis of runs R2 and R3, 
we also present results from the three $2 \times 2$
submatrices.

In each run, 96 low precision (LP) measurements were made on each of
the 443 configurations that are separated by 10 RHMC trajectories.  In
three of the four runs (R1, R3 and R4), we also carried out 3 high
precision (HP) measurements on each configuration to correct for
possible bias in the LP calculation. As shown later, we find no
significant indication of a bias in any of these three calculations,
so we did not perform HP measurements in the case of R2 and give the
mean values obtained from just the LP measurements as our final
estimates.

\begin{table*}
\centering
\begin{ruledtabular}
\begin{tabular}{l|c|c|c|c|c|c}
ID       & Method   & Analysis    &  Smearing Parameters                  & $t_{\rm sep}$   & LP      & HP    \\
\hline
R1       & AMA      & 2-state     &  $\{5,60\}$                           & 10,12,14,16,18  & 96      & 3     \\
\hline                                                                                      
R2       & LP       & VAR         &  $\{3,22\}$, $\{5,60\}$, $\{7,118\}$  & 12              & 96      &       \\
\hline                                                                                       
R3       & AMA      & VAR         &  $\{5,46\}$, $\{7,91\}$, $\{9,150\}$  & 12              & 96      & 3     \\
\hline                                                                                       
R4       & AMA      & 2-state     &  $\{9,150\}$                          & 10,12,14,16,18  & 96      & 3     \\
\end{tabular}
\end{ruledtabular}
\caption{Description of the four calculations (R1--R4) done to understand the
  dependence of the analysis on the smearing size $\sigma$, the efficacy of the
  variational method and the quality of the convergence of the 2-state
  fit using data at multiple source-sink separation $t_{\rm sep}$. The smearing 
  parameters are $\{\sigma, N_{\rm GS}\}$ as described in the text. AMA
  indicates that the bias in the LP measurements was corrected using 3
  HP measurements and Eq.~\protect\eqref{eq:2-3pt_AMA}.  VAR indicates
  that the full $3 \times 3$ matrix of correlation functions was
  calculated and a variational analysis performed as described in the
  text. Analysis using Eq.~\protect\eqref{eq:2pt_3pt} to fit data at
  multiple $t_{\rm sep}$ simultaneously is labeled ``2-state fit to data at multiple $t_{\rm sep}$ ''. 
  }
  \label{tab:4runs}
\end{table*}

\begin{figure}[tb]
     \includegraphics[width=0.95\linewidth]{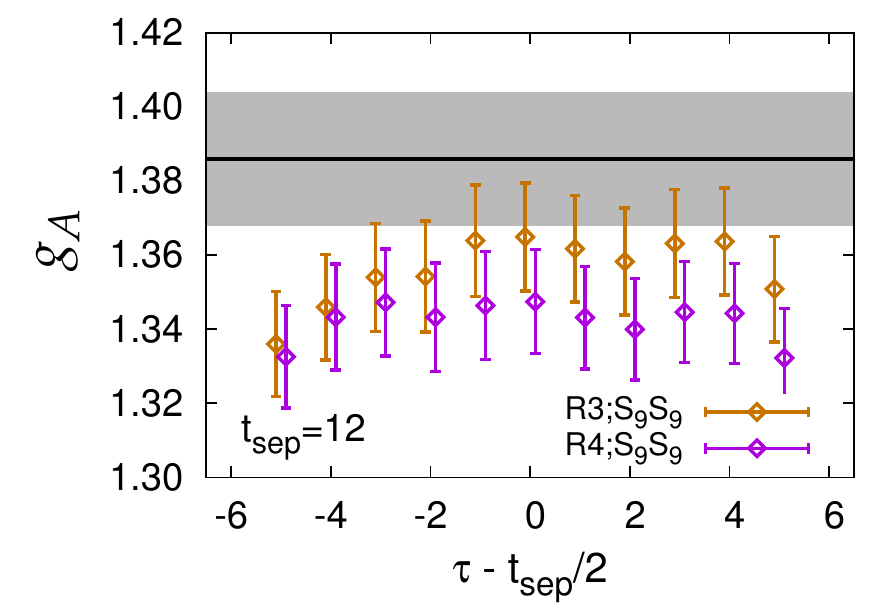}
 \caption{Comparison of estimates of unrenormalized $g_A$ obtained
   with $\sigma = 9$ and $t_{\rm sep} = 12$ from the two different
`   runs R3 and R4. The choice of the 96 LP source positions on each
   configuration is different in the two runs and the resulting
   difference in estimates is consistent with our estimate of
   statistical errors.  The gray error band and the solid line within
   it is the V579 variational estimate discussed in the text.  }
\label{fig:S9compareStats}
\end{figure}

We caution the reader that some of the measurements have been made
more than once in the different runs. The two calculations with
smearing parameters $\{5,60\}$ (R1) and $\{5,60\}$ (part of R2) are
identical.  The two sets $\{5,60\}$ and $\{7,118\}$ (part of R2) and
$\{5,46\}$ and $\{7,91\}$ (part of R3) differ in the number of
iterations $N_{\rm GS}$ of the Klein-Gordon smearing operator
and the choice of the location of the LP sources. Over the range
investigated, we find that the results are insensitive to the value of
$N_{\rm GS}$ and henceforth characterize the smearing by the single
parameter $\sigma$. Different choices of the 96 randomly selected LP
source positions on each configuration implies a different average
over the gauge fields and the resulting difference provides a check on
our estimation of the statistical errors. This is illustrated in
Fig.~\ref{fig:S9compareStats} using data for $g_A$ with $\sigma = 9$
and $t_{\rm sep} = 12$ obtained from runs R3 and R4. Our final
variational result V579 is also shown for comparison. We find that the
difference in results from the two choices of LP source positions is
comparable to our estimate of the statistical errors in the two
measurements and the error in the V579 estimate, i.e., our estimation
of errors is realistic.

\subsection{Correlation Functions}
\label{sec:CorrelationFunctions}

The interpolating operator $\chi$ used to create and annihilate the nucleon
state is 
\begin{align}
 \chi(x) = \epsilon^{abc} \left[ {q_1^a}^T(x) C \gamma_5 
            \frac{(1 \pm \gamma_4)}{2} q_2^b(x) \right] q_1^c(x)
\label{eq:nucl_op}
\end{align}
with color indices $\{a, b, c\}$, charge conjugation matrix $C = \gamma_0 \gamma_2$, and
the two different flavors of light quarks $q_1$ and $q_2$.
The non-relativistic projection $(1 \pm \gamma_4)/2$ is inserted to
improve the signal, with the plus and minus sign applied to the
forward and backward propagation in Euclidean time, respectively.

The 2-point and 3-point nucleon correlation functions at zero momentum 
are defined as 
\begin{align}
{\mathbf C}_{\alpha \beta}^{\text{2pt}}(t)
  &= \sum_{\mathbf{x}} 
   \langle 0 \vert \chi_\alpha(t, \mathbf{x}) \overline{\chi}_\beta(0, \mathbf{0}) 
   \vert 0 \rangle \,, 
\label{eq:corr_funs2}
\\
{\mathbf C}_{\Gamma; \alpha \beta}^{\text{3pt}}(t, \tau)
  &= \sum_{\mathbf{x}, \mathbf{x'}} 
  \langle 0 \vert \chi_\alpha(t, \mathbf{x}) \mathcal{O}_\Gamma(\tau, \mathbf{x'})
  \overline{\chi}_\beta(0, \mathbf{0}) 
   \vert 0 \rangle \,,
\label{eq:corr_funs3}
\end{align}
where $\alpha$ and $\beta$ are the spinor indices. The source
time slice $t_i$ is translated to $t_i=0$; $t_f=t$ is the sink
time slice; and $\tau$ is the time slice at which the bilinear operator
$\mathcal{O}_\Gamma^q(x) = \bar{q}(x) \Gamma q(x)$ is inserted. The
Dirac matrix $\Gamma$ is $1$, $\gamma_4$, $\gamma_i \gamma_5$ and
$\gamma_i \gamma_j$ for scalar (S), vector (V), axial (A) and tensor
(T) operators, respectively.
Here, subscripts $i$ and $j$ on gamma matrices run over $\{1,2,3\}$, 
with $i<j$. 

The charges $g_\Gamma^q$ in the nucleon state $\vert N(p, s) \rangle$ are  defined as 
\begin{align}
 \langle N(p, s) \vert \mathcal{O}_\Gamma^q \vert N(p, s) \rangle
 = g_\Gamma^q \bar{u}_s(p) \Gamma u_s(p)
\end{align}
with spinors satisfying
\begin{align}
 \sum_s u_s(\mathbf{p}) \bar{u}_s(\mathbf{p})  = {\qslash{p} + m_N} \,.
\end{align}

To analyze the data, we construct the projected 2- and 3-point correlation functions
\begin{align}
C^{\text{2pt}}(t) & = {\langle \Tr [ \mathcal{P}_\text{2pt} {\mathbf C}^{\text{2pt}}(t) ] \rangle} \nonumber \\
C_{\Gamma}^{\text{3pt}}(t, \tau)  & = \langle \Tr [ \mathcal{P}_{\rm 3pt} {\mathbf C}_{\Gamma}^{\text{3pt}}(t, \tau) ]\rangle \, .
 \label{eq:3pt_2pt_proj}
\end{align}
%
The operator $\mathcal{P}_\text{2pt} = (1+\gamma_4)/2$ is
used to project on to the positive parity contribution for the nucleon
propagating in the forward direction. For the connected 3-point
contributions, $\mathcal{P}_{\rm 3pt} =
\mathcal{P}_\text{2pt}(1+i\gamma_5\gamma_3)$ is used.  Note that the
3-point function in Eq.~\eqref{eq:3pt_2pt_proj} becomes zero if $\Gamma$
anti-commutes with $\gamma_4$, so only $\Gamma = 1$, $\gamma_4$,
$\gamma_i \gamma_5$ and $\gamma_i \gamma_j$ elements of the Clifford
algebra survive.  To
extract the charges, we make 2-state fits to the 2- and 3-point
correlation functions defined in Eq.~\eqref{eq:3pt_2pt_proj} 
as described next.

%
\subsection{Behavior of the Correlation Functions}
\label{sec:2state}

Our goal is to extract the matrix elements of the various bilinear
quark operators between ground state nucleons. The lattice operator
$\chi$, given in Eq.~\eqref{eq:nucl_op}, couples not only to the
nucleon but to all its excitations and multiparticle states with the same
quantum numbers that are allowed on the lattice. The correlation
functions, therefore, get contributions from all these intermediate
states. Using spectral decomposition, the behavior of the 2- and
3-point functions is given by the expansion:
\begin{align}
C^\text{2pt}
  &(t_f,t_i) = \nonumber \\
  &{|{\cal A}_0|}^2 e^{-M_0 (t_f-t_i)} + {|{\cal A}_1|}^2 e^{-M_1 (t_f-t_i)}  \nonumber \\
  &+ \ldots \,,\nonumber \\
C^\text{3pt}_{\Gamma}&(t_f,\tau,t_i) = \nonumber\\
  & |{\cal A}_0|^2 \langle 0 | \mathcal{O}_\Gamma | 0 \rangle  e^{-M_0 t_{\rm sep}} +{}\nonumber\\
  & |{\cal A}_1|^2 \langle 1 | \mathcal{O}_\Gamma | 1 \rangle  e^{-M_1 t_{\rm sep}} +{}\nonumber\\
  & {\cal A}_0{\cal A}_1^* \langle 0 | \mathcal{O}_\Gamma | 1 \rangle  e^{-M_0 (\tau-t_i)} e^{-M_1 (t_f-\tau)} +{}\nonumber\\
  & {\cal A}_0^*{\cal A}_1 \langle 1 | \mathcal{O}_\Gamma | 0 \rangle  e^{-M_1 (\tau-t_i)} e^{-M_0 (t_f-\tau)} \nonumber \\
  & + \ldots \,,
\label{eq:2pt_3pt}
\end{align}
where we have shown all the contributions from the ground and one
excited state. For simplicity, all the source positions are
shifted to $t_i=0$, and in 3-point functions, the source-sink
separation $t_f - t_i \equiv \tsep$.  The states $|0\rangle$ and
$|1\rangle$ represent the ground and ``first'' excited nucleon states,
respectively. Throughout the paper it will be understood that, in
practice, fits using Eq.~\eqref{eq:2pt_3pt} lump the contributions of
all excited states into these two states, so demonstrating convergence
of the estimates with respect to $t_{\rm sep}$ is important.

To extract the charges $g_A$, $g_S$, $g_T$ and $g_V$, we
only need operator insertion at zero momentum, in which case ${\cal
  A}_0$ and ${\cal A}_1$ are real and the matrix element $\langle 0 |
\mathcal{O}_\Gamma | 1 \rangle = \langle 1 | \mathcal{O}_\Gamma | 0
\rangle$.\footnote{The charge $g_V$ is one for a conserved vector
  current.  The local vector operator we are using is not conserved
  and only $Z_V g_V = 1$.  In many of the calculations of interest we
  construct ratios $Z_\Gamma/Z_V$ and $g_\Gamma/g_V$ as they have a 
  better signal due to the cancellation of some of the systematic 
  errors~\protect\cite{Bhattacharya:2015wna}. We therefore include $g_V$
  in the analysis.} Thus, keeping one excited state in the analysis
requires extracting seven parameters from fits to the 2- and 3-point
functions.\footnote{Including a second excited state would introduce
  five additional parameters, $M_2$, ${\cal A}_2$, $\langle 0 |
  \mathcal{O}_\Gamma | 2 \rangle$, $\langle 1 | \mathcal{O}_\Gamma | 2
  \rangle$ and $\langle 2 | \mathcal{O}_\Gamma | 2 \rangle$.} We use
Eqs.~\eqref{eq:2pt_3pt} for the analysis of all the charges and
form factors and call it the ``2-state fit''.

Five of the seven parameters, $M_0$, $M_1$ and the three matrix
elements $\langle 0 | \mathcal{O}_\Gamma | 0 \rangle \equiv g_\Gamma$,
$\langle 0 | \mathcal{O}_\Gamma | 1 \rangle $ and $ \langle 1 |
\mathcal{O}_\Gamma | 1 \rangle$ are physical provided the
discretization errors and higher excited-state contaminations have
been removed.  The amplitudes ${\cal A}_0$ and ${\cal A}_1$ depend on
the choice of the interpolating nucleon operator and/or the smearing
parameters used to generate the smeared sources.  It is evident from
Eq.~\eqref{eq:2pt_3pt} that the ratio of the amplitudes, ${\cal
  A}_1/{\cal A}_0$, is the quantity to minimize in order to reduce
excited-state contamination as it determines the relative size of the
overlap of the nucleon operator with the first excited
state.\footnote{With increasing precision of data, we will be able to
  add additional states to the ansatz. The goal will then be to reduce all
  the higher state amplitudes, ${\cal A}_n/{\cal A}_0$, by tuning
  the nucleon interpolating operator.}

We first estimate the four parameters, $M_0$, $M_1$, ${\cal A}_0$ and
${\cal A}_1$ from the 2-point function data and then use these as
inputs in the extraction of matrix elements from fits to the 3-point
data.  Both of these fits, to 2- and 3-point data, are done within the
same jackknife process to take into account the correlations between
the errors.  We performed both correlated and uncorrelated fits to the
nucleon 2- and 3-point function data. In all cases in which the
correlated fits were stable under changes in the fit ranges the two
fits gave overlapping estimates.  The final analysis of the 2-point
function data used correlated $\chi^2$ fits. Since correlated fits to
3-point functions with multiple $t_{\rm sep}$ did not work in some
cases, we used uncorrelated $\chi^2$ for 3-point fits for uniformity.
The errors in both 2- and 3-point correlation functions have been
calculated using a single elimination jackknife method.

To extract the three matrix elements $\langle 0 | \mathcal{O}_\Gamma |
0 \rangle \equiv g_\Gamma$, $ \langle 1 | \mathcal{O}_\Gamma | 0
\rangle$ and $ \langle 1 | \mathcal{O}_\Gamma | 1 \rangle$ from the
3-point functions for each operator $\mathcal{O}_\Gamma =
\mathcal{O}_{A,S,T,V}$ insertion, we make one overall fit using the
data at all values of the operator insertion time $\tau$ and the
various source-sink separations $t_{\rm sep}$ using
Eq.~\eqref{eq:2pt_3pt}.  In practice, in all the fits, we neglect the
data on the 3 points on either end, adjacent to the source and the
sink, of the 3-point functions for each $t_{\rm sep} $ as they have
the largest excited-state contamination.  To the extent that the
central values of $\tau$ dominate the 2-state fit,
Eq.~\eqref{eq:2pt_3pt}, to data at a single $\tsep$, the contributions 
of all higher states vanish in the limit $t_{\rm sep} \to
\infty$. We extract this limit using the 2-state fit to data 
at multiple values of $\tsep$ in the range 0.8--1.4~fm. Also, as is
evident from Eq.~\eqref{eq:2pt_3pt}, the contribution of the matrix
element $\langle 1 | \mathcal{O}_\Gamma | 1 \rangle$ cannot be
isolated from fits to 3-point function data obtained at a single
finite value of $t_{\rm sep}$.

Post facto, using Eq.~\eqref{eq:2pt_3pt} and reliable estimates of
$\langle 0 | \mathcal{O}_\Gamma | 1 \rangle $, $ \langle 1 |
\mathcal{O}_\Gamma | 1 \rangle$, the mass gap $M_1-M_0$, and the ratio
${\cal A}_1/{\cal A}_0$ one can bound the size of the excited-state
contamination at central values of $\tau$ for a given source-sink
separation $t_{\rm sep}$.

\subsection{The variational Method}
\label{sec:var}

One can also reduce excited-state contamination by implementing a
variational analysis (see~\cite{Dragos:2015ocy} and references therein
for previous use of the variational method for calculating nucleon
matrix elements).\footnote{A different version of the variational
  method, in which the sequential propagator is calculated starting at
  the point of insertion of the operator, is discussed
  in~\protect\cite{Owen:2012ts}.  We have not explored the
  cost effectiveness of that approach.}  This can be done by
calculating 2-point and 3-point functions in two ways by (i) using a
basis of nucleon interpolating operators with different overlap with
the ground and excited states. The operator given in
Eq.~\eqref{eq:nucl_op} is one such operator.  (ii) Constructing
multiple correlation functions with the same interpolating operator
but defined with smeared quark fields using a number of different
smearing sizes. In this work, we explore the second method in runs R2
and R3. In each of these two runs, the calculation is done using three
different smearing parameters $S_i$ summarized in
Table~\ref{tab:4runs}.  The 2-point correlation function for the
nucleon at each time $t$ is then a $3 \times 3$ matrix, $G^{\rm
  2pt}_{ij}(t)$, made up of correlation functions defined in
Eqs.~\eqref{eq:corr_funs2},~\eqref{eq:corr_funs3}
and~\eqref{eq:2pt_3pt} with source smearing $S_i$ and sink smearing $
S_j$.  The best overlap with the ground state is given by the
eigenvector corresponding to the largest eigenvalue $\lambda_0$
obtained from the generalized eigenvalue relation~\cite{Fox:1981xz}:
\begin{eqnarray}
G^{\rm 2pt}(t+\Delta t) u_i = \lambda_i G^{\rm 2pt}(t) u_i \,, 
\label{eq:EV}
\end{eqnarray}
where $u_i$ are the eigenvectors with eigenvalues $\lambda_i$.  The
matrix $G^{\rm 2pt}(t)$ at each $t$ should be symmetric up to
  statistical fluctuations, so we symmetrize it by averaging the
  off-diagonal matrix elements.

\begin{figure*}
\centering
 \includegraphics[width=0.98\linewidth]{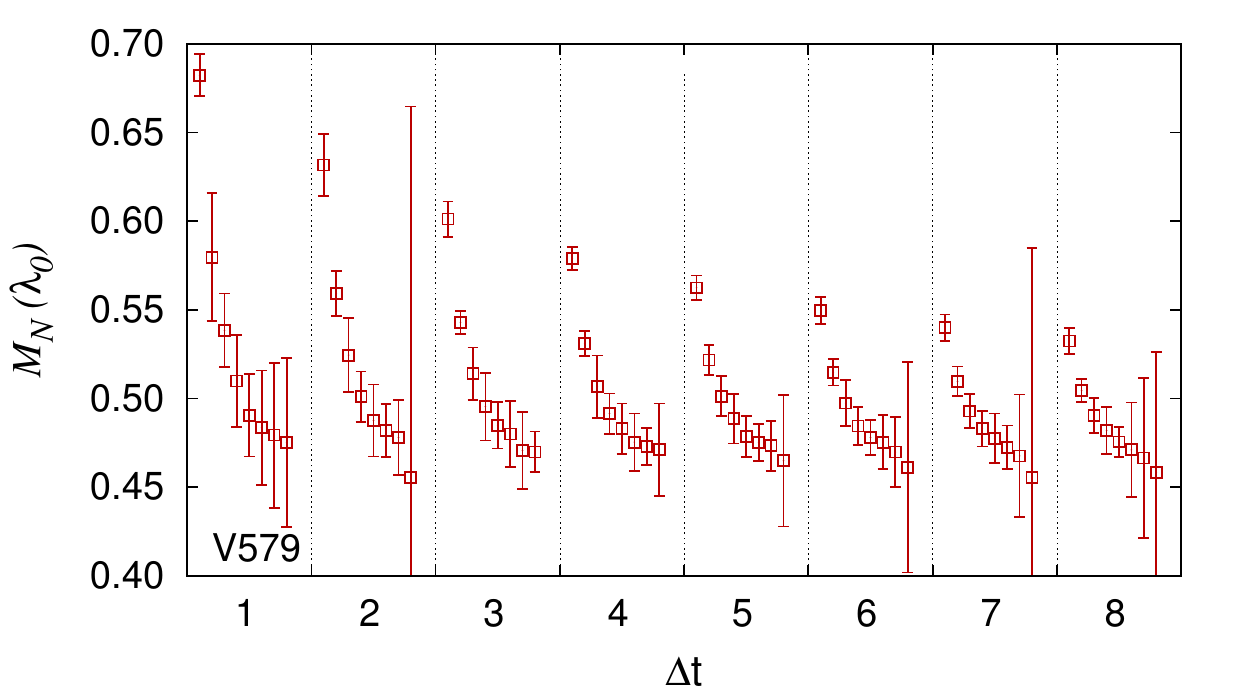}
\caption{Estimates of the nucleon mass from the largest eigenvalue of
  the $3 \times 3$ matrix V579 as a function of $t$ and $\Delta t$. 
  For clarity, the vertical dashed lines separate the sets of
  eight ($ t=1-8$) estimates for a given value of $\Delta t$. Data
  show that the asymptotic estimate $M_0 \approx 0.47$ given in
  Table~\protect\ref{tab:2ptfits8} is reached only for $t > 5$ and
  there is a significant decrease in the errors for $\Delta t > 2$.  }
  \label{fig:Mfromeigenvalue}
\end{figure*}

To select the $t$ and $\Delta t$ to use in the analysis, we show in
Fig.~\ref{fig:Mfromeigenvalue} the nucleon mass $M_N(\lambda_0) = -
(\ln \lambda_0)/\Delta t$ obtained from the ground state eigenvalue
$\lambda_0$ for a range of combinations.  The criteria we used for
choosing the $t$ and $\Delta t$ used in the final analysis are: (i)
the interval should be sensitive to both the ground and the excited
states, (ii) the correlation functions should exhibit a good
statistical signal over this range, (iii) the estimate of $M_N$ from
$\lambda_0$ should be close to the final estimate of the ground state
mass, and (iv) the resulting 2-state fit to the projected 2-point
function should have a small value for the ratio ${\cal A}_1/{\cal
  A}_0$.  Data in Fig.~\ref{fig:Mfromeigenvalue} show that $M_N$
starts to plateau towards its asymptotic value for $t \gsim 5$ and the
errors show a significant decrease for $\Delta t > 2$. These trends
still leave a number of ``equally'' good choices based on our four
criteria, for example, $t=6$ and $\Delta t=3$ or $t=5$ and $\Delta
t=4$.  We selected $t=6$ and $\Delta t = 3$.

\begin{figure*}[tb]
  \subfigure{
     \includegraphics[width=0.75\linewidth,trim={0 1.10cm 0 0},clip]{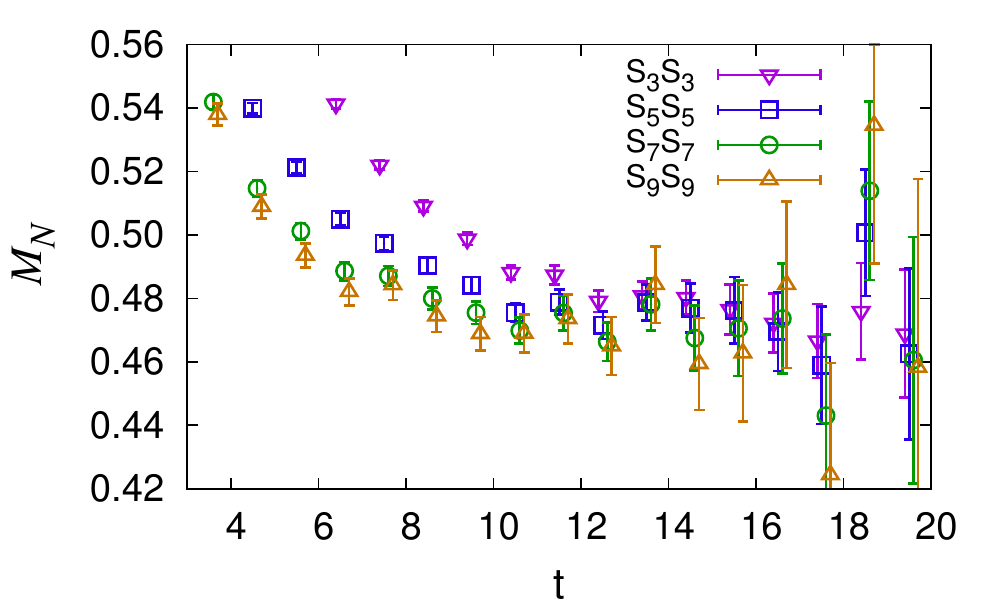}
  }
  \subfigure{
     \includegraphics[width=0.75\linewidth,trim={0 1.10cm 0 0},clip]{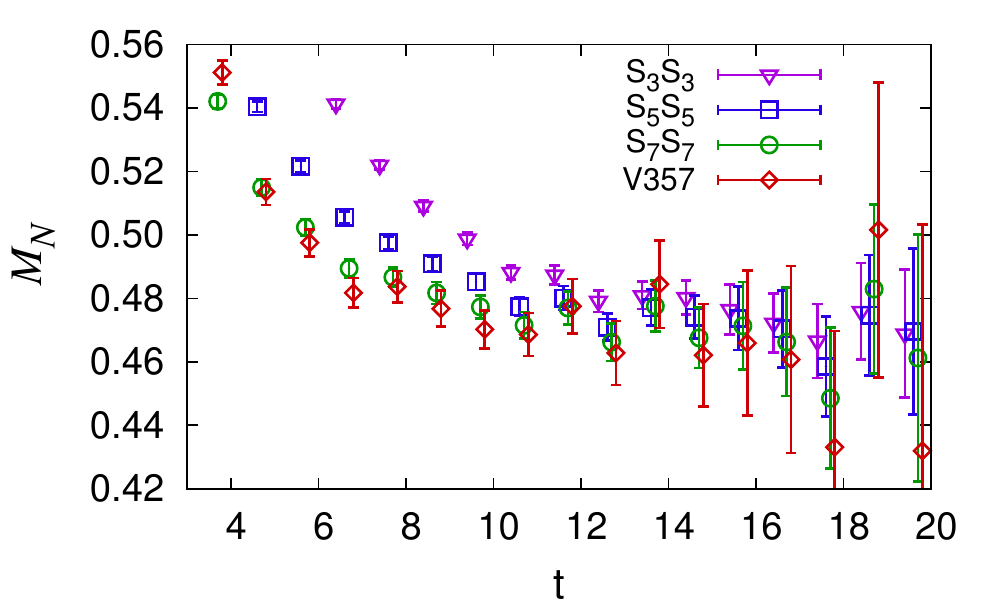}
  }
    \vspace{-0.01\linewidth}
  \subfigure{
     \includegraphics[width=0.75\linewidth]{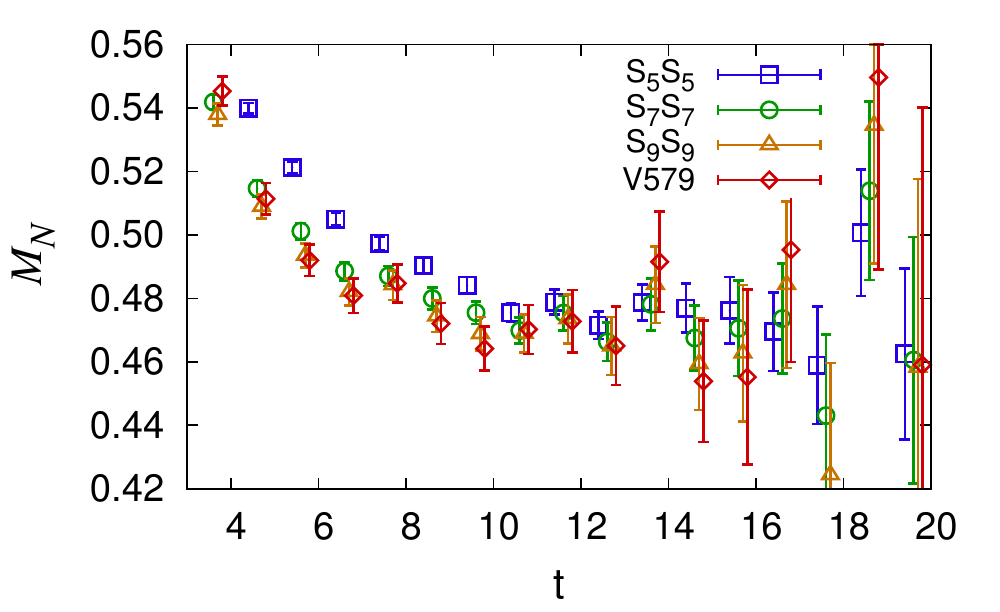}
  }
 \caption{Nucleon effective mass at zero momentum as a function of
   Euclidean time $t$.  (Top) Results for smearing size $\sigma = 3,
   5, 7, 9$; (Middle) comparison of $\sigma = 3, 5, 7$ single smearing
   data with the variational data V357; and (Bottom) comparison of
   $\sigma = 5, 7, 9$ single smearing data with the variational data
   V579.  }
\label{fig:N2ptV357}
\end{figure*}

\begin{figure}
\centering
 \includegraphics[width=0.98\linewidth]{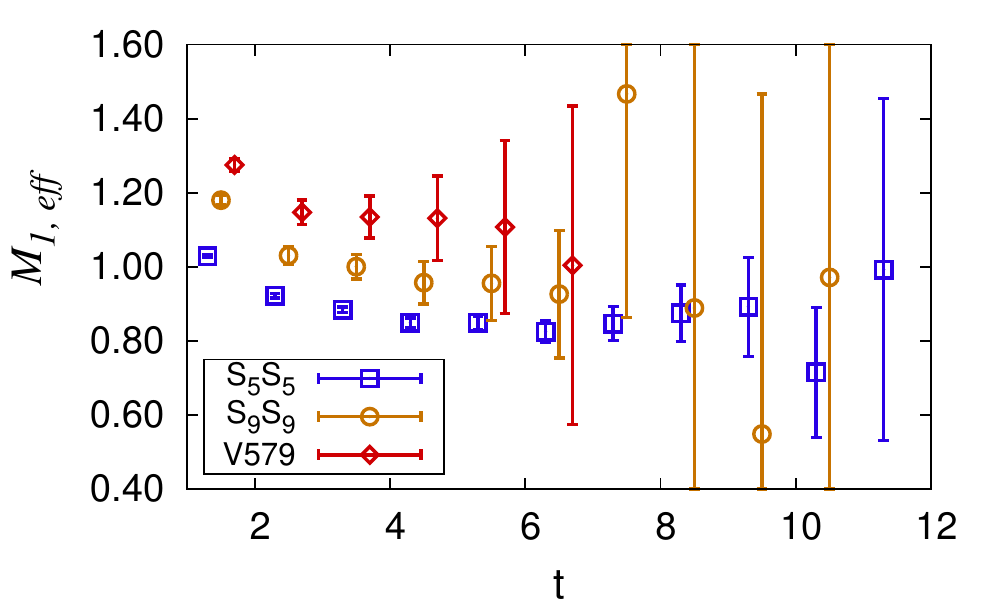}
\caption{Plot of the effective mass for the excited state evaluated
  from the $S_5 S_5$, $S_9 S_9$ and V579 nucleon correlation functions
  after subtraction of the respective ground state fit.  }
  \label{fig:M1eff}
\end{figure}

With a good estimate of $u_0$, the expectation is that the ground
state, in the projected functions $u_0^T G^{\rm 2pt}(t) u_0$ dominates
at earlier $t$. In Fig.~\ref{fig:N2ptV357}, we compare the behavior of
the nucleon effective mass obtained from correlation functions with
different smearing and with the projected variational V357 and V579
data.  We find that as the smearing size $\sigma$ is increased, the
plateau sets in at earlier time (top panel). The V357 data are a
little below $S_7 S_7$ (middle panel) while V579 overlap with $S_9
S_9$ (bottom panel).  In Fig.~\ref{fig:M1eff}, we compare the
effective mass plot for the excited state, i.e., that obtained by
subtracting the ground state result from the nucleon correlation
function.  Estimates of $M_1$ increase from $S_5 S_5$ to $S_9 S_9$ to
V579, indicating that the contribution of higher excited states
becomes larger as more of the first excited state is removed. Also,
the excited state signal in V579 dies out by $t \approx 8$. This
behavior of $M_1$ informed our choice $t =6$ and $\Delta t=3$ with which
we estimated the eigenvectors $u_i$.

\begin{figure*}
\centering
 \includegraphics[width=0.98\linewidth]{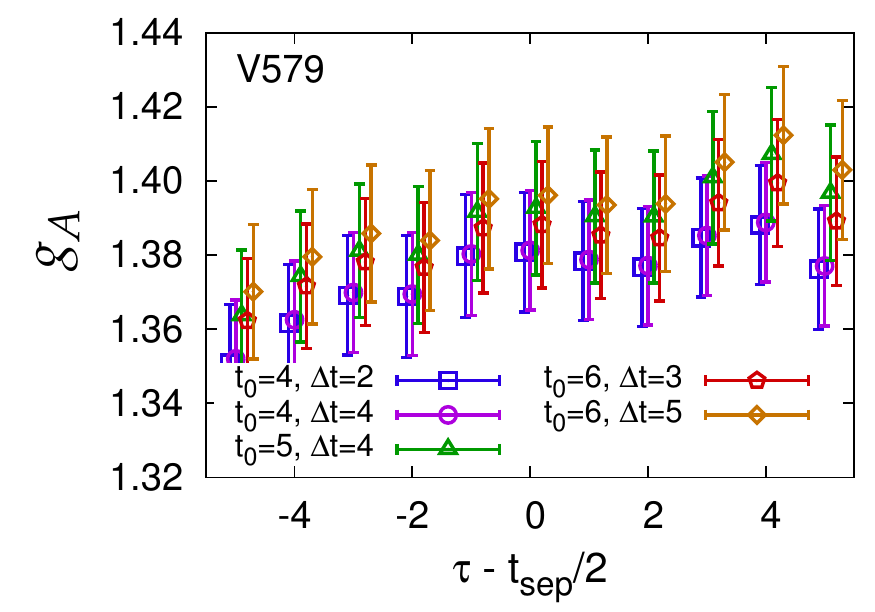}
\caption{Estimates of the unrenormalized $g_A$ from the V579 analysis for five
  representative values of $t$ and $\Delta t$ used to diagonalize the
  $3 \times 3$ 2-point matrix correlation function and obtain the
  eigenvector used in the diagonalization of the 3-point correlation
  matrix. All final results are obtained with the choice $t=6$ and
  $\Delta t=3$.  }
  \label{fig:gAvsDeltat}
\end{figure*}

Similarly, in the variational analysis for the 3-point functions
$C^{\rm 3pt}(\tau,t_{\rm sep})$, from which various charges are
extracted, the data at each $\tau$ and $\tsep$ are $3\times 3$
matrices. The ground state estimate is obtained by projecting these
matrices $G^{\rm 3pt}$ using the $u_0$ estimated from the 2-point
variational analysis, Eq.~\eqref{eq:EV}, i.e., $u_0^{T} G^{\rm
  3pt}(\tau,t_{\rm sep}) u_0$. We use the eigenvectors determined with
$t=6$ and $\Delta t=3$ for projecting the 3-point data at all
$\tau$. These projected data define the variational 3-point
function that is then fit using the 2-state ansatz given in
Eq.~\eqref{eq:2pt_3pt}, but with the $ \langle 1 | \mathcal{O}_\Gamma
| 1 \rangle$ term set to zero, to obtain the charges.  Note that the
eigenvectors $u_i$ do not depend on $t_{\rm sep}$. Also, we use the
same $u_0$ for all $\tau$.

To understand the sensitivity of this projected 3-point data to our
choice $t=6$ and $\Delta t=3$ for estimating $u_0$, we show $g_A$ data
for 5 representative combinations, that satisfy our selection criteria, 
in Fig.~\ref{fig:gAvsDeltat}.  We find that all five give estimates
are consistent and have errors of roughly the same size. Estimates from
the combination $\{t,\Delta t\} = \{4,2\}$ and $\{4,4\}$ are about
$0.5\sigma$ below the other three, $\{5,4\}$, $\{6,3\}$ and $\{6,5\}$. 
We consider the latter three to be equally good choices.

In the variational analysis carried out using data at a single $t_{\rm sep}$, 
the signal for a reduction in the excited-state
contamination in the projected correlation function 
is a larger flatter plateau, i.e., it 
should show less dependence on the operator
insertion time $\tau$ compared to a correlation function with the same
$t_{\rm sep}$ but with a single smeared source. We illustrate this
feature using the data from R2 for $g_A$ in
Fig.~\ref{fig:R2R3compVar}.\footnote{Note that the residual
  contribution of the matrix element $ \langle 1 | \mathcal{O}_\Gamma
  | 1 \rangle$ cannot be isolated from $ \langle 0 |
  \mathcal{O}_\Gamma | 0 \rangle$ by the 2-state fit to data at a
  single $t_{\rm sep}$. The effect of a non-zero $ \langle 1 |
  \mathcal{O}_\Gamma | 1 \rangle$ is to raise or lower all the data
  points but not change the curvature.}  The four variational
estimates have a larger plateau and a larger value compared to $S_5
S_5$ with $t_{\rm sep} = 12$. This improvement is less obvious when
comparing V579 to the $S_9 S_9$ data because, as discussed in
Sec.~\ref{sec:excited}, $S_9 S_9$ has much smaller contributions from
the excited states and has a plateau comparable in extent to V579.

If $ \langle 0 | \mathcal{O}_\Gamma | 1 \rangle$ is the dominant
contamination, one can also set up and solve an optimization condition
using the $3 \times 3$ matrix of 3-point data $M(\tau) \equiv
\Tr[{\cal P}_\Gamma C_\Gamma^{\rm 3pt}(t_{\rm sep}, \tau)]$. In this
case, one needs to determine the projection vector $\zeta$ such that
$\zeta^{T} M(\tau) \zeta$ is insensitive to $\tau$.  Again, to be
sensitive to excited states in the determination of $\zeta$, one needs
to choose $\tau$ in a region where the excited-state effect is
significant. Also, a good estimate of $\zeta$ should make the
projected correlation function flatter.  This analysis, in general,
needs to be done separately for each charge.  We have not carried out
this more elaborate analysis.

\subsection{Test of the coherent sequential source method}
\label{sec:coherent}

The coherent sequential source method is a technique to reduce
computational cost in the connected 3-point
functions~\cite{Bratt:2010jn}. It relies on the observation that for a
large enough lattice independent measurements can be made using a
distributed array of sources. Then, instead of calculating a separate
sequential propagator from each sink, a single coherent sequential
propagator may be calculated from the sum of all the sink source points.

\begin{figure}
\centering
 \includegraphics[width=0.98\linewidth]{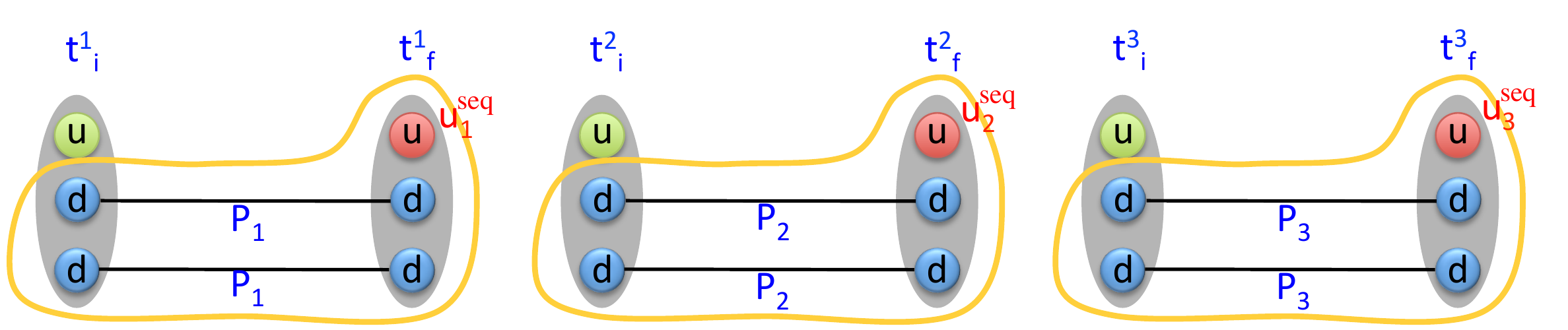}
\caption{Illustration of the construction of the sequential sources,
  $u_i$, for each spin and color component of the $u$ quark in three
  well-separated regions of the lattice. The insertion of the neutron 
  at each of the three sink time slices $t_f$ is done using 
  quark propagators $P_i$ generated independently from three initial time slices
  $t_i$. The three sources, $u_i^{\rm seq}$, are then added to produce
  the coherent sequential source.  }
  \label{fig:3sources}
\end{figure}

In our calculations on the $a081m312$ lattices, the signal in the
nucleon 2-point function becomes poor for $t > 16$ as shown in
Fig.~\ref{fig:N2ptV357}. We, therefore, partition the lattice with
Euclidean time extent $T=64$ into three sublattices of length $21$
($(T/3)_{\rm int}$)~\cite{Yamazaki:2009zq}.  We calculate the 2- and 3-point functions on the
three sublattices of a given lattice in a single computer job. We
start by calculating three quark propagators from randomly selected source
positions on the time slices $t_i = r,\ r + 21$ and $ r + 42$,
where $r \in \{1-21\}$. (To decrease correlations, $r$ is offset by 9 time slices
between successive configurations). The three measurements of the
2-point functions are made using these three independently calculated
propagators.  The calculation of the 3-point functions is done by
inserting a zero-momentum nucleon state at Euclidean times $t_f = t_i
+ t_{\rm sep}$ using these propagators and the interpolating operator
given in Eq.~\eqref{eq:nucl_op}. These nucleon states at the three
sink time slices $t_f$ have uncontracted spin and color indices,
associated with either the $u$ or the $d$ quark in the nucleon
interpolating operator. These states are used as sources to generate the
corresponding $u$ and $d$ sequential propagators. An illustration 
of the construction of these three sources in different parts of the 
lattice is shown in Fig.~\ref{fig:3sources}.

\begin{figure}
\centering
 \includegraphics[width=0.98\linewidth]{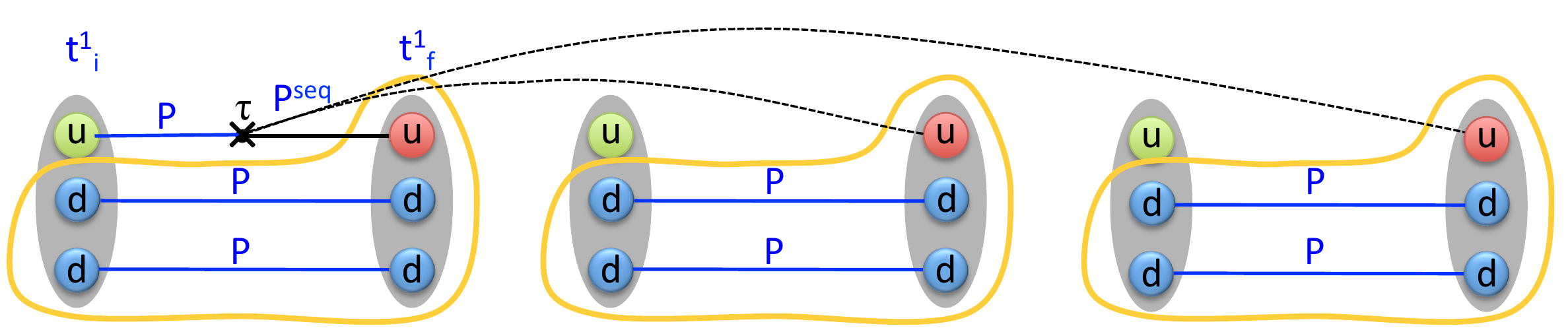}
\caption{Illustration of the construction of the 3-point function in
  the first of the three regions using the coherent sequential source
  propagator, $P^{\rm seq}$. The original propagator, $P$, from the
  source $u$ at $t_i^1$ is contracted with the quark bilinear operator at
  an intermediate time $\tau$  marked with a cross and $P^{\rm seq}$ from
  the sequential source $u$ at $t_f^1$. The contributions of the
  other two sources to $P^{\rm seq}$ are shown by the black dotted
  lines and average to zero by gauge invariance because the 3 sources are 
not connected by either gauge links or quark lines. }
  \label{fig:coherent}
\end{figure}

To obtain the 3-point function, this sequential propagator from $t_f =
t_i + t_{\rm sep}$ and the original propagator from $t_i$ are then
contracted with the operator at all intermediate time slices $\tau$
between $t_i$ and $t_f$ using Eq.~\eqref{eq:3pt_2pt_proj}.

In the coherent sequential source method the three regions of the
lattice are regarded as independent. Under this assumption, the three
$u$ ($d$) sources with nucleon insertion at $r+t_{\rm sep}$,
$r+21+t_{\rm sep}$ and $r+42+t_{\rm sep}$ can be added before the
inversion for creating the sequential $u$ ($d$) propagators,
respectively.  Such a summed source is called a coherent
source~\cite{Bratt:2010jn} and using it reduces the computational
cost from $N_{\rm meas}+2\times N_{\rm meas}$ to $N_{\rm meas} + 2$ inversions
when $N_{\rm meas}$ measurements are done at the same time on
different parts of the lattice.  

The contributions of a coherent
source in the region, say $r \le t \le r+t_{\rm sep}$, is illustrated
in Fig.~\ref{fig:coherent}. The contributions from the other two
sources, shown by dotted lines, to gauge invariant correlation
functions are formally zero on gauge averaging, however, they can
increase the statistical fluctuations. Therefore, one has to
demonstrate that for a finite statistical sample, the extra noise
introduced is small so that there is an overall reduction in
computational cost.  The magnitude of the noise, for fixed statistics,
is reduced by increasing the distance between the sources, which we
accomplish by choosing $N_{\rm meas}=3$ partitions on a lattice with $T=64$.

To validate the assumption that with our coherent source construction
and finite statistics, the measurements in the region, for example, $r
\le t \le r+t_{\rm sep}$ do not have significantly enhanced errors due
to contributions from the nucleon sources at $r+21 + t_{\rm sep}$ and
$r+42 + t_{\rm sep}$, i.e., their contribution averages to zero and
there is no significant increase in the error estimates, we simulated
100 configurations with the same parameters and source/sink locations
as Run 4 but without using the coherent source trick.  The data for
the four charges, summarized in Table~\ref{tab:coherent}, show that (i)
the difference in the mean values for the 3-point function data,
averaged over these 100 configurations, is smaller than the
statistical errors in all cases and (ii) there is no significant
difference in the error estimates with the coherent source
trick. Parenthetically, we remark that in the case of correlation
functions at large momenta (needed for the form factor calculations),
the differences in the means are as large as $30\%$, however, the
statistical errors in these data are $O(1)$. Note that any difference
or any additional noise in any of the correlation functions due to the
coherent source trick is even smaller in our final analysis with
the full set of 443 configurations.  

\begin{table}
\centering
\begin{ruledtabular}
\begin{tabular}{c|cccc}
Analysis   &  $g_A$       & $g_S$      & $g_T$     & $g_V$        \\
\hline                                                       
Coherent    &  1.368(50)   & 1.34(23)   & 1.132(44)  & 1.217(32)     \\
No Coherent &  1.377(47)   & 1.33(25)   & 1.138(44)  & 1.199(33)     \\
\end{tabular}
\end{ruledtabular}
\caption{Comparison of estimates for the four charges with and without the coherent 
  sequential source trick using the $S_9 S_9$ setup with a subset of 100 configurations. }
  \label{tab:coherent}
\end{table}

Our overall conclusion is that with a judicious partitioning of the
lattice with a large $T$ extent, the coherent sequential source method
does not give rise to a detectable increase in the statistical errors
for the charges. The reduction in the computational cost is significant: it
reduces the number of inversions from $N_{\rm meas} + 2 \times
N_{t_{\rm sep}} \times N_{\rm meas}$ to $N_{\rm meas} + 2 \times
N_{t_{\rm sep}}$, which for $N_{\rm meas}=3$ and $N_{t_{\rm sep}}=5$
is a reduction by a factor of 2.5.

\subsection{The AMA Method for High Statistics}
\label{sec:AMA}

To increase the statistics, given a fixed number of configurations,
the calculation was carried out using the all-mode-averaging (AMA)
technique~\cite{Blum:2012uh} with $96$ low precision (LP) and $3$ high
precision (HP) measurements, respectively. Also, the calculations used
the coherent sequential source method discussed in
Sec.~\ref{sec:coherent} to reduce the computational cost.  To implement
these methods, we carried out three measurements on a given
configuration at the same time. As discussed in
Sec.~\ref{sec:coherent}, the three starting source points were placed
on three time slices $t_i = r$, $r+21$ and $r+42$ and offset by 9
time slices between successive configurations to improve
decorrelations.

The locations of the 32 LP source points on each of these three time
slices $t_i$ were selected as follows to reduce correlations: the
first point was selected randomly and the remaining 31 points were
offset by multiples of $N_x=16$, $N_y=8$ and $N_z=8$. The resulting 96
LP estimates for 2- and 3-point functions from these sources are, 
{\it a priori}, biased since the Dirac matrix is inverted with a low
precision stopping criterion. To remove this bias, we place an
additional high precision (HP) source on each of the 3 time slices
from which we calculate both LP and HP correlation functions. Thus, in
our implementation of the AMA method, $93+3$ LP and 3 HP measurements
were done on each configuration for runs R1, R3 and R4. In R2, no HP
measurements were made and the results are averages over the 96 LP
measurements.

Using HP and LP correlators on each configuration, the
bias corrected 2- and 3- point functions are given by
\begin{align}
 C^\text{AMA}& 
 = \frac{1}{N_\text{LP}} \sum_{i=1}^{N_\text{LP}} 
    C_\text{LP}(\mathbf{x}_i^\text{LP}) \nonumber \\
  +& \frac{1}{N_\text{HP}} \sum_{i=1}^{N_\text{HP}} \left[
    C_\text{HP}(\mathbf{x}_i^\text{HP})
    - C_\text{LP}(\mathbf{x}_i^\text{HP})
    \right] \,,
  \label{eq:2-3pt_AMA}
\end{align}
where $C_\text{LP}$ and $C_\text{HP}$ are the correlation functions
calculated in LP and HP, respectively, and $\mathbf{x}_i^\text{LP}$
and $\mathbf{x}_i^\text{HP}$ are the two kinds of source positions.
The bias in the LP calculation (first term) is corrected by the second
term provided the correlation functions are translationally invariant, which the
2- and 3-point functions are.  If the algorithm used to invert the
Dirac matrix handles all modes well, $i.e.$ the HP and LP calculations
from the same source point are correlated, then the error in the AMA
estimate is dominated by the LP measurement and the bias
correction term does not significantly increase the error.

We used the multigrid algorithm for inverting the Dirac
matrix~\cite{Babich:2010qb} and set the low-accuracy stopping
criterion $r_{\rm LP} \equiv |{\rm residue}|_{\rm LP}/|{\rm source}| =
10^{-3}$ and the HP criterion to $r_{\rm HP} = 10^{-10}$.  To quantify
the bias, we have compared the AMA and LP estimates for both the 2-
and 3-point correlation functions themselves and for the seven fit
parameters $M_0$, $M_1$, ${\cal A}_0$, ${\cal A}_1$, $\langle 0 |
\mathcal{O}_\Gamma | 0 \rangle$, $\langle 0 | \mathcal{O}_\Gamma | 1
\rangle $ and $ \langle 1 | \mathcal{O}_\Gamma 1 \rangle$.  In each
case we find that the difference between the two is a tiny fraction
(few percent) of the statistical error in either.

\begin{figure*}[tb]
  \subfigure{
     \includegraphics[width=0.46\linewidth]{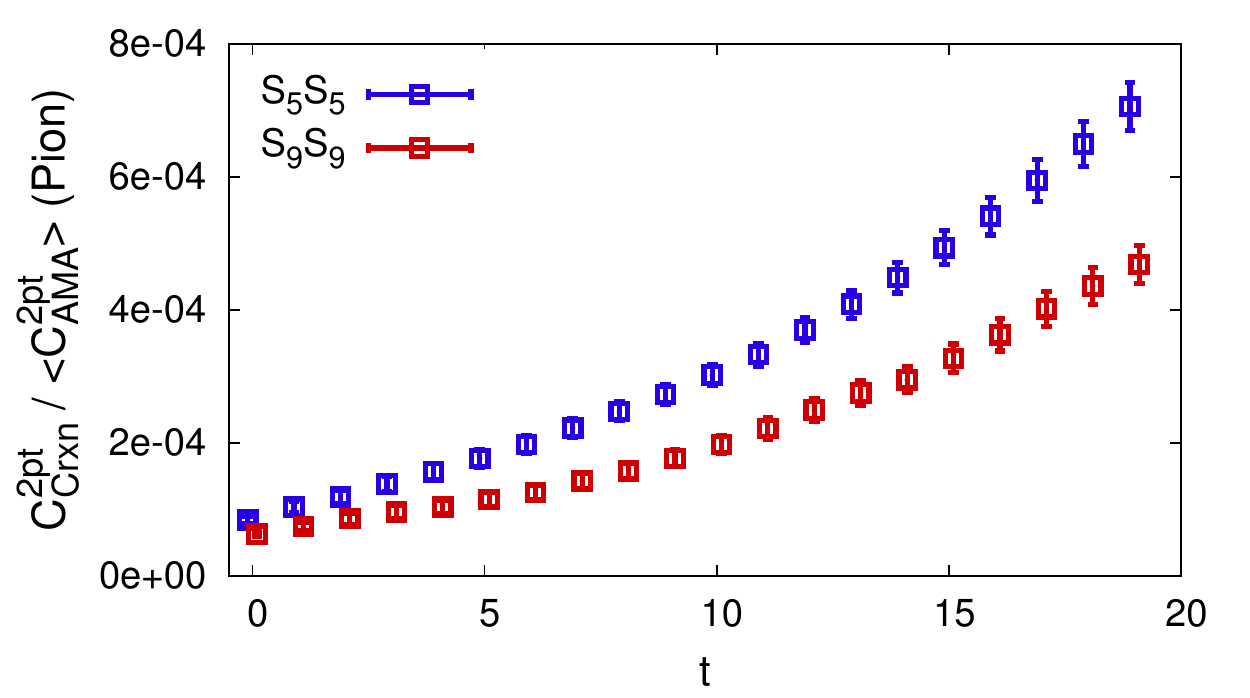}
     \includegraphics[width=0.46\linewidth]{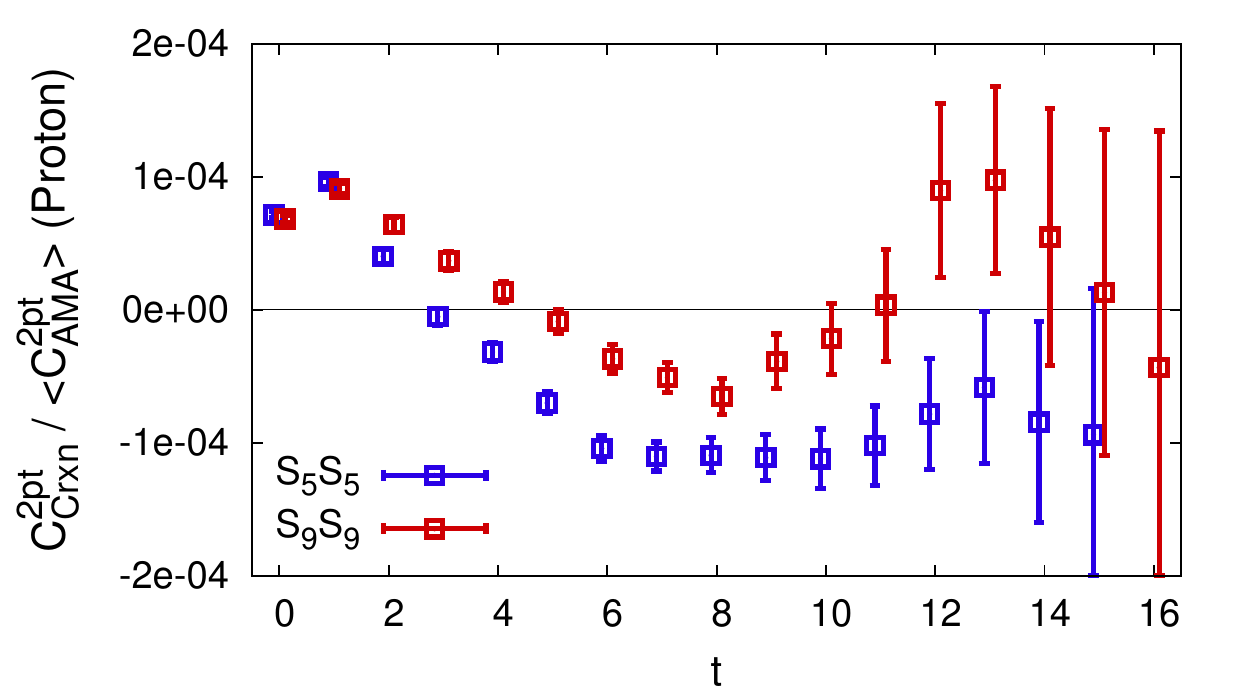}
  }
 \caption{The ratio of the bias correction term defined in
   Eq.~\protect\eqref{eq:2-3pt_AMA} to the AMA correlator as a
   function of Euclidean time $t$ for (left) the pion and (right)
   nucleon 2-point functions. The data are from runs R1 and R4 with
   $S_5 S_5$ and $S_9 S_9$, respectively.  }
\label{fig:bias2pt}
\end{figure*}

\begin{figure}[tb]
     \includegraphics[width=0.96\linewidth]{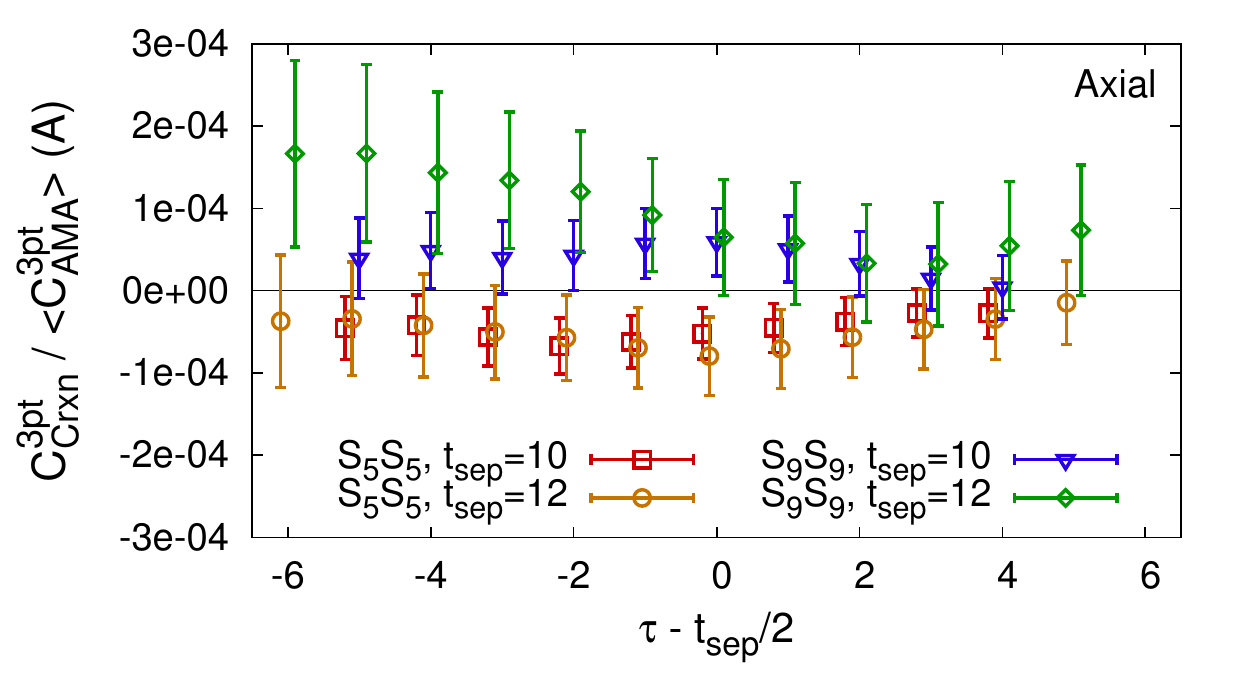}
     \includegraphics[width=0.96\linewidth]{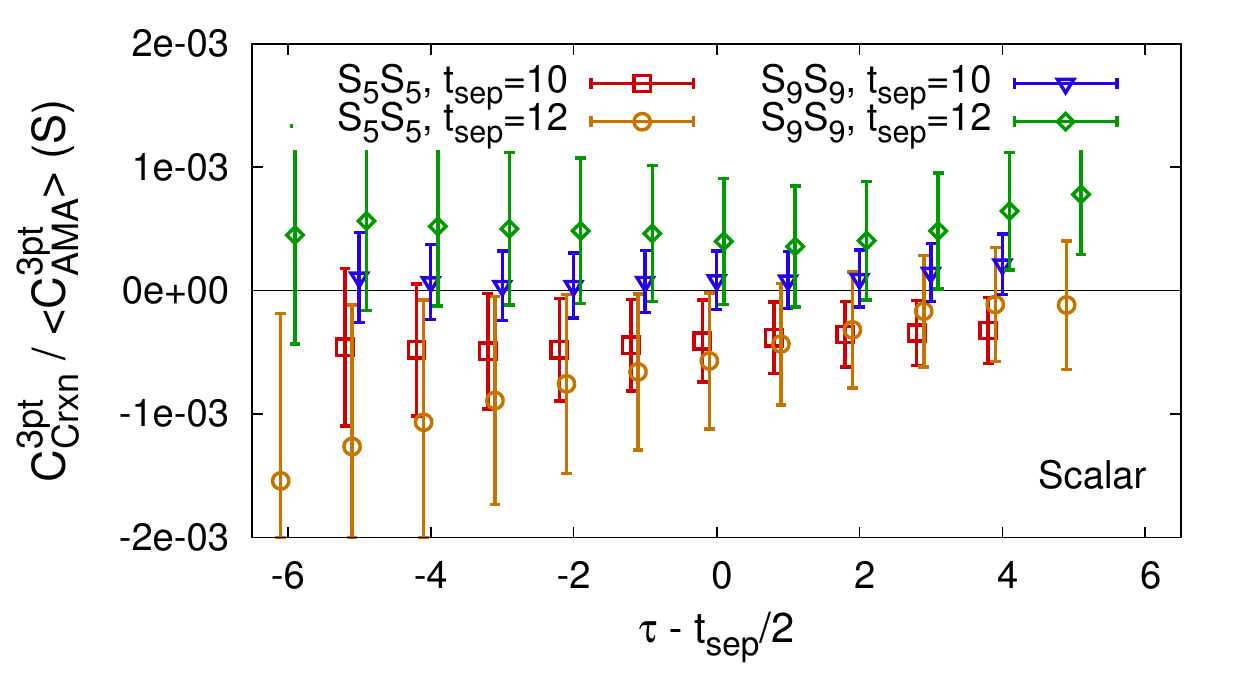}
     \includegraphics[width=0.96\linewidth]{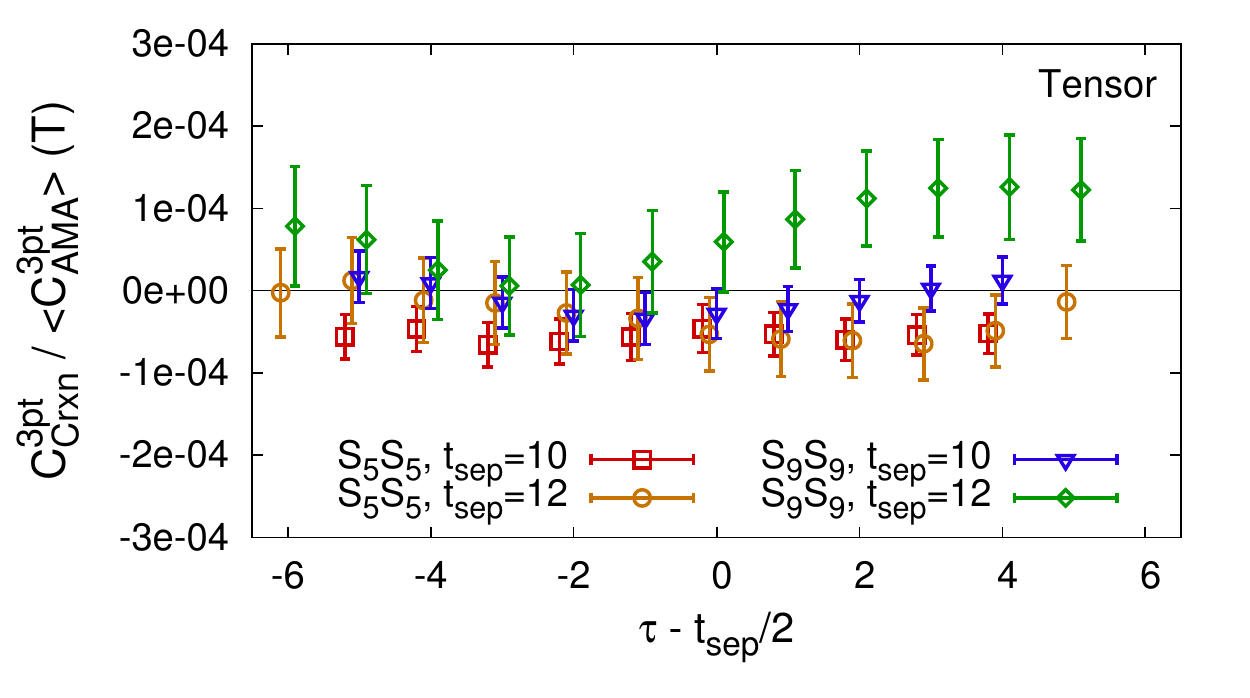}
     \includegraphics[width=0.96\linewidth]{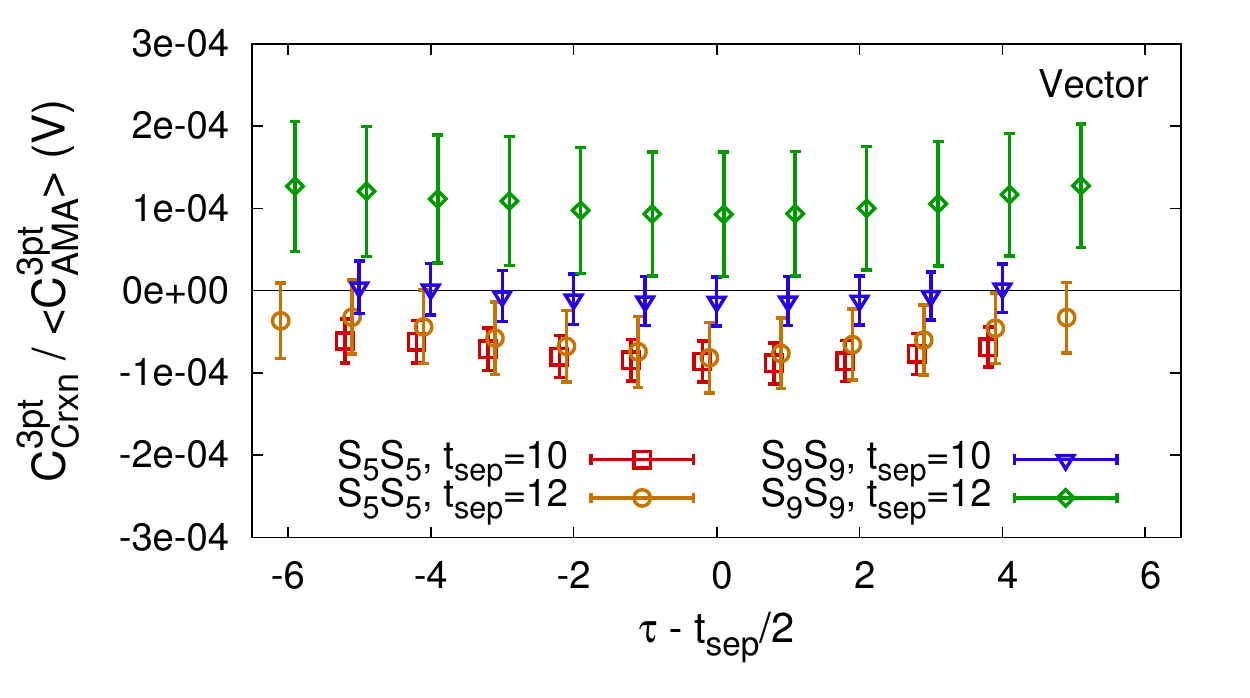}
 \caption{The ratio of the bias correction term defined in
   Eq.~\protect\eqref{eq:2-3pt_AMA} to the AMA correlator as a
   function of operator insertion time $\tau$ for the four charges. We show data
   from both runs R1 and R4. }
\label{fig:bias3pt}
\end{figure}

We illustrate the size and behavior of the bias correction term in the
pion and nucleon 2-point correlators as a ratio to the signal in
Fig.~\ref{fig:bias2pt}. In the case of the nucleon 2-point function we
find that the bias correction term is $\lsim 10^{-4}$ of the signal
for all $t$.  In the case of the pion 2-pt
function, which has the smallest errors and whose signal does not
degrade with $t$, the correction term grows with $t$ but remains $<
10^{-3}$ for $t < 25$. In Fig.~\ref{fig:bias3pt}, we show the data for
the four charges. In the cases of $g_A$, $g_T$ and $g_V$, the effect
is again $O(10^{-4})$. It is $O(10^{-3})$ for $g_S$ but in this case
the statistical errors are also correspondingly larger.  In
Table~\ref{tab:results}, we show that the results for the
unrenormalized charges with and without the bias correction term are
essentially identical. Based on such comparisons that have been
carried out for all the correlation functions, we conclude that any
possible bias in the LP calculations is negligible compared to our
current statistical errors.

%
%

\begin{figure*}[tb]
  \subfigure{
     \includegraphics[width=0.46\linewidth]{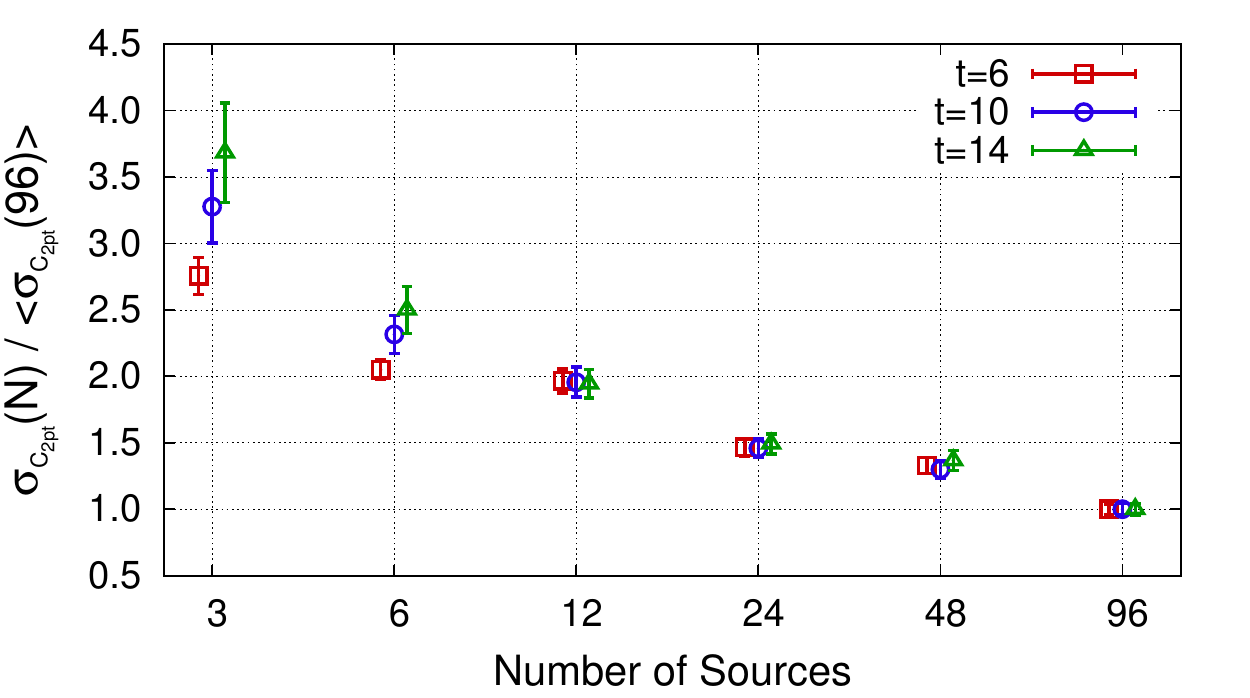}
     \includegraphics[width=0.46\linewidth]{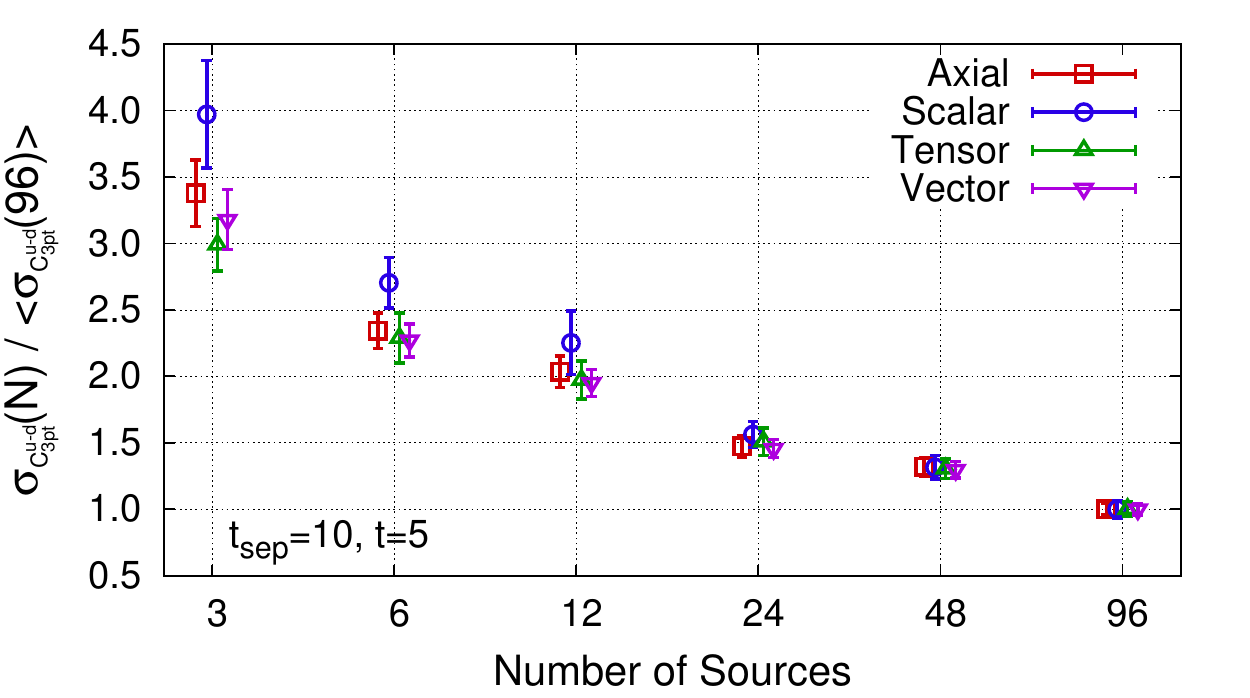}
  }
 \caption{(Left) The reduction in errors in the nucleon 2-point
   correlator as a function of the number of LP sources averaged per
   configuration. The data are shown for three different source-sink
   separations $t=6, 10, 14$. (Right) The ratio of errors in the four
   unrenormalized charges as a function on the number of LP sources analyzed. 
   The data are from run R4 with $S_9 S_9$. The data shown are at the midpoint
   $\tau=5$ of the $t_{\rm sep} = 10$ calculation. 
   In both figures, the error estimates from $N$ LP measurements
   are normalized by those from 96 LP measurements.  }
\label{fig:scaling}
\end{figure*}

In current lattice QCD simulations of nucleon charges and form
factors, the most computationally expensive part is the generation of
lattices. Thus one wants to extract the most precise results from a
fixed number of gauge configurations by having a large number of LP
measurements on each configuration. To consider the cost effectiveness
of the AMA method, we use the data presented in this work to compare
the decrease in errors with 96 LP + 3 HP measurements versus 35 HP
measurements. These two calculations have the same computational cost
on these lattices because one HP measurement takes the same time as
three LP ones.  As discussed above, since there is no detectable
difference in the values or errors between LP and HP measurements, we,
therefore, use the more extensive LP data to make this comparison. In
Fig.~\ref{fig:scaling}, we show the decrease in errors with the number
of LP measurements made on each gauge configuration for both the
2-point nucleon correlation function and the four charges. These
errors were calculated by first averaging over randomly selected 3, 6,
12, 24 or 48 of the 96 measurements on each configuration and then
performing a jackknife analysis over the 443 configurations.  We find
that the errors decrease by $\approx 1.4$ between LP=35 and 96, i.e.,
a gain in statistics by a factor of 2. The continued reduction in
errors up to $96$ LP measurements is what gives a factor of 2 saving
with the 96 LP + 3 HP over 35 HP measurements. This, post facto,
justifies using $O(100)$ measurements on each configuration. In a
related study~\cite{Bhattacharya:2016}, we found that at the physical
pion mass, one HP measurement costs as much as 17 LP ones with the
multigrid inverter. Thus, the cost effectiveness of the AMA method
increases very significantly as the light quark masses are lowered
towards their physical value.

Lastly, as discussed earlier, a second feature we incorporate in the
AMA calculation to improve statistical precision by reducing
correlations between measurements is to choose the source points
randomly within and between configurations.

Our conclusion is that already on $M_\pi=300$~MeV lattices, the AMA
method is a cost effective way to increase the statistics.  Our
results suggest a stronger statement for the calculation of nucleon
charges and form factors: with an inverter such as multigrid that does
not exhibit critical slowing down and becomes more efficient as the
quark mass is reduced, using $r_{\rm LP} = 10^{-3}$ as the stopping
criteria does not give rise to any significant bias compared to the 
statistical errors estimated from $O(100,000)$ measurements. The LP
measurement should, therefore, be considered unbiased at this level of
statistical precision and performing $O(100)$ measurements per 
configuration is cost effective. 

\section{Statistical Errors}
\label{sec:statistics}

In this section, we study the size of errors in 2- and
3-point correlation functions as a function of the smearing size
$\sigma$ and the source-sink separation $t_{\rm sep}$ and compare them
to those in the variational estimates.

\subsection{Statistical Errors in 2-point Functions}
\label{sec:2ptstats}

The nucleon 2-point correlation function was calculated 8 times over
the course of the four runs. The resulting values of the two masses
$M_0$ and $M_1$ and the amplitudes ${\cal A}_0$ and ${\cal A}_1$ are
given in Table~\ref{tab:2ptfits8} along with the fit range $t_{\rm min}-t_{\rm max}$. 
All the estimates for $M_0$ are consistent
within errors. Note that the two sets of $S_5 S_5$ and $S_7 S_7$ measurements
from runs R2 and R3 are different because different LP source
positions were used, i.e., the average over gauge field fluctuations is
different.  In both cases we find that the difference in the estimates
is smaller than the quoted statistical errors in either measurement. 

\begin{table*}
\centering
\begin{ruledtabular}
\begin{tabular}{c|c|cccc|c|c}
Type & Fit Range  & $aM_0$      & $aM_1$     & ${\cal A}_0^2$  & ${\cal A}_1^2$   & ${\cal A}_1^2/{\cal A}_0^2$  & $\chi^2/{\rm d.o.f.}$ \\
\hline
$S_5 S_5$  & 4--15  & 0.4717(38)  & 0.850(40)  & 2.85(13)e-08  & 3.45(19)e-08  & 1.212(59)  &  0.86    \\
\hline                                                                                              
$S_3 S_3$  & 6--20  & 0.4720(50)  & 0.844(41)  & 6.01(41)e-07  & 1.54(17)e-06  & 2.57(19)   &  0.79    \\
$S_5 S_5$  & 4--15  & 0.4717(38)  & 0.850(40)  & 2.85(13)e-08  & 3.45(19)e-08  & 1.211(59)  &  0.86    \\
$S_7 S_7$  & 4--15  & 0.4696(44)  & 0.855(83)  & 5.50(30)e-12  & 4.14(53)e-10  & 0.752(78)  &  0.60    \\
V357       & 2--14  & 0.4736(25)  & 1.194(47)  & 6.43(14)e-11  & 9.82(61)e-11  & 1.526(81)  &  0.59    \\
\hline                                                                                              
$S_5 S_5$  & 4--15  & 0.4709(40)  & 0.849(40)  & 2.80(14)e-08  & 3.41(18)e-08  & 1.219(60)  &  0.99    \\
$S_7 S_7$  & 4--15  & 0.4683(46)  & 0.854(83)  & 5.38(31)e-12  & 4.14(52)e-10  & 0.769(77)  &  0.67    \\
$S_9 S_9$  & 3--15  & 0.4700(32)  & 1.031(84)  & 4.70(15)e-12  & 4.48(66)e-12  & 0.95(12)   &  0.60    \\
V579       & 2--14  & 0.4710(27)  & 1.148(55)  & 1.316(32)e-12 & 1.73(13)e-12  & 1.316(83)  &  0.60    \\
\hline                                                                                             
$S_9 S_9$  & 4--15  & 0.4652(52)  & 0.87(12)   & 4.42(29)e-12  & 3.25(73)e-12  & 0.74(13)   & 0.81     \\
$S_9 S_9$  & 3--15  & 0.4682(35)  & 0.986(83)  & 4.59(17)e-12  & 4.27(57)e-12  & 0.93(10)   &  0.84    \\
$S_9 S_9$  & 2--15  & 0.4701(27)  & 1.061(48)  & 4.70(12)e-12  & 4.93(27)e-12  & 1.05(5)    &  0.88   \\
\end{tabular}
\end{ruledtabular}
\caption{Estimates of the masses $M_0$ and $M_1$ and the amplitudes
  ${\cal A}_0$ and ${\cal A}_1$ extracted from the fits to the 2-point
  correlation functions using the 2-state ansatz given
  in~\eqref{eq:2pt_3pt} and using the variational method. The data are
  organized by the four separate runs described in the text and
  Table~\protect\ref{tab:4runs}. The notation $S_3 S_3$ labels a
  nucleon correlation function with source and sink constructed using
  smearing parameter $\sigma=3$.  V357 stands for a $3 \times 3$
  variational analysis with smearings $\sigma=3,5,7$. We also give the
  $\chi^2/{\rm d.o.f.}$ for these fits obtained using the full covariance
  matrix. For $S_9 S_9 $ from R4, we give results with three different
  fit ranges to show sensitivity to $t_{\rm min}$.  }
  \label{tab:2ptfits8}
\end{table*}

In Fig.~\ref{fig:N2ptV357}, we compare the estimates for the effective
mass, $M_{N,eff}(t+0.5) = \ln (C^{\rm 2pt}(t)/C^{\rm 2pt}(t+1))$,
obtained from runs with different smearing parameters and with the
variational estimates.  Together with the results given in
Table~\ref{tab:2ptfits8}, we note that
\begin{itemize}
\item
the excited-state contamination decreases with $\sigma$ over the range studied and the
plateau sets in at earlier time slices, however, the errors in the data
increase with $\sigma$.
\item
The V357 estimate of $M_{N, {\rm eff}}$ lies below $S_7 S_7$ data and
the V579 values overlap with the $S_9 S_9$ data.  The errors in the
V357 variational data shown in Fig.~\ref{fig:N2ptV357} are larger than
in $S_7 S_7$ but the results of the fits shown in
Table~\ref{tab:2ptfits8} have smaller errors. The same is true for V579
versus the $S_9 S_9$ data. This is because, to get the final
estimates, the V357 and V579 data are fit with a smaller $t_{\rm min}$
as shown in Table~\ref{tab:2ptfits8}. 
\item
Estimates of $M_0$, using the 2-state fit and the variational
analysis, agree within errors in all cases as shown in
Table~\ref{tab:2ptfits8}.
\item
Estimates of $M_1$ from the individual 2-state fits agree, however,
the variational ansatz gives a significantly larger value.  This
feature is found to be independent of our choice of $t$ and $\Delta t$
in the construction of the variational ans\"atze.  This is because the
estimates are being extracted with a smaller $t_{\rm min}$, so the
contributions of the higher states are larger. One can see a similar
behavior in the $S_9 S_9$ estimates shown for three different fit
ranges in Table~\ref{tab:2ptfits8}. Also note that the errors in estimates 
from fits with a smaller $t_{\rm min}$ are smaller.
\item
The ratio ${\cal A}_1^2 / {\cal A}_0^2$, reducing which reduces the
excited-state contamination, is found to decrease on increasing the
smearing size from $\sigma=3$ to $\sigma=7$.  Our estimate for $S_9
S_9$ with our best fit-range 3---15 is larger than that for $S_7 S_7$,
but on using a common fit range, $4-15$, one finds a leveling off for
$\sigma \gsim 7$. This stabilization leads us to conclude that $\sigma
\approx 7$, or $\sigma \approx 0.57$~fm in physical units, is the best
compromise choice between reducing the ratio ${\cal A}_1 / {\cal A}_0$
and keeping the statistical errors small. 
\item
Two-state fits to the variational correlation functions are done with
an earlier starting time slice, as they have little sensitivity to the
excited-states beyond $t=6$ and become unstable for
$t_{\rm min} \gsim 4$.  Comparing the two variational runs, we note
that the ratio ${\cal A}_1^2/{\cal A}_0^2$ for V579 is smaller than
for V357, similar to the trend seen in the 2-state fit.
\item
The data in Table~\ref{tab:2ptfits8} show that ${\cal A}_1^2/{\cal
  A}_0^2$ increases as $t_{\rm min}$ is decreased. While, this pattern
is clear for each method, it is not obvious how to compare the values
between methods. Even for the same fit range, the value from the
variational method is significantly larger than that from the single
smearing 2-state fit even though the data suggest that the overall
excited state contamination in $M_0$ and the charges is smaller.  The
most likely explanation is that the contributions of higher states is
larger at small $t$ but these die off faster due to their larger
masses.
\end{itemize}

The bottom line is that the errors in $M_0$, $M_1$, ${\cal A}_0$ and
${\cal A}_1$ shown in Table~\ref{tab:2ptfits8} are sensitive to the fit
range, which in turn depends on $\sigma$. As the excited-state
contamination is reduced, fits can be made with an earlier starting
time $t_{\rm min}$ and the errors in $M_0$ and $M_1$ become
smaller. However, with a smaller $t_{\rm min}$, the estimated $M_1$
and the ratio of amplitudes ${\cal A}_1^2/{\cal A}_0^2$ is larger,
most likely due to the larger contribution of the higher excited states at
short Euclidean times.  To get estimates for $M_1$ and ${\cal A}_1$
that are insensitive to the fit range will require much more precise
data to which a 3-state fit can be made.

\subsection{Statistical Errors in the 3-point Functions}
\label{sec:3ptstats}

\begin{figure*}[tb]
  \subfigure{
     \includegraphics[width=0.65\linewidth,trim={0 1.22cm 0 0},clip]{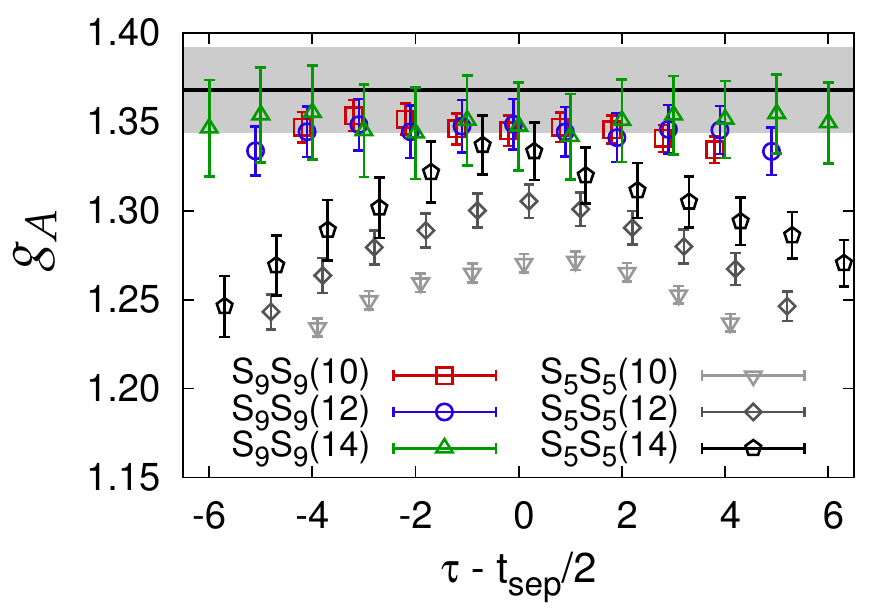}
  }
  \subfigure{
     \includegraphics[width=0.65\linewidth,trim={0 1.22cm 0 0},clip]{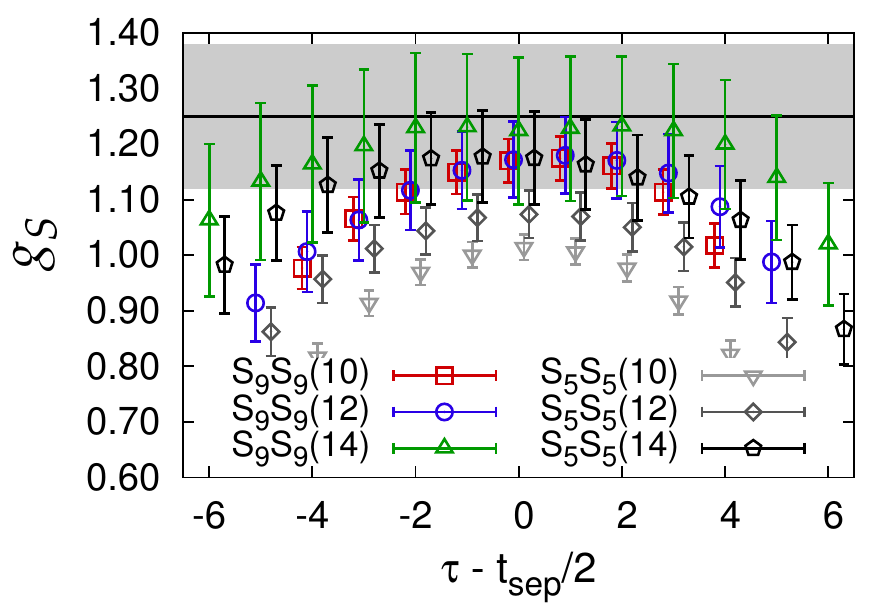}
  }
  \subfigure{
     \includegraphics[width=0.65\linewidth]{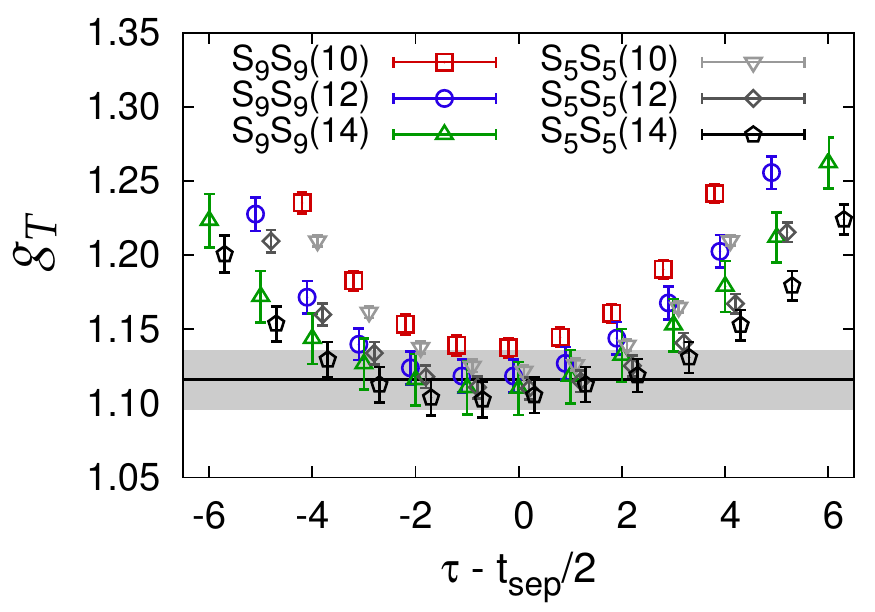}
  }
 \caption{Ratio plot comparing the statistical errors and excited-state
   contamination in the three unrenormalized isovector charges between runs 
   R1 ($S_5 S_5$) and R4 ($S_9 S_9$) for values of $t_{\rm sep} = 10,
   12, 14$ shown within parenthesis. The error band and the solid line
   within it are the $t_{\rm sep} \to \infty$ results of fits to the
   $S_9 S_9$ data. In most cases, the data with the two different
   smearings start to overlap by $t_{\rm sep} = 14$. The errors in the 
   data with $t_{\rm sep} = 16$,  shown in Fig.~\protect\ref{fig:S5S9}, 
   are too large to confirm the convergence. }
\label{fig:3pterrors}
\end{figure*}

The errors in the charges are a combination of the statistical errors
in the data for the correlation functions and the uncertainty in the
fits used ($n-$state, fit-range, $\cdots$) to extract the matrix
elements. This is true in both methods: the 2-state fit and the
variational analysis. To exhibit the behavior of the charges as a
function of $\tsep$ and $\tau$, we show in Fig.~\ref{fig:3pterrors},
and in all similar figures henceforth, the data for the 3-point function
divided by the result of the 2-point fit, ${\cal A}_0^2 \exp{(-M_0
  \tsep)} + {\cal A}_1^2 \exp{(-M_1 \tsep)}$.  This construction of
the ``ratio'' plot is a variant of the standard method in which the
data for the 2-point function at appropriate $\tsep$, and not the
result of the fit, are used for the normalization.

In Fig.~\ref{fig:3pterrors}, we compare the $t_{\rm sep} = 10, 12, 14$
data for the isovector charges between R1 ($\sigma = 5$) and R4
($\sigma = 9$) runs.  We find that the excited-state contamination in
$g_A$ and $g_S$ is significantly reduced in the data with $\sigma =
9$, however, the errors are about 50\% larger on each $t_{\rm sep}$
when compared to the $\sigma = 5$ data. In the case of $g_T$, the
excited-state contamination at central values of $\tau$ is smaller
than $ 5\%$ in both cases with the $\sigma = 5$ data showing a
slightly smaller effect and smaller statistical errors.  The data also
show that the statistical errors increase by about $80\%$ for every
two units of $t_{\rm sep}$.  To first approximation, this holds for
all four charges and for both smearing sizes.  Thus, to reduce
computational cost, the goal is to tune methods to get the $t_{\rm
  sep} \to \infty $ estimate from simulations with the smallest
$t_{\rm sep}$.

\begin{figure*}
\centering
  \subfigure{
    \includegraphics[width=0.47\linewidth,trim={0 1.22cm 0 0},clip]{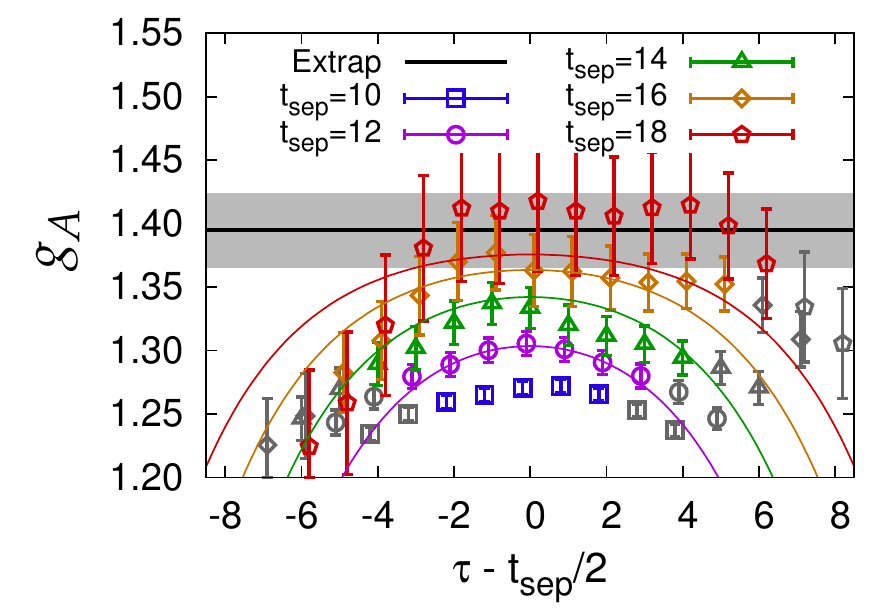}
  }
  \quad
  \subfigure{
    \includegraphics[width=0.43\linewidth,trim={0.8cm 1.22cm 0 0},clip]{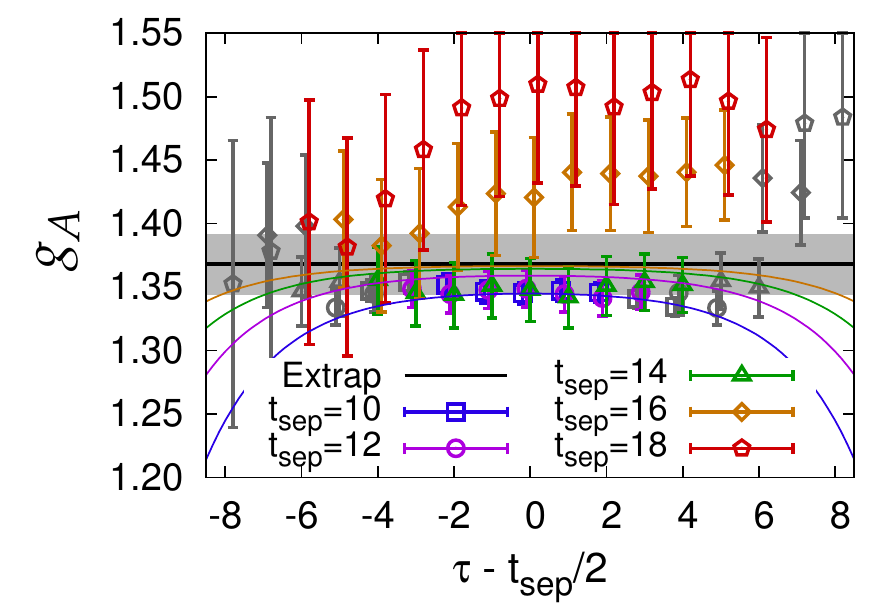}
  }
  \subfigure{
    \includegraphics[width=0.47\linewidth,trim={0 1.22cm 0 0},clip]{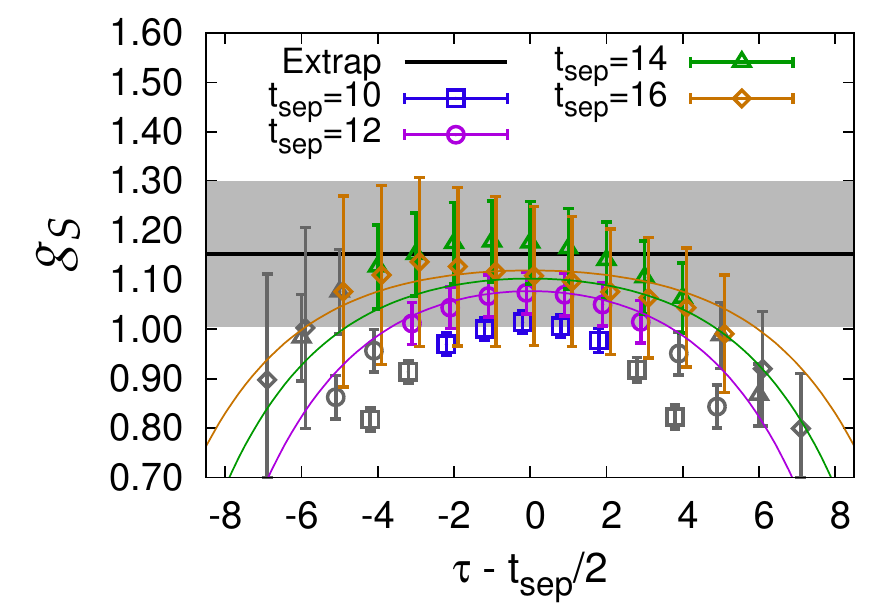}
  }
  \quad
  \subfigure{
    \includegraphics[width=0.43\linewidth,trim={0.8cm 1.22cm 0 0},clip]{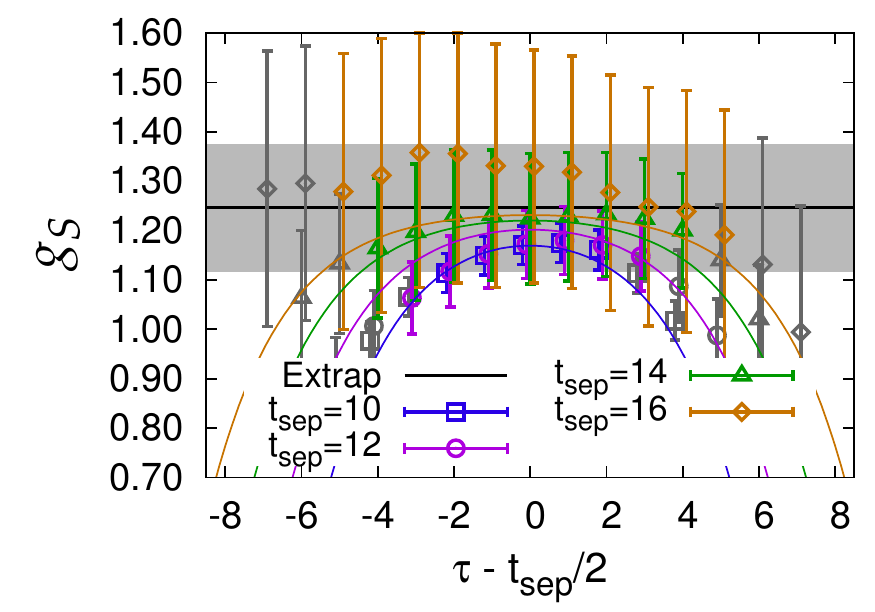}
  }
  \subfigure{
    \includegraphics[width=0.47\linewidth,trim={0 1.22cm 0 0},clip]{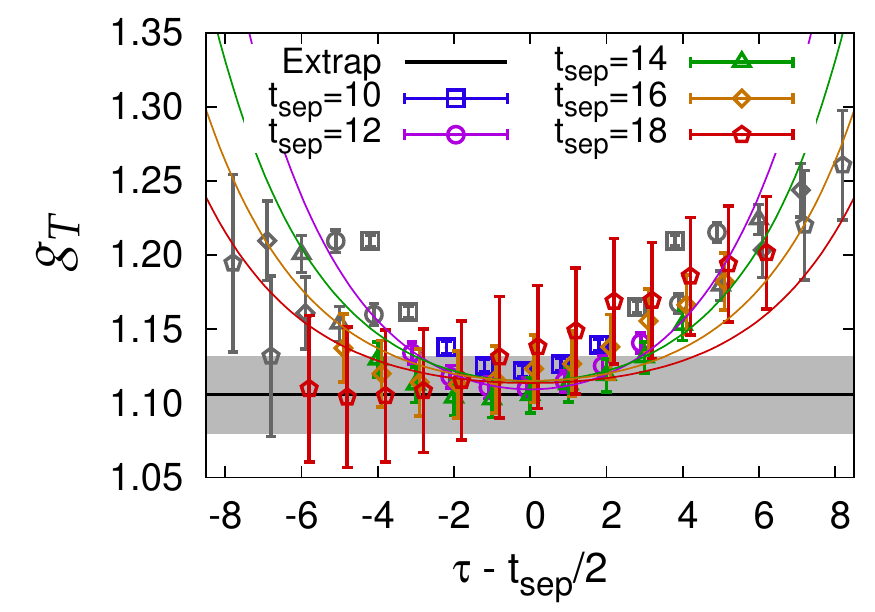}
  }
  \quad
  \subfigure{
    \includegraphics[width=0.43\linewidth,trim={0.8cm 1.22cm 0 0},clip]{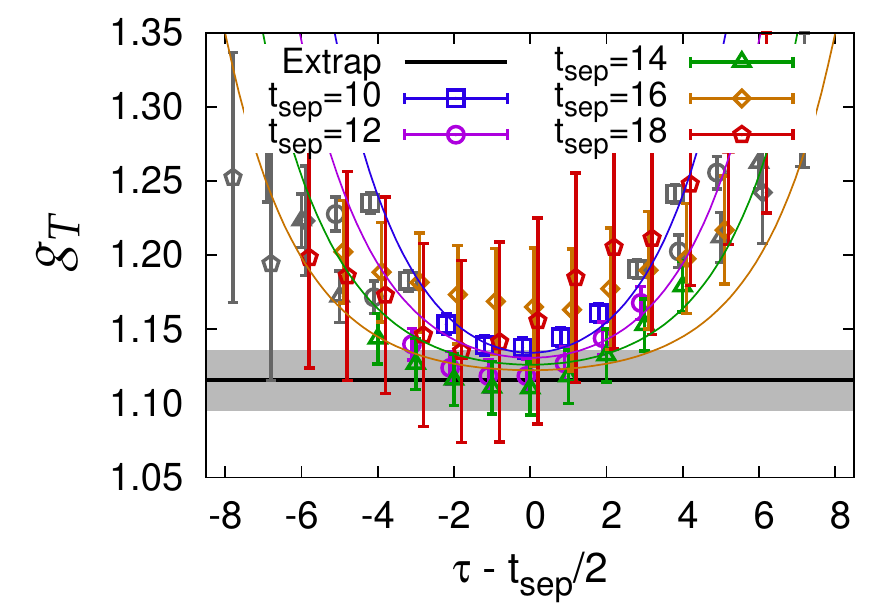}
  }
  \subfigure{
    \includegraphics[width=0.47\linewidth,trim={0 0 0 0},clip]{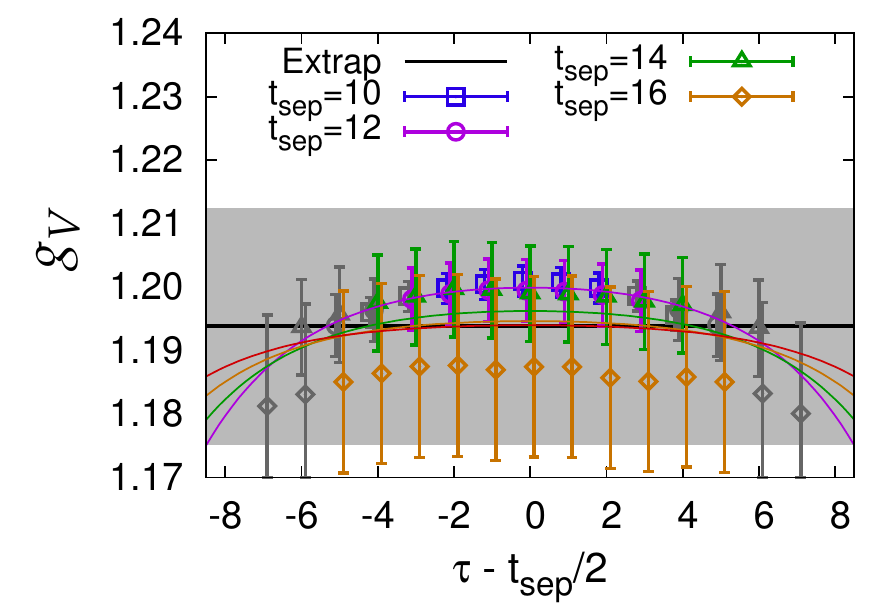}
  }
  \quad
  \subfigure{
    \includegraphics[width=0.43\linewidth,trim={0.8cm 0 0 0},clip]{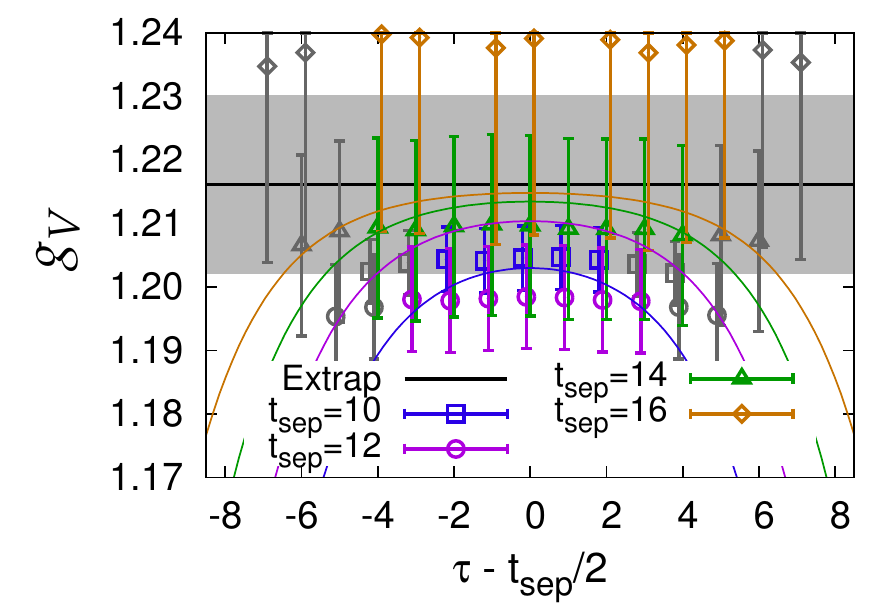}
  }
\caption{Data and fits for the unrenormalized isovector charges from
  runs R1 with $\sigma = 5$ (left) and R4 $\sigma = 9$ (right).  In
  each case, the gray band and the solid line within it is the $t_{\rm
    sep} \to \infty$ estimate obtained using the 2-state fit. The $S_5
  S_5$ fit used the $t_{\rm sep} = [12, 14, 16, 18]$ data while the $S_9
  S_9$ fit used the $t_{\rm sep} = [10, 12, 14, 16]$ data. To significantly
  improve the 2-state fit to $S_9 S_9$ requires at least doubling the
  statistics to reduce errors in the
  $t_{\rm sep} = 16$ and $ 18$ data.  }
\label{fig:S5S9}
\end{figure*}

In Fig.~\ref{fig:S5S9}, we extend this comparison to include the
results of the 2-state fit. We find that the two $t_{\rm sep} \to
\infty$ estimates, $S_5 S_5$ and $S_9 S_9$, overlap for all four
charges, and the final error estimates are comparable even though the
errors in the 3-point data $C^{\rm 3pt}(\tau,\tsep)$ for $S_9 S_9$ are
larger. Based on the observation that the 2-state fit to the $S_5 S_5$
data gives a reliable $t_{\rm sep} \to \infty $ estimate even though
the excited-state contamination is significant whereas the $S_9 S_9$
data show much smaller excited-state contamination but the fit is less
reliable as the data overlap and have larger errors, we again
conclude that $\sigma \approx 7$ is the best compromise choice for
reducing the excited-state contamination in these charges and having
small enough errors in the data at different $\tsep$ to give
confidence in a 2-state fit.  To improve the estimates from such
2-state fits, the statistical errors in the larger $t_{\rm sep}$ data
need to be reduced.

%

\section{Excited-State Contamination}
\label{sec:excited}

The overall goal is to get the best $t_{\rm sep} \to \infty$ estimates
on each ensemble for a given computational cost.  In this Section, we
investigate the efficacy of using different smearing parameters, the
2-state fit with data at multiple $t_{\rm sep}$ and a variational
analysis towards this goal. The final results for the charges are
given in Table~\ref{tab:results}. The overall observation is that 
for each of the four charges, all four estimates agree within $1 \sigma$, however,
the errors in the estimates from the variational analysis V357 (V579)
are about $60\%$ ($35\%$) smaller than those from $S_5 S_5$ ($S_9 S_9$),
respectively.

\begin{table}
\centering
\begin{ruledtabular}
\begin{tabular}{c|cccc}
Analysis   &  $g_A$       & $g_S$      & $g_T$     & $g_V$        \\
\hline
$S_5 S_5$  &  1.395(29)   & 1.15(15)   & 1.106(26)  & 1.194(19)     \\
$S_5 S_5$* &  1.395(29)   & 1.15(15)   & 1.106(26)  & 1.194(19)     \\
\hline                                                       
$S_9 S_9$  &  1.368(24)   & 1.25(13)   & 1.116(20)  & 1.216(14)     \\
$S_9 S_9$* &  1.369(24)   & 1.25(13)   & 1.116(20)  & 1.216(14)     \\
\hline                                                        
V35        &  1.365(13)   & 1.173(60)  & 1.123(10)  & 1.213(8)     \\
V37        &  1.375(15)   & 1.183(66)  & 1.114(11)  & 1.206(8)     \\
V57        &  1.381(16)   & 1.189(70)  & 1.112(12)  & 1.204(9)     \\
V357       &  1.386(16)   & 1.185(75)  & 1.116(13)  & 1.205(10)     \\
\hline                                                        
V57        &  1.373(16)   & 1.166(78)  & 1.108(13) & 1.207(10)     \\
V59        &  1.382(17)   & 1.202(84)  & 1.113(14) & 1.209(10)     \\
V79        &  1.385(18)   & 1.214(86)  & 1.115(15) & 1.210(11)     \\
V579       &  1.386(18)   & 1.220(87)  & 1.116(15) & 1.210(11)     \\
V579*      &  1.386(18)   & 1.220(87)  & 1.116(15) & 1.210(11)     \\
\end{tabular}
\end{ruledtabular}
\caption{Estimates of the unrenormalized charges from the four
  analyses.  The $S_5 S_5$ data are with fits to $t_{\rm sep} = 12,
  14, 16, 18$ and the $S_9 S_9$ data are with fits to $t_{\rm sep} =
  10, 12, 14, 16$. The variational results are from the analyses of
  the $3 \times 3$ V357 and V579 and their $2\times 2 $ subsets. The
  results marked with an asterisk are obtained from just the LP data
  and given here to show that the bias correction term in the 2- and
  3-point functions has negligible impact on final estimates of the
  charges.}
  \label{tab:results}
\end{table}

\begin{figure*}
\centering
  \subfigure{
    \includegraphics[width=0.47\linewidth,trim={0 1.22cm 0 0},clip]{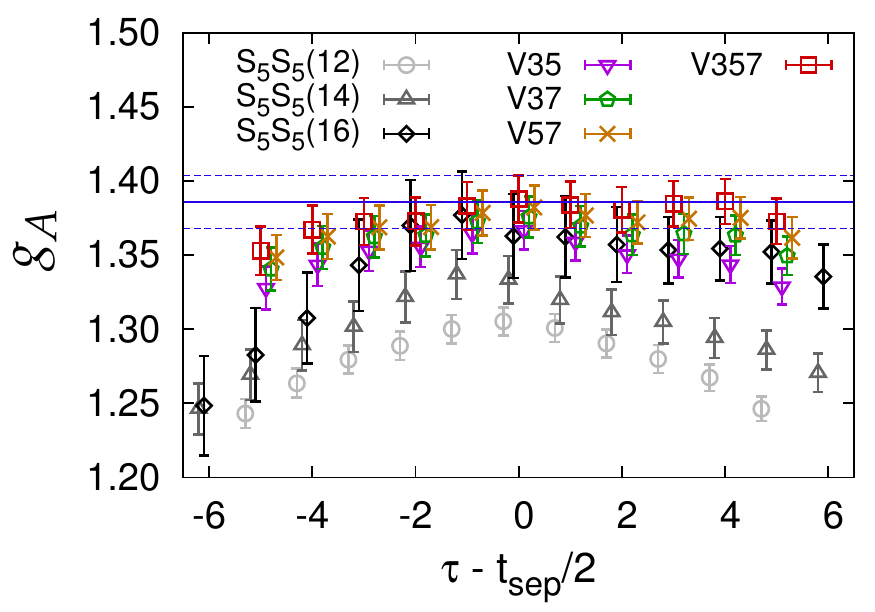}
  }
  \quad
  \subfigure{
    \includegraphics[width=0.43\linewidth,trim={0.8cm 1.22cm 0 0},clip]{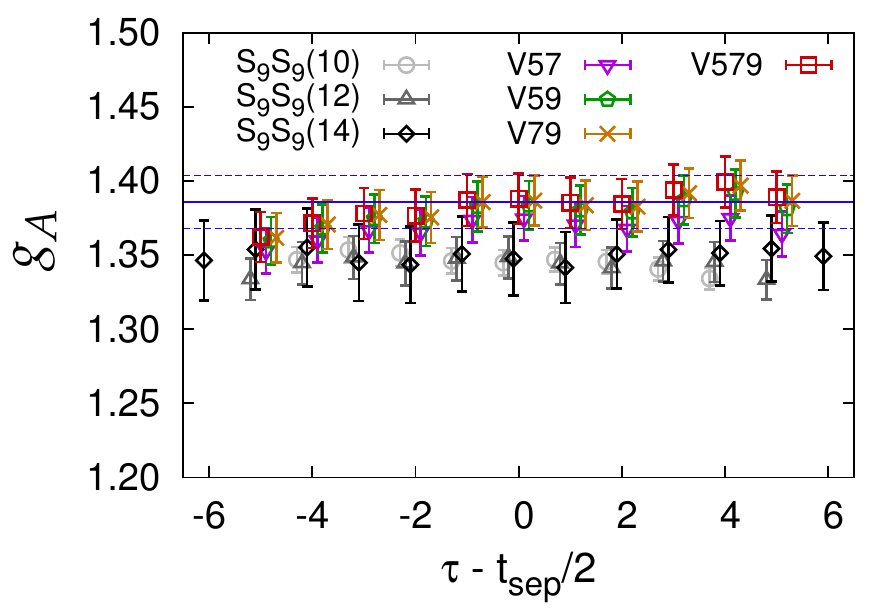}
  }
  \subfigure{
    \includegraphics[width=0.47\linewidth,trim={0 1.22cm 0 0},clip]{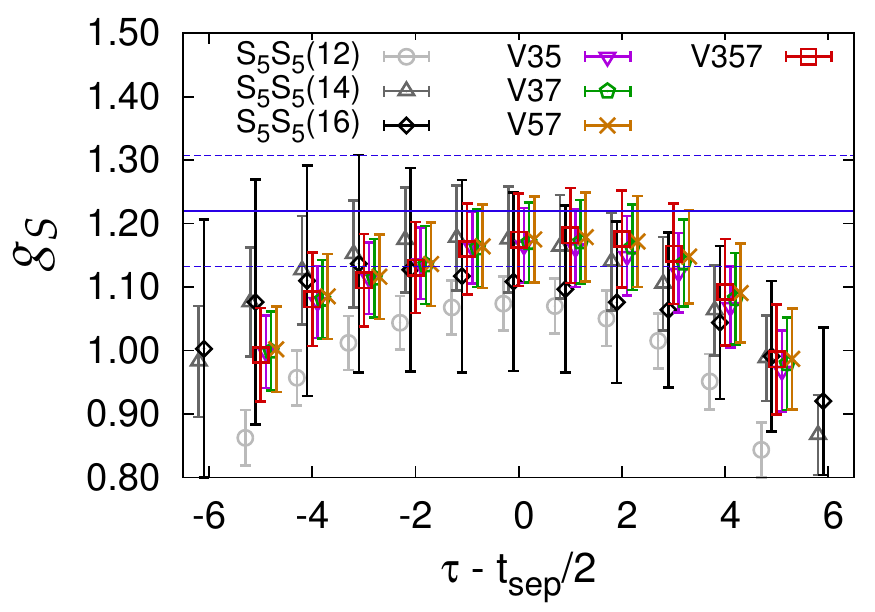}
  }
  \quad
  \subfigure{
    \includegraphics[width=0.43\linewidth,trim={0.8cm 1.22cm 0 0},clip]{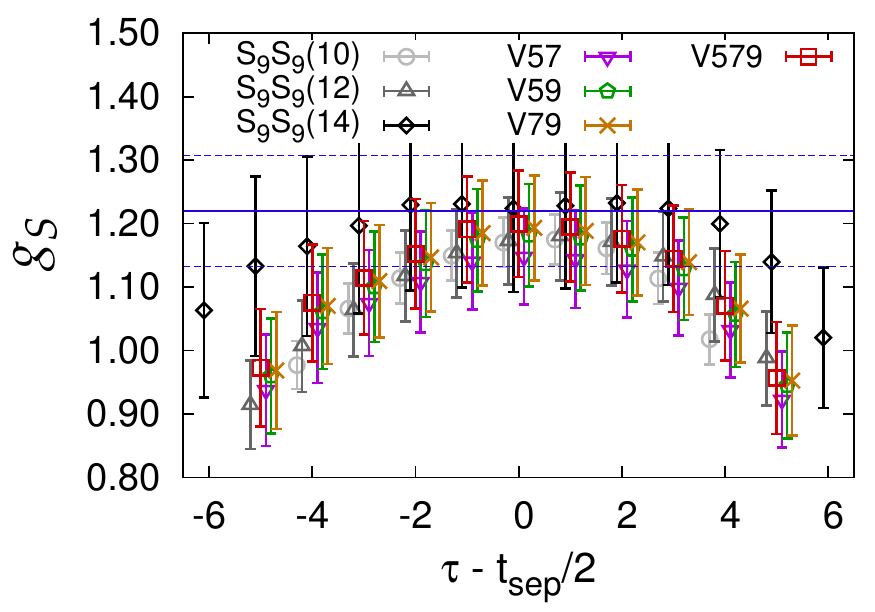}
  }
  \subfigure{
    \includegraphics[width=0.47\linewidth,trim={0 1.22cm 0 0},clip]{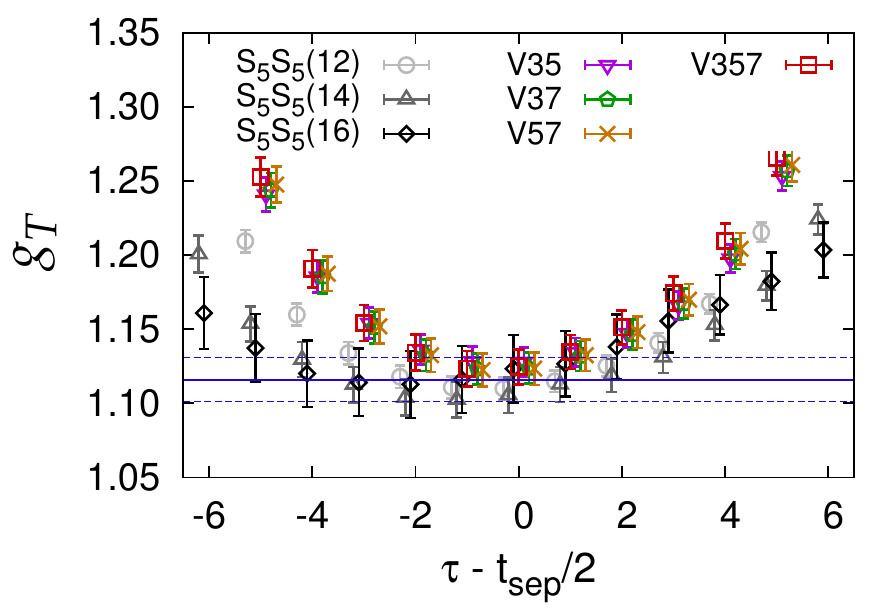}
  }
  \quad
  \subfigure{
    \includegraphics[width=0.43\linewidth,trim={0.8cm 1.22cm 0 0},clip]{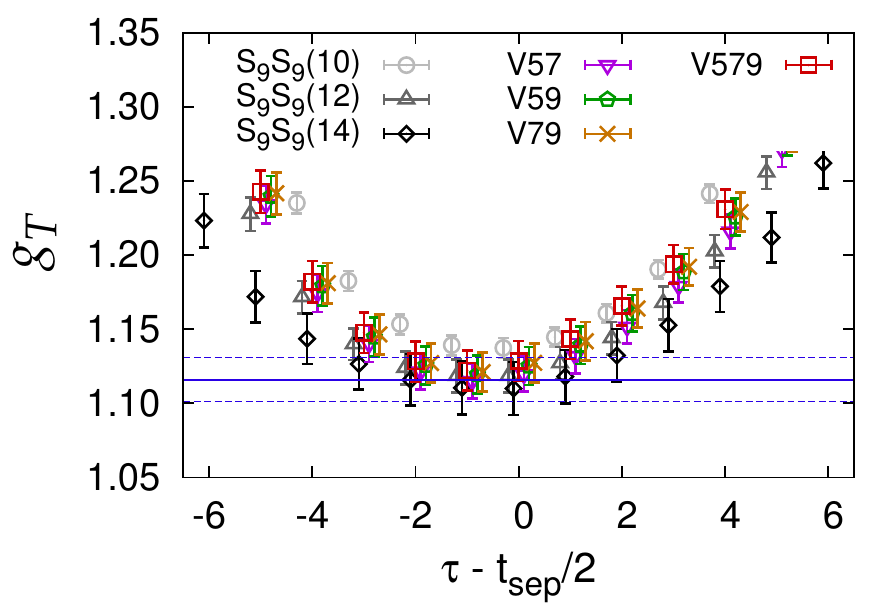}
  }
  \subfigure{
    \includegraphics[width=0.47\linewidth,trim={0 0 0 0},clip]{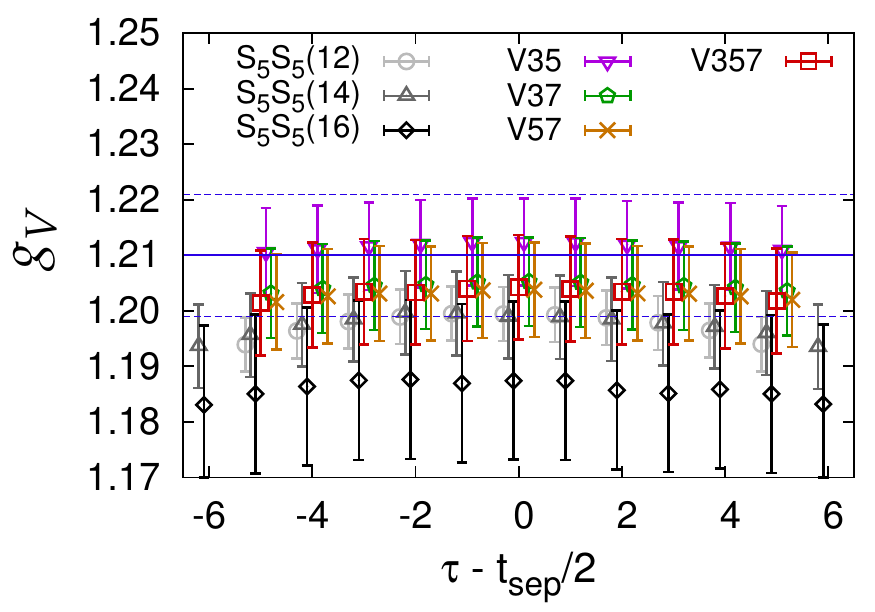}
  }
  \quad
  \subfigure{
    \includegraphics[width=0.43\linewidth,trim={0.8cm 0 0 0},clip]{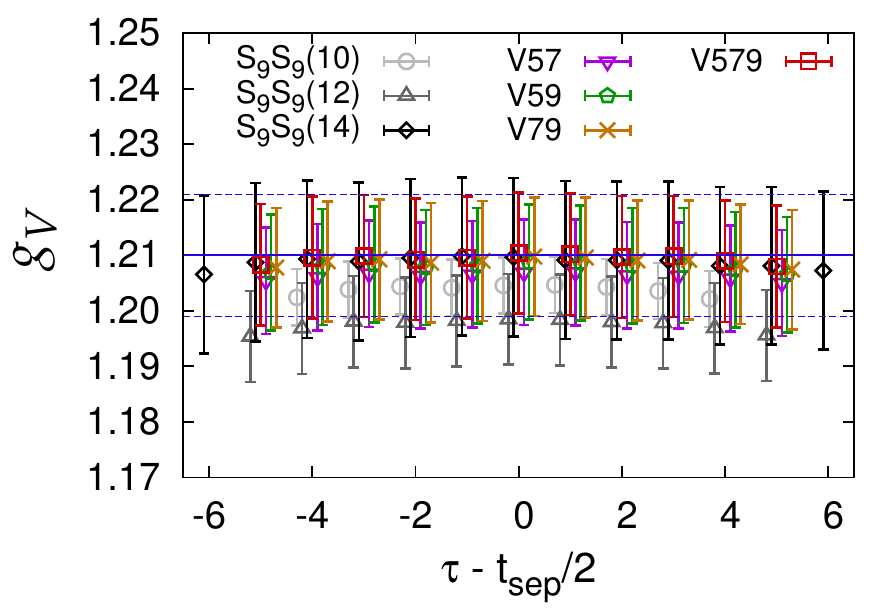}
  }
\caption{Comparison of the variational estimates for the
  unrenormalized isovector charges $g_A$, $g_S$, $g_T$ and $g_V$ using
  data from run R2 (left) and R3 (right).  We also show the data from
  R1 for $S_5 S_5$ with $t_{\rm sep} = 12, 14, 16$ (left) and from
  R4 for $S_9 S_9$ with $t_{\rm sep} = 10, 12, 14$ (right). Only the V357
  data for $g_A$ show a larger plateau compared to the $S_5 S_5$ data,
  indicative of smaller excited-state contamination. The horizontal
  lines in both sets of figures are the results of the 2-state fit
  with $ \langle 1 | \mathcal{O}_\Gamma | 1 \rangle \equiv 0$ to the
  V579 data. The seven data points at each $\tau$ are displaced
  slightly along the x axis for clarity. }
  \label{fig:R2R3compVar}
\end{figure*}

In Fig.~\ref{fig:R2R3compVar}, we compare the variational estimates
for the unrenormalized charges from runs R2 (left) and R3 (right).  We
also show the R1 $\sigma=5$ data with $t_{\rm sep} = 12, 14, 16$
(left) and R4 $\sigma=9$ data with $t_{\rm sep} = 10, 12, 14$ (right).
We observe the following features in the variational estimates:
\begin{itemize}
\item
The V357 and V579 estimates overlap for all the charges. The errors in
the V579 estimates are marginally larger than those in V357.
\item
The size of the errors in the V357 and V579 data for all four charges 
agree with those from $S_5 S_5$ with $t_{\rm sep} = 12$ and lie in between those 
in the $S_9 S_9$ data with $t_{\rm sep} = 12$ and $14$. 
\item
$g_A$: The data  converge from below and the
  variational data also show a small increase between $V35 \to V37 \to
  V57 \to V357$ with the V35 estimates being about $1\sigma$ below
  V357.  Thus, to get estimates to within $1\%$ accuracy, we estimate
  that a three smearing variational analysis is needed. 
\item
$g_S$: All the variational estimates overlap while the single smearing
  data converge from below. The significant curvature in the data from
  both methods suggests that the $ \langle 1 | \mathcal{O}_S | 0
  \rangle$ matrix element dominates the excited-state contamination.
  The errors in all the data and estimates for $g_S$ are about a
  factor of 5 larger than those in $g_A$ or in $g_T$.
\item
$g_T$: The data for the four combinations, V35, V37, V57 and V357 (or
  V59, V59, V79 and V579) overlap but the curvature in the data again
  points to a significant contribution from $ \langle 1 |
  \mathcal{O}_T | 0 \rangle$.  The data for 
  $g_T$ converge from above. The small downward trend in $S_9 S_9$
  data with increasing $t_{\rm sep}$ leads to a $t_{\rm sep} \to
  \infty$ value that is about $0.5\sigma $ smaller than the
  variational estimates.
\item
$g_V$: No significant trends indicating excited-state contamination
  are observed. Statistical fluctuations dominate the error. All the
  estimates are consistent within errors that are $\approx 1\%$.
\end{itemize}

Our conclusion on the variational method, looking especially at the
data for $g_A$, is that one needs the full $3 \times 3$ variational
ansatz V357 if the smearing size is restricted to $\sigma \le 7$. In
the case of V579, one finds that V79 and V579 give consistent
estimates, so a $2 \times 2$ analysis may be sufficient. The
conservative approach, in the absence of detailed information on the
smearing sizes to use, would be to use a $3 \times 3$ variational
ansatz if results with $< 2\%$ total uncertainty are desired.

\begin{figure*}
\centering
  \subfigure{
    \includegraphics[width=0.48\linewidth,trim={0 1.22cm 0 0},clip]{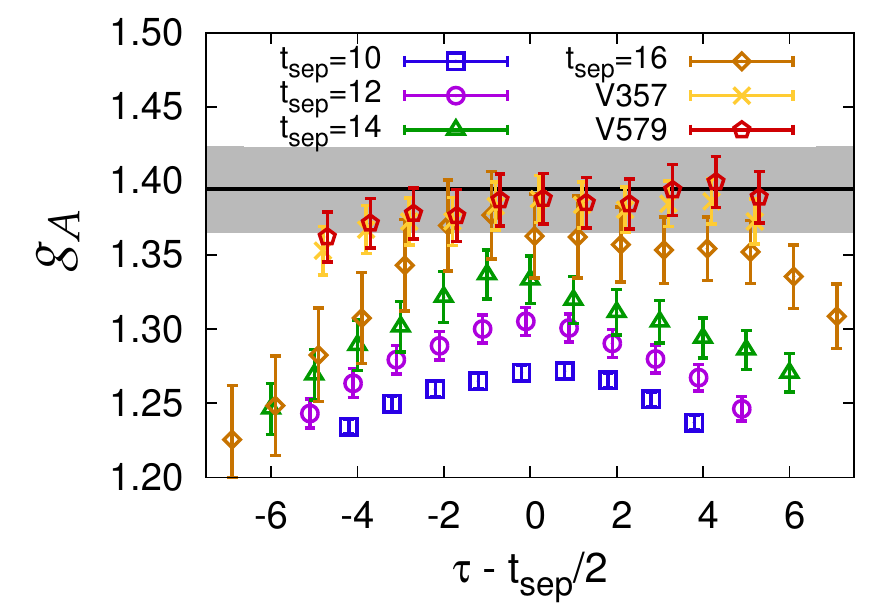}
  }
  \quad
  \subfigure{
    \includegraphics[width=0.43\linewidth,trim={1.00cm 1.22cm 0 0},clip]{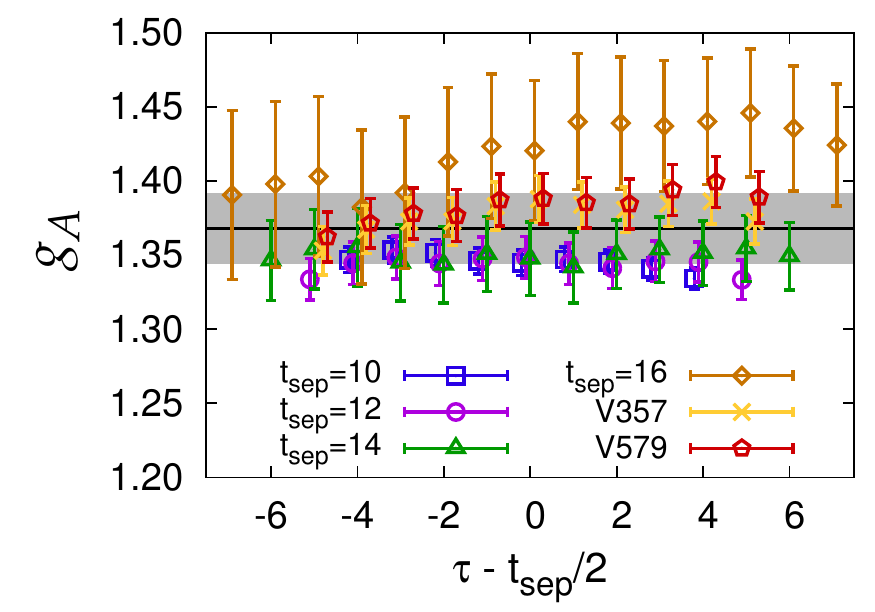}
  }
  \subfigure{
    \includegraphics[width=0.48\linewidth,trim={0 1.22cm 0 0},clip]{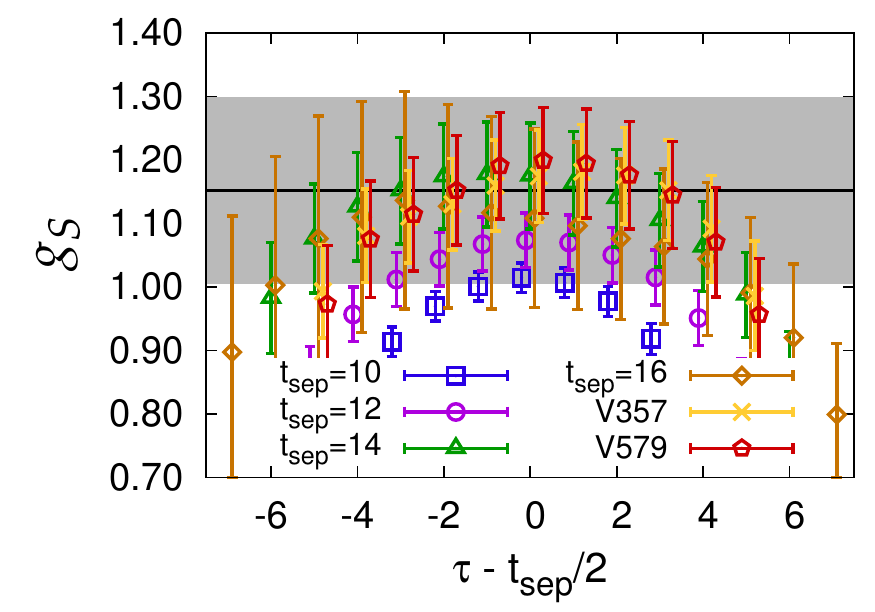}
  }
  \quad
  \subfigure{
    \includegraphics[width=0.43\linewidth,trim={1.00cm 1.22cm 0 0},clip]{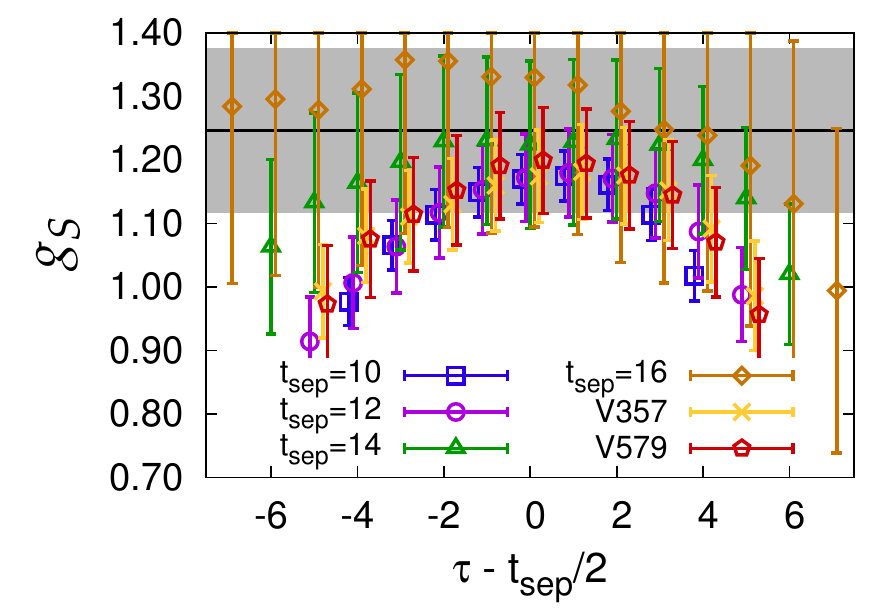}
  }
  \subfigure{
    \includegraphics[width=0.48\linewidth,trim={0 1.22cm 0 0},clip]{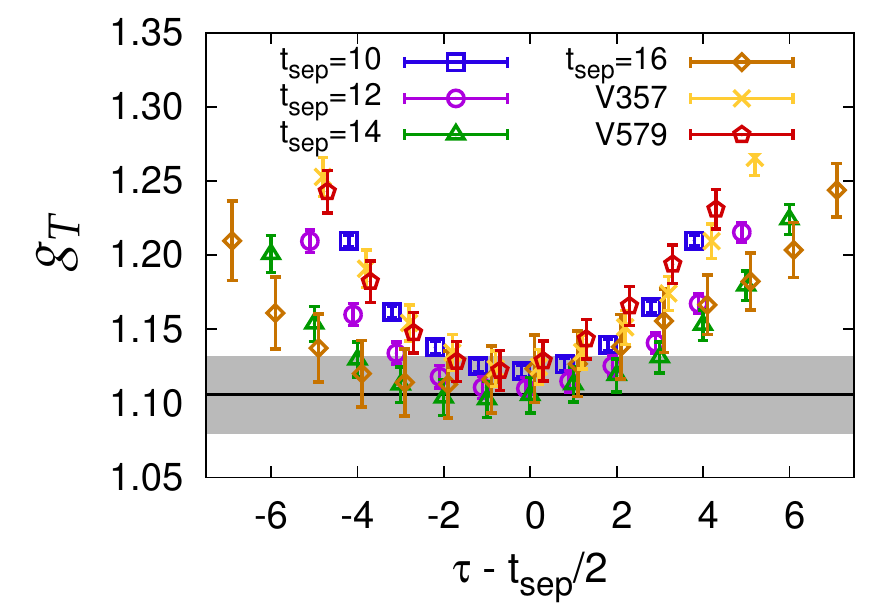}
  }
  \quad
  \subfigure{
    \includegraphics[width=0.43\linewidth,trim={1.00cm 1.22cm 0 0},clip]{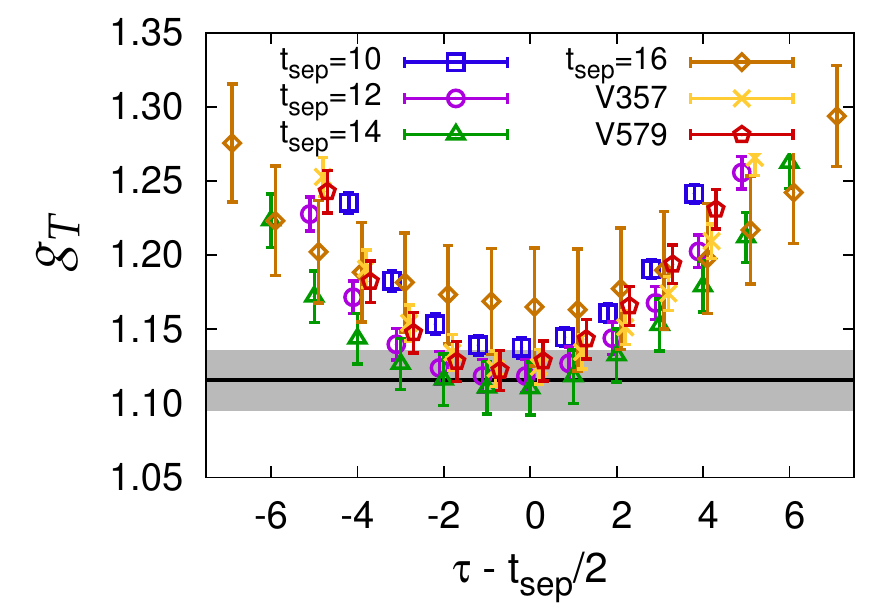}
  }
  \subfigure{
    \includegraphics[width=0.48\linewidth,trim={0 0 0 0},clip]{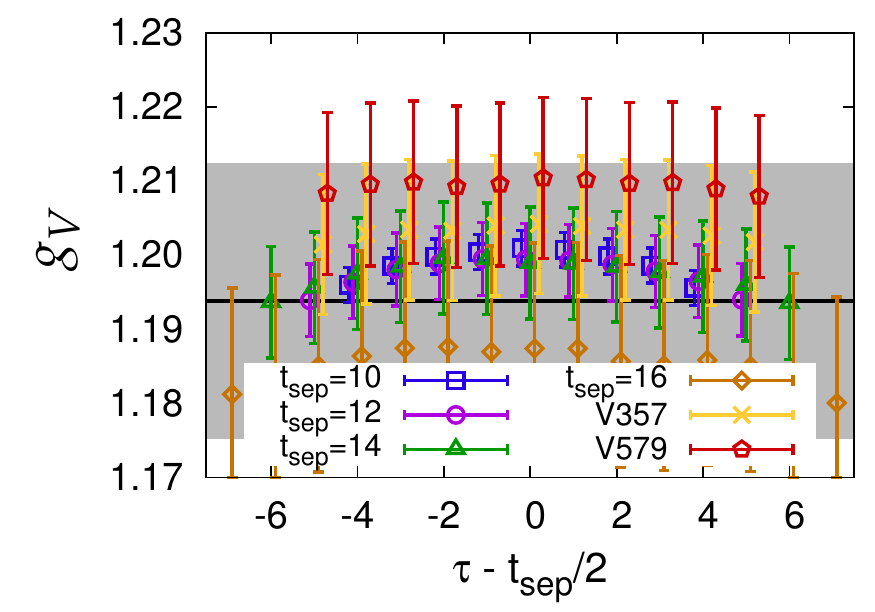}
  }
  \quad
  \subfigure{
    \includegraphics[width=0.43\linewidth,trim={1.00cm 0 0 0},clip]{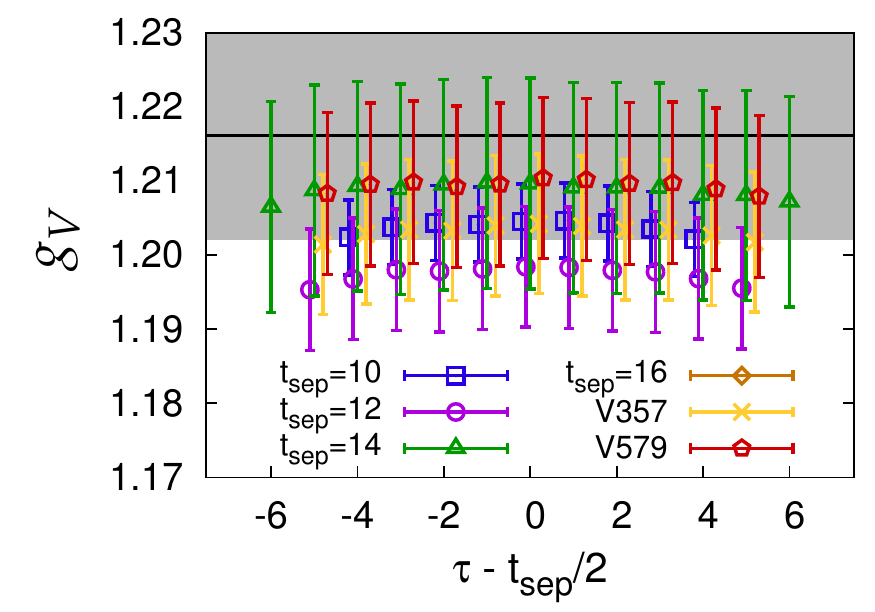}
  }
 \caption{Comparison of estimates for the unrenormalized isovector
   charges $g_A$, $g_S$, $g_T$ and $g_V$ from the variational analysis
   V357 (R2) and V579 (R3) with (left) R1 data with
   $\sigma=5$ and (right) R4 data with $\sigma=9$. The gray error
   band and the solid line within it is the $t_{\rm sep} \to \infty$
   estimate from the 2-state fit using (left) $S_5 S_5$ data with $t_{\rm sep} = [12, 14, 16, 18]$ and 
   (right) $S_9 S_9$ using data with $t_{\rm sep} = [10, 12, 14, 16]$.  }
\label{fig:S5S9vsVar}
\end{figure*}

In Fig.~\ref{fig:S5S9vsVar}, we compare the estimates for the
unrenormalized isovector charges $g_A$, $g_S$, $g_T$ and $g_V$
obtained from the 2-state fit to R1 data with $\sigma=5$ and $t_{\rm
  sep} = [12, 14, 16, 18]$ with the R4 data with $\sigma=9$ and $t_{\rm
  sep} = [10, 12, 14, 16]$. We also show the $3 \times 3$ variational
estimates  V357 (R2) and V579 (R3) obtained using $t_{\rm sep} = 12$. Comparing the 
two methods we find:

\begin{itemize}
\item
The excited-state effect in $g_A$ in the $S_5 S_5$ data is large but
the 2-state ansatz fits the data and gives a $t_{\rm sep} \to \infty$
estimate that agrees with the V357 and V579 values.
\item
The excited-state contamination in $g_A$ is much smaller in the $S_9
S_9$ data. However, since the data with $t_{\rm sep} = 10, 12, 14$
overlap, the fit gives a $t_{\rm sep} \to \infty$ estimate that is
about $1\sigma $ below the V357 and V579 estimates. (It is also about $1\sigma$ 
below the estimate from the fit to $S_5 S_5$ data as shown in Fig.~\ref{fig:S5S9}.) The
combined one sigma difference between the overlapping $t_{\rm sep}=10,
12, 14$ data and the $t_{\rm sep}=16$ data reduces the confidence in
the 2-state fit.  This case highlights a generic problem: for the
2-state fit to give the $t_{\rm sep} \to \infty$ estimate with $< 1\%$
error, the statistics have to be large enough that the trend in the
data is resolved at at least three values of $t_{\rm sep}$ .
\item
In lattice calculations with dynamical fermions, the factor limiting
the statistics is the number of independent gauge configurations
available. For a fixed statistical sample, the errors in our data
increase by $\approx 80\%$ with each two units of $\tsep$ as discussed
previously. Consequently, the error in the 2-state fit estimate
increases as data at larger $\tsep$ are included in the multiple
$\tsep$ analysis to get the $t_{\rm sep} \to \infty$ value. For
example, the estimates for $g_A$, using R1 with $S_5 S_5$, are
1.353(18), 1.366(20), 1.378(22), 1.382(25), 1.395(29), and 1.424(44)
with fits to $\tsep = [10, 12, 14]$, $[10, 12, 14, 16]$, $[10,
12, 14, 16, 18]$, $[12, 14, 16]$, $[12, 14, 16, 18]$ and $[14,
16, 18]$ data, respectively. Our best estimate, 1.395(29), is
obtained by neglecting the data at $\tsep = 10$, which have the
largest excited state contamination.  In comparison, the V357
variational result with $\tsep = 12$ is 1.386(16). We anticipate that
the error in the variational method would also increase with $\tsep$.
\item
For $g_S$, the overall trend in the $S_5 S_5$ data with $t_{\rm sep} =
10, 12, 14, 16$ is it converges from below and show significant
excited-state contamination. The $S_9 S_9$ data at each $t_{\rm sep} =
10, 12, 14, 16$ agree with V357 and V579 data. The excited-state
contamination is manifest in all the data as the curvature with
$\tau$. The 2-state fit to $S_5 S_5$ and $S_9 S_9$ gives an estimate
consistent with V357 and V579.
\item
For $g_T$, the 2-state fits to $S_5 S_5$ and $S_9 S_9$ data with
$t_{\rm sep} = [12, 14, 16, 18]$ and $t_{\rm sep} = [10, 12, 14, 16]$,
respectively, give consistent results and are about $1\sigma$ below
V357 and V579. Surprisingly, the $S_5 S_5$ data show smaller
curvature than $S_9 S_9$ data.  Overall, excited-state contamination is smaller than 
in $g_A$ and $g_S$ with the total variation with $\tsep$ at
the central value of $\tau$ being $\lsim 5\%$.
\item
All estimates for $g_V$ are consistent within $ 1\%$ uncertainty. The
largest difference is between the $S_5 S_5$ and $S_9 S_9$ estimates,
which is about $1\sigma$.  
\end{itemize}
To summarize, our comparison shows that once good choices of the
smearing sizes and $t_{\rm sep} $ are known, the two methods give
reliable and consistent results but the variational estimates have
smaller errors with the same statistics because they were obtained 
from a smaller value of $\tsep$.


\begin{figure*}[tb]
  \subfigure{
    \includegraphics[width=0.46\linewidth]{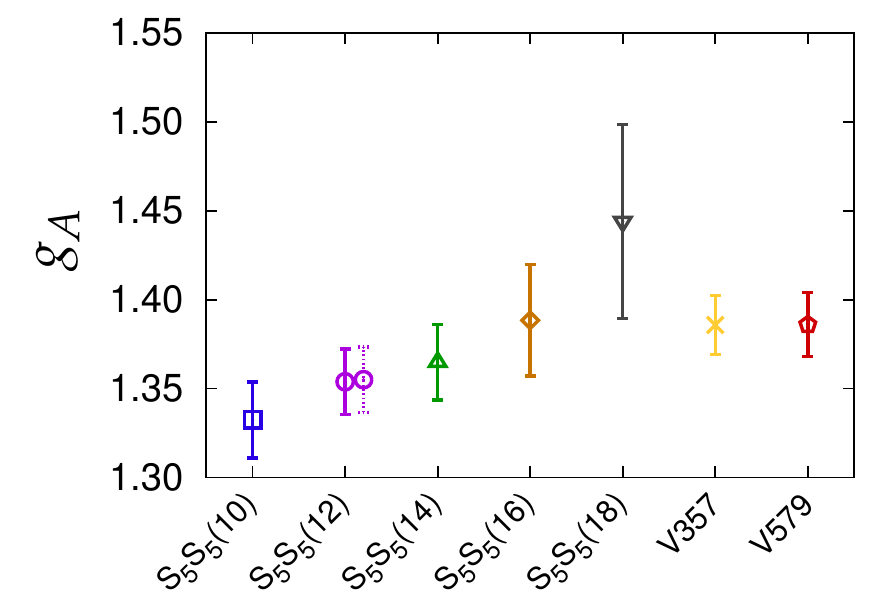}
    \includegraphics[width=0.46\linewidth]{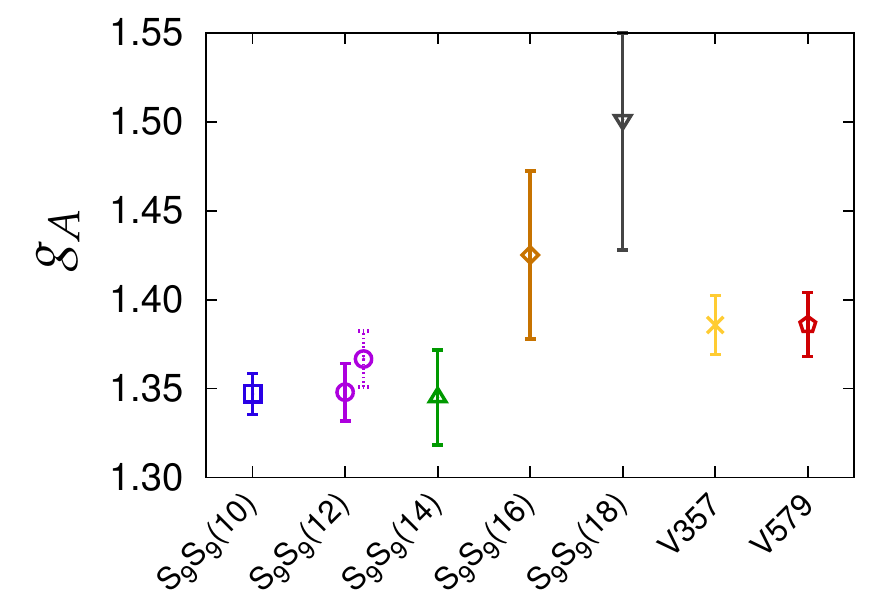} }
  \subfigure{
     \includegraphics[width=0.46\linewidth]{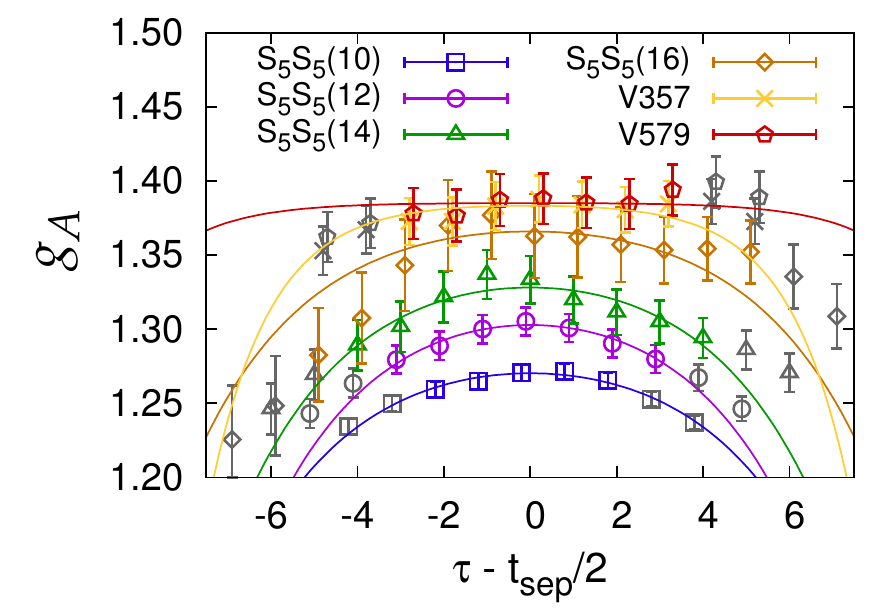}
     \includegraphics[width=0.46\linewidth]{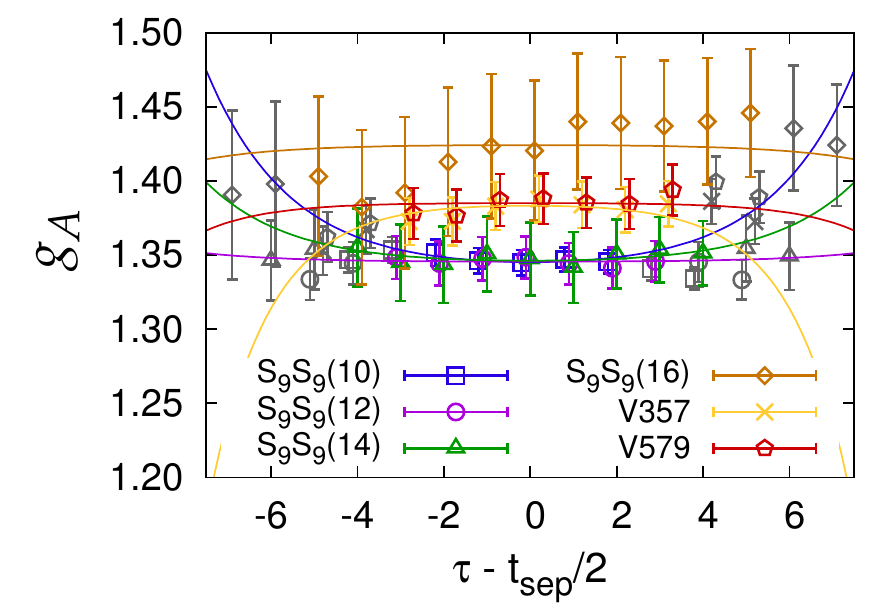}
  }
 \caption{Comparison of estimates of the unrenormalized $g_A$ from the
   $S_5 S_5$ data (left) and the $S_9 S_9$ data (right) for different
   values of $\tsep$ with V357 and V579.  In each case, the fit is
   made to data from a single $t_{\rm sep}$, given within parentheses,
   using the 2-state ansatz with $ \langle 1 | \mathcal{O}_\Gamma | 1
   \rangle =0 $.  We show the data and the fits (bottom panels) and a
   comparison of the resulting estimating of $g_A$ (top panels).  For
   clarity, the $\tsep=18$ data shown in Fig.~\protect\ref{fig:S5S9},
   are not reproduced here. The data points in black on either end in
   the variable $\tau$ are not used in the fits. The second estimates
   for $S_5 S_5(12)$ and $S_9 S_9(12)$ from R3 are shown with dotted
   error bars. The $1\sigma$ difference between the two $S_9 S_9(12)$
   estimates is discussed in the text and the data shown in
   Fig.~\protect\ref{fig:S9compareStats}.  }
\label{fig:gA_one_tsep_fit}
\end{figure*}

The important question is the following: does the consistency of the
four analyses confirm that the $\tsep \to \infty$ value has been
obtained?  In Fig.~\ref{fig:gA_one_tsep_fit}, we compare the trends in
the estimates of $g_A$ and their errors by making independent 2-state
fits with $\langle 1 | \mathcal{O}_\Gamma | 1 \rangle =0 $ to data at
a fixed value of $t_{\rm sep}$.  We find that the estimates from the $S_5 S_5$
data increase with $\tsep$.  This behavior is consistent with the
general trend observed in all the data---the estimate of $g_A$
converge from below. Even though the total variation between $\tsep=10$ and $18$ 
estimates is less than 3$\sigma$, taken at face value, this trend would indicate
that the V357 and V579 results are underestimates. On the other hand,
the incremental increase with $\tsep$ has to go to zero at
sufficiently large $\tsep$. Unfortunately, the errors in the $\tsep
\ge 16$ estimates, crucial to determining the value of $\tsep$ by
which the asymptotic value is reached, are too large.

The situation is not resolved by the $S_9 S_9$ data as they do not
show a uniform trend---the data with $\tsep=10,\ 12$ and $14$ are flat
and below V357 and V579, whereas the $\tsep=16$ and $18$ data are
above but their significance is less as they have large errors. Since
the differences are about one combined sigma, it is hard to quantify trends
with current statistics. For example, as shown in the top right panel
of Fig.~\ref{fig:gA_one_tsep_fit}, the two $S_9 S_9(12)$ data points
from R3 and R4 (see Fig.~\ref{fig:S9compareStats} for the data versus
$\tau$) differ by $1\sigma$.  If we use the $S_9 S_9(12)$ point from
R3 (shown with the dotted error bar) to determine the trend, we would
conclude that the $S_9 S_9$ data also show a rising trend and
the observed consistency of $\tsep=10,\ 12$ and $14$ estimates from R4
is a statistical fluctuation.  No such fluctuation is seen in the two
$S_5 S_5(12)$ data points from R1 and R3 plotted in the top left panel
of Fig.~\ref{fig:gA_one_tsep_fit}.

\begin{figure*}[tb]
     \includegraphics[width=0.98\linewidth]{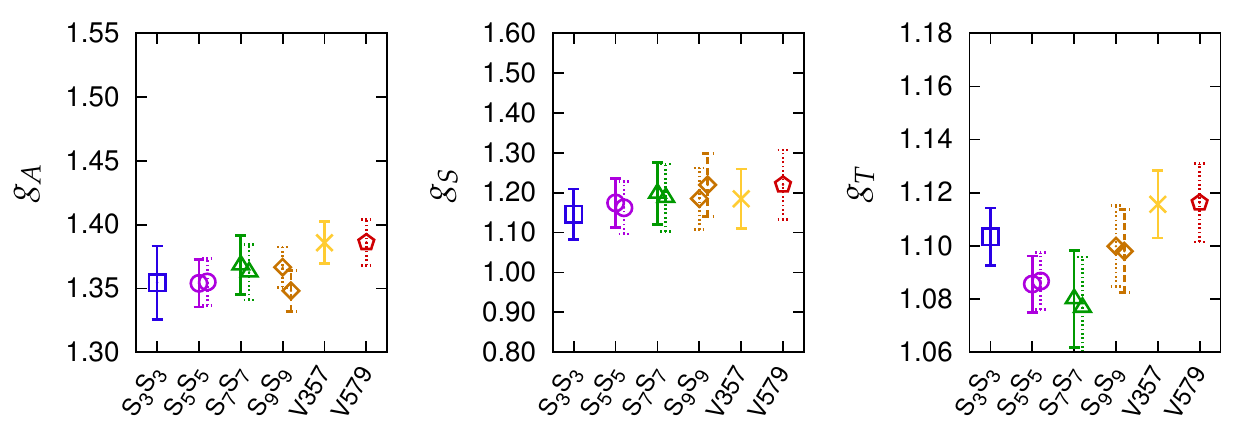}
 \caption{Comparison of the estimates of the unrenormalized charges
   obtained using a 2-state fit with $ \langle 1 | \mathcal{O}_\Gamma
   | 1 \rangle =0 $ to all the $\tsep=12$ data. The data point with
   solid error bars are from R2, dotted from R3 and dashed from R4.  }
\label{fig:compare12}
\end{figure*}

A comparison of the estimates in Fig.~\ref{fig:compare12}, where we
plot all the results obtained from data with $\tsep=12$, shows that
the errors in the V357 (V579) result are comparable to those in $S_5
S_5(12)$ ($S_9 S_9(12)$) with the same statistics but with less
excited-state contamination. Equally important, the trends in the data
in Figs.~\ref{fig:gA_one_tsep_fit} and~\ref{fig:compare12} show that
the error estimates in the $t_{\rm sep} \to \infty$ values for $S_5
S_5$ and $S_9 S_9$, given in Table~\ref{tab:results}, are reasonable
and cover the uncertainties discussed here.

\begin{figure*}[tb]
  \subfigure{
     \includegraphics[width=0.46\linewidth]{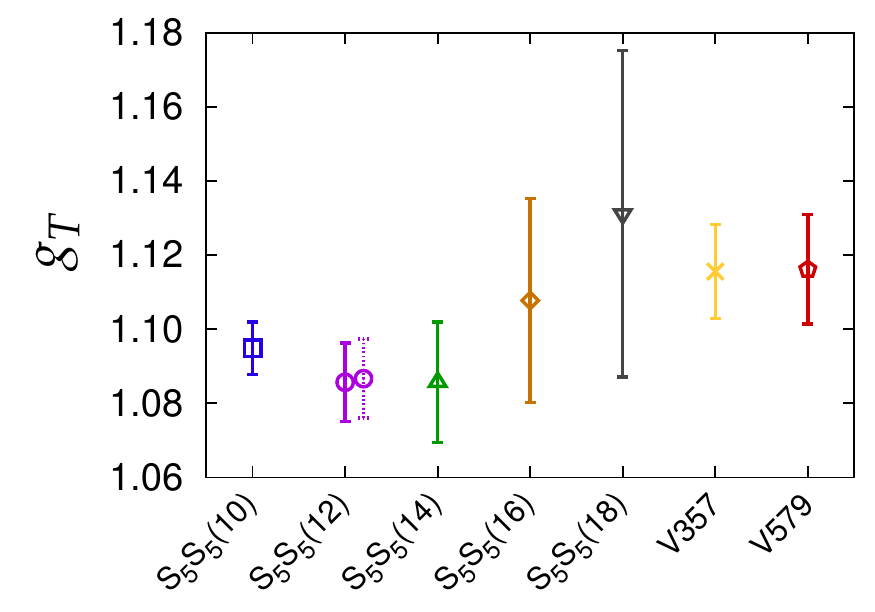}
     \includegraphics[width=0.46\linewidth]{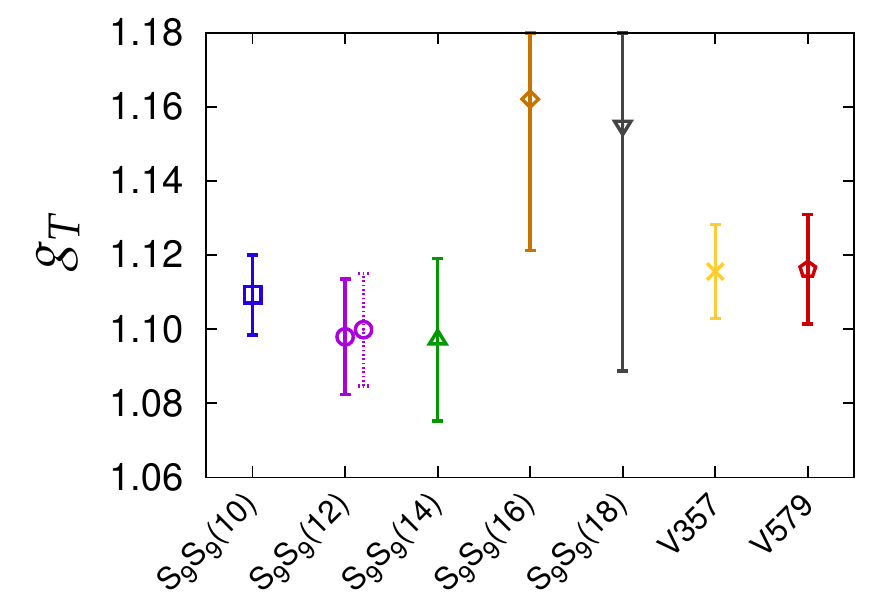}
  }
  \subfigure{
     \includegraphics[width=0.46\linewidth]{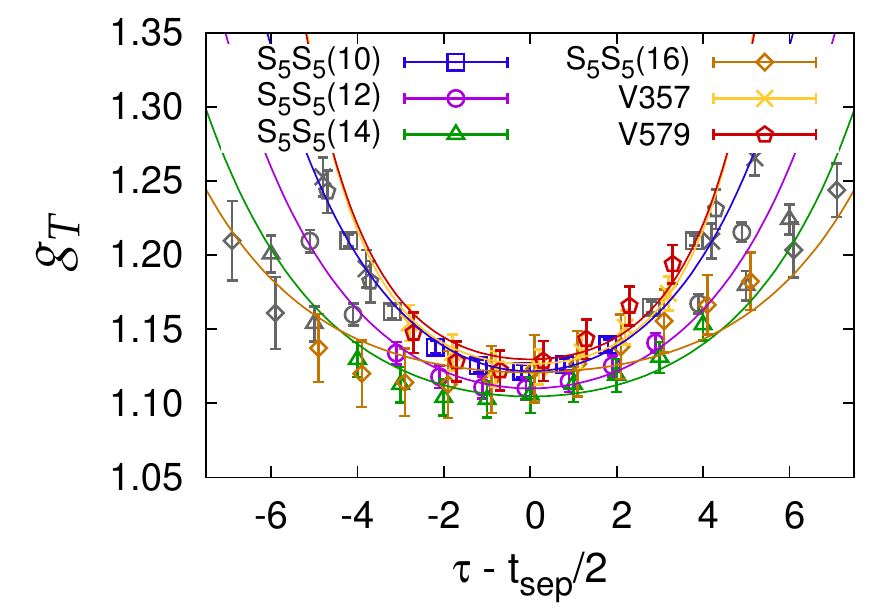}
     \includegraphics[width=0.46\linewidth]{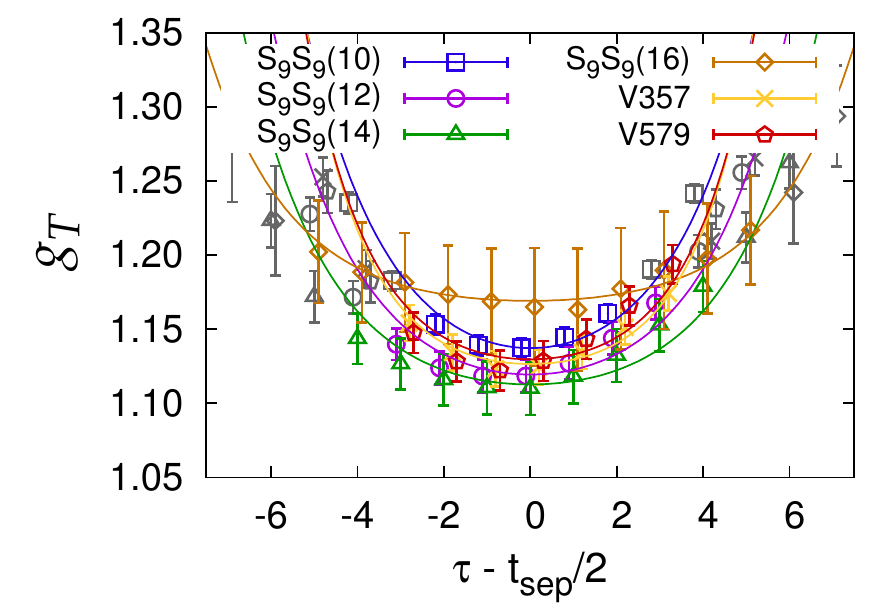}
  }
 \caption{Comparison of estimates of the unrenormalized $g_T$ from the
   $S_5 S_5$ data (left) and the $S_9 S_9$ data (right) for different
   values of $\tsep$ with V357 and V579.  The rest is the same as in
   Fig.~\protect\ref{fig:gA_one_tsep_fit}.  
  }
\label{fig:gT_one_tsep_fit}
\end{figure*}

We compare the behavior of $g_T$ in Figs.~\ref{fig:compare12}
and~\ref{fig:gT_one_tsep_fit}.  The overall trend, that $g_T$ converges from
above, would imply that the $\tsep=10,\ 12$ and $14$ estimates from
both the $S_5 S_5$ and $S_9 S_9$ data are better estimates of the
$t_{\rm sep} \to \infty$ value and lie about $1\sigma$ below V357
and V579 results.  On the other hand, with current statistics, all the
estimates are consistent within one combined $\sigma$.  Note that,
unlike $g_A$, the two sets of results for $S_9 S_9(12)$ and also those 
for $S_5 S_5(12)$ and $S_7 S_7(12)$, obtained using different source
positions, are in very good agreement.

\begin{figure*}[tb]
  \subfigure{
     \includegraphics[width=0.46\linewidth]{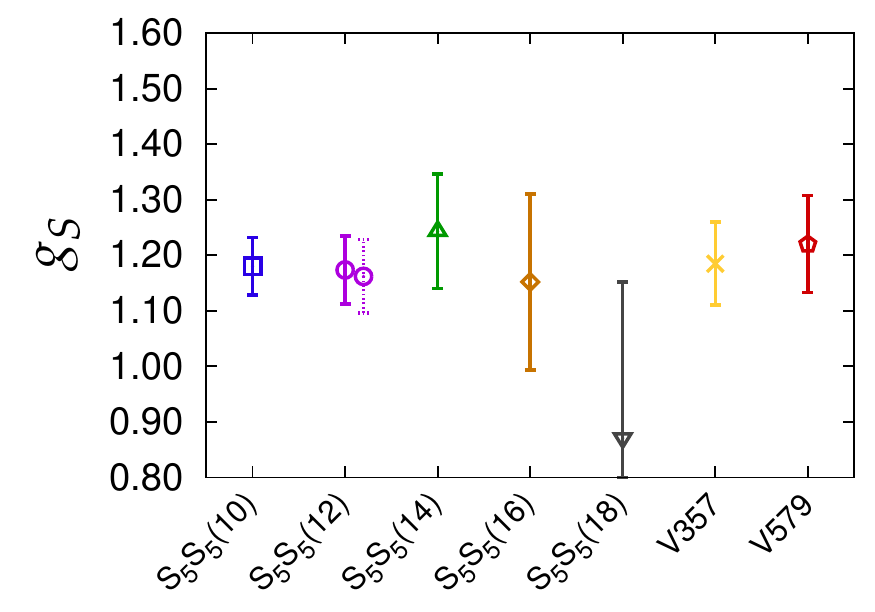}
     \includegraphics[width=0.46\linewidth]{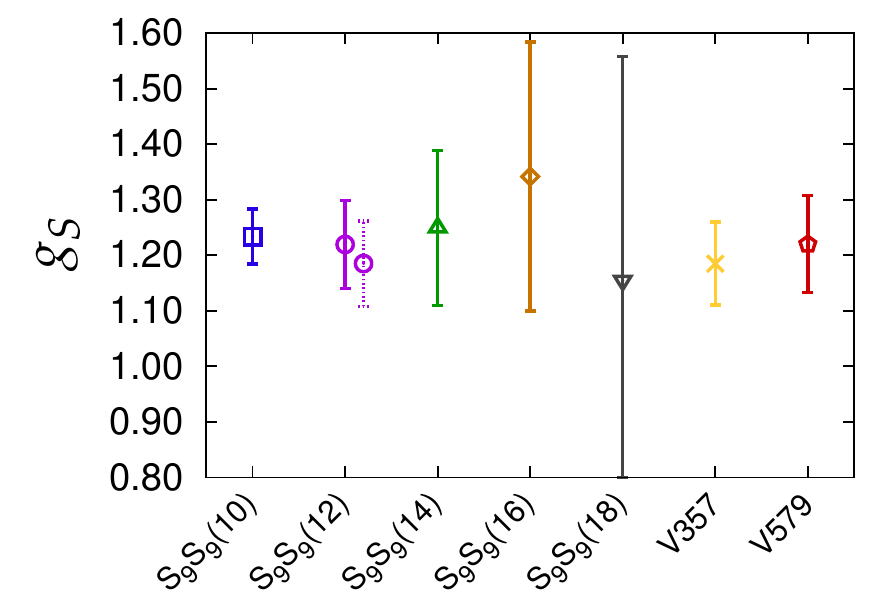}
  }
  \subfigure{
     \includegraphics[width=0.46\linewidth]{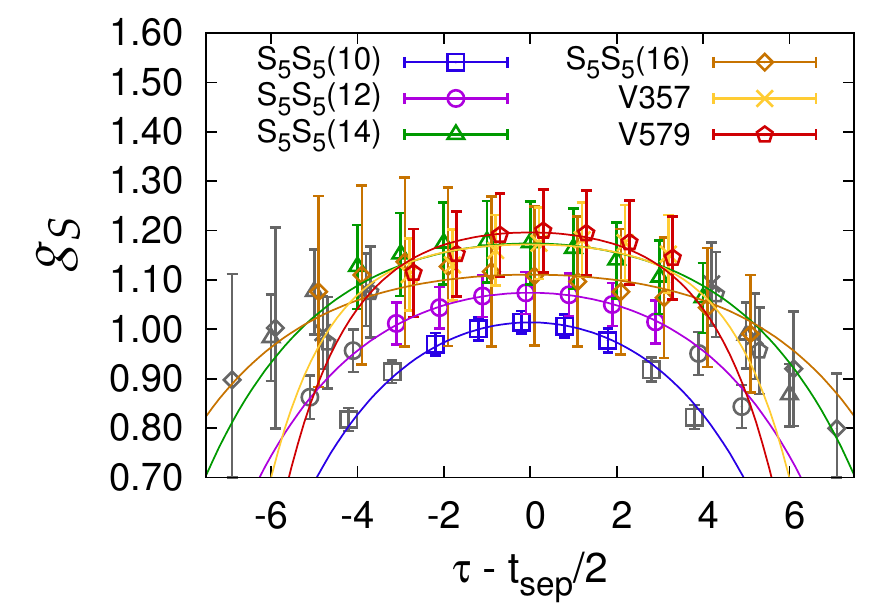}
     \includegraphics[width=0.46\linewidth]{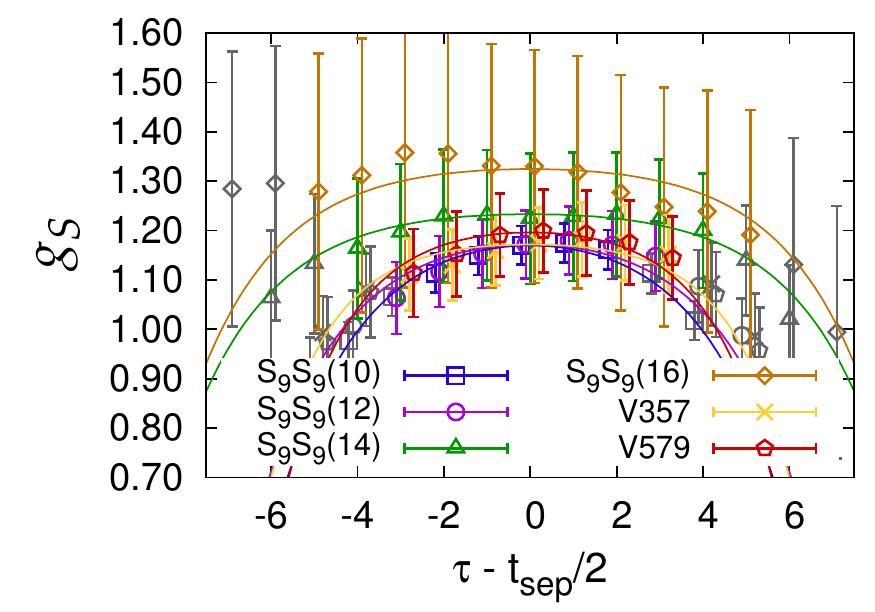}
  }
 \caption{Comparison of estimates of the unrenormalized $g_S$ from the
   $S_5 S_5$ data (left) and the $S_9 S_9$ data (right) for different
   values of $\tsep$ with V357 and V579.  The rest is the same as in
   Fig.~\protect\ref{fig:gA_one_tsep_fit}.  
  }
\label{fig:gS_one_tsep_fit}
\end{figure*}

The comparison of the scalar charge $g_S$ is shown in
Figs.~\ref{fig:compare12} and~\ref{fig:gS_one_tsep_fit}. The data are
consistent within their much larger error estimates and no trend with
$\tsep$ is apparent. Also, similar to the case of $g_T$, the two
independent estimates of $g_S$ from $S_5 S_5(12)$, $S_7 S_7(12)$ and
$S_9 S_9(12)$ are in very good agreement.

Our overall conclusion, based on the data shown in
Figs.~\ref{fig:gA_one_tsep_fit},~\ref{fig:compare12},~\ref{fig:gT_one_tsep_fit}
and~\ref{fig:gS_one_tsep_fit} that compare results from fixed $\tsep$
analyses, is that the errors in the V357 (V579) estimates are similar
to those in the $S_5 S_5 $ ($S_9 S_9 $) values with the same $t_{\rm
  sep} = 12$, but the excited-state contamination in $g_A$ is
smaller. In the case of $g_S$, the errors are large and all the
estimates are consistent.  There is a small but consistent trend
indicating an increase in the estimates of $g_A$ and $g_S$ towards the
$t_{\rm sep} \to \infty$ value with $\tsep$.  The situation with $g_T$
is less clear. Considering the results for all the three charges, we
again conclude that a smearing size $\sigma \approx 7$ is optimal for
a 2-state fit analysis with multiple $\tsep$. In the variational 
analysis, there is no significant difference between V357 and V579.

Lastly, we briefly comment on the similar behavior of excited-state contamination 
observed in the calculation
of nucleon matrix elements and its dependence on smearing parameters and $\tsep$ 
by other lattice QCD
collaborations~\cite{Bali:2014nma,Abdel-Rehim:2015owa,Capitani:2015sba,Djukanovic:2015hnh}.
These three collaborations first use different amounts of APE smearing
to smooth the links and then construct smeared sources using Wuppertal
(Gaussian) smearing. A detailed comparison of their results with our
analysis is not straightforward because each collaboration has used
different smearing methods, smearing sizes and values of $\tsep$ on different
ensembles. For example, translating RQCD collaboration's
~\cite{Bali:2014nma} parameters would give smearing sizes between
0.7--0.9~fm on their various ensembles. The smearing size used by the
ETMC collaboration is $\approx 0.5$~fm and they report similar excited-state
contamination in the extraction of all the
charges~\cite{Abdel-Rehim:2015owa}. The Mainz
group~\cite{Capitani:2015sba,Djukanovic:2015hnh} also tunes the
smearing size to $\approx 0.5$~fm in their study of electric and
magnetic form factors.  Our work shows that the size of the
excited-state contamination in the extraction of various charges and
form factors is sensitive to the smearing parameters and values of
$\tsep$ simulated. It is, therefore, important to demonstrate 
that the $\tsep \to \infty$ value has been obtained and compare errors 
in this limit.

\section{Which Method is More Cost Effective?}
\label{sec:Cost}

In the previous Sec.~\ref{sec:excited}, we showed that both the
2-state fit with data at multiple $t_{\rm sep}$ and the variational
analysis with multiple smearings can be made essentially equally
effective in reducing excited-state contamination and give overlapping
estimates.  The errors in the variational analysis are, however,
35--60\% smaller compared to the estimates from the $S_9 S_9$ 2-state
analysis with multiple $\tsep$ as shown in Table~\ref{tab:results}.
Comparing data at fixed $\tsep=12$ shows that the variational method
yields estimates closer to the asymptotic value for $g_A$, while for
$g_S$ the two estimates $S_7 S_7 (12)$ and $S_9 S_9(12) $ are as
good. The trend in $g_T$ is not clear, but if the convergence from
above is validated by higher precision data, then $S_7 S_7 (12)$ would
be the preferred estimate.  Being able to obtain the $t_{\rm sep} \to
\infty$ estimate from the smallest value of $\tsep$ is important
because the errors grow by $\approx 80\%$ for every two units of
$t_{\rm sep}$.

To decide between the two methods---variational versus the 2-state fit
to data at multiple $t_{\rm sep}$, we present a cost-benefit analysis
assuming that the best value of the smearing parameter $\sigma$ (for
example, $\sigma=7$ in this work) has already been determined using
trial runs.  Also, based on the discussion in Sec.~\ref{sec:excited},
we will mostly use $g_A$, and its extrapolation to $t_{\rm sep} \to
\infty$, to compare the two methods as it shows large excited-state
contamination. Next, based on the $S_5 S_5$ and $S_9 S_9$ data, we
assume that the errors in a 2-state fit to $S_7 S_7$ data with
$\tsep=10,\ 12,\ 14$ and $16 $ will be about $50\%$ larger than
those from V579.  Lastly, we assume that the sequential $u$ and $d$
propagators are calculated using the coherent sequential source trick
with $N_{\rm meas}$ source locations being processed simultaneously on
each configuration. Keeping in mind that the goal is to get the best
estimate for the $t_{\rm sep} \to \infty$ value with a fixed
computational cost, we count the number of inversions of the Dirac
matrix required for the minimum computation in each case as follows.

\begin{itemize}
\item
A 2-state fit with $N_{\rm tsep}$ values of $t_{\rm sep}$ requires
$N_{\rm meas} + 2 \times N_{\rm tsep}$ inversions: Our analyses
indicate that $N_{\rm tsep}=3$ is sufficient and $N_{\rm tsep} = 4$
allows for validation.  Typical values of $N_{\rm meas}$ on lattice
sizes currently being used are either 3 or 4. For $N_{\rm meas}=3$,
one needs 9 inversions for $N_{\rm tsep}=3$ and 11 for $N_{\rm
  tsep}=4$.  Doubling the statistics to improve the fit would increase
the cost to 22 inversions for $N_{\rm tsep}=4$. However, recognizing
that the reduction in errors is required mainly in our $\tsep =16$
data, doubling its statistics would increase the cost to 16 inversions.
\item
A variational analysis with $N_{\rm smear}$ smearings requires $N_{\rm
  smear} \times N_{\rm meas} + (2 \times N_{\rm smear} \times N_{\rm
  smear}) \times N_{\rm tsep}$ inversions if all combinations of the
source and sink 3-point functions are calculated.  Our analysis
suggests that $N_{\rm smear}=3$ is needed for high precision.  In that
case, for $N_{\rm meas}=3$ and $N_{\rm tsep}=1$ one needs 27
inversions.

This cost can be reduced significantly if a good estimate of the
eigenvector $u_0$ used for constructing the projected variational
correlation function is known before starting the calculation of the
3-point functions.  In that case the dot product of the $N_{\rm smear}
\times N_{\rm smear}$ matrix of zero-momentum nucleon sources at the
sink with $u_0$ can be taken before the final inversion to construct
the sequential propagators. This trick would reduce the number of
sequential propagators to calculate from $ 2 \times N_{\rm smear}
\times N_{\rm smear}$ to $2 \times N_{\rm smear}$.  For each of the
$N_{\rm smear}$ projected sources, the coherent source can be constructed in the
same way as before, i.e., by repeating the operation on the $N_{\rm
  meas}$ time slices and adding the sources after projection using $u_0$.
With this simplification, the cost is reduced to $N_{\rm smear} \times
N_{\rm meas} + (2 \times N_{\rm smear} \times N_{\rm tsep})$
inversions, which for $N_{\rm meas}=N_{\rm smear}=3$ and $N_{\rm
  tsep}=1$ is 15 inversions and increases to 21 for $N_{\rm
  tsep}=2$. Lastly, we anticipate, based on the $S_5 S_5$ and $S_9
S_9$ analyses showing that the errors increase by a factor of $\approx 0.8$ for
increase in $\tsep$ by two units, that a similar increase would be
present in the variational analysis, i.e., errors in a $\tsep = 14$
variational calculation, done to confirm that the $\tsep \to \infty $
value has been obtained, would be larger by a factor of $\approx 1.8$.
\end{itemize}

In Sec.~\ref{sec:excited}, we found that estimates of $g_S$, $g_T$,
and $g_V$ from a 2-state fit to just the $t_{\rm sep} = 16$ data are
also compatible with those from the variational analysis but the
errors are larger by a factor of about two.  To raise the precision of the
2-state fit with $N_{\rm tsep} = 4$ to the level of the variational
result, i.e., achieve comparable errors, we would need to roughly
double the statistics. In
this scenario, the computational cost of a $3 \times 3$ variational
analysis with a good estimate of $u_0$ would be more cost effective
(15 versus $2 \times 11=22$ inversions). However, if the statistics for
only the $\tsep=16$ data is doubled, then the 2-state fit is equally 
cost effective (15 versus 16 inversions). 

In the most conservative approach, assuming two values of  $t_{\rm
  sep}$ need to be simulated in the variational approach to
demonstrate convergence to the $t_{\rm sep} \to \infty $ estimate, as
indicated by the discussion in Sec.~\ref{sec:excited}, or one needs double 
the statistics in the 2-state state fit with $N_{\rm tsep}=4$, the
two methods are again equally cost effective (21 versus 22 inversions).

The cost effectiveness of the 2-state fit method increases as the
quark mass is reduced and the lattice size $T$ is increased. On our
$64^3 \times 128$ lattices at $M_\pi \approx 200$~MeV we can use
$N_{\rm meas}=5$ or even $6$ since the signal in the nucleon 2-point
correlation function dies out by $t \approx 20$.  For $N_{\rm
  meas}=5$, the 2-state fit with $N_{\rm tsep} = 4$ would cost 13
inversions, while a $ N_{\rm smear}=3$ variational analysis with
$t_{\rm sep}=1$ and known $u_0$ would cost 21 inversions.

A somewhat different conclusion is reached in
Ref.~\cite{Dragos:2015ocy}, in which the authors claim that the
variational method offers a more efficient and robust method for the
determination of the nucleon matrix elements.  Some of the reasons for
their conclusion that the variational method is decidedly better are:
\begin{itemize}
\item
Their calculation was done on a finer lattice with $a=0.074$~fm. Thus,
to first approximation, all our length scales should be multiplied by
$1.1$ when comparing with their analysis.
\item
The much higher statistical precision of our calculation (42,528
versus 1050 measurements) allows us to better resolve the trends in both
methods.
\item
Their variational analysis was done with three smearing sizes, $\sigma
\approx 4.1, 5.8$ and $ 8.3$.  These three sizes cover the value
$\sigma=7.7$ corresponding to $\sigma \approx 0.57$~fm we consider
optimal.  Thus, we expect their analysis to give a good estimate with
$t_{\rm sep} =13$, which, in physical units, is equivalent to the
$t_{\rm sep} =12$ used in our variational analysis.
\item
Their 2-state fits were based on data with $\sigma \approx 4.1$
($N_{\rm GS}=32$), for which the excited-state contamination is very
large as shown in this work. With such an unoptimized value of
$\sigma$ and given that their data for $g_A$ with $t_{\rm sep} = 16,
19$ and $22$ has large errors, it is not surprising that their $t_{\rm
  sep} \to \infty$ estimate from a 2-state fit has much larger errors
compared to their variational estimate.  For the same reasons, we
suspect that their 2-state fit slightly underestimates the $t_{\rm
  sep} \to \infty$ value.
\item
They do not provide a cost estimate for the two analyses. Assuming
that they constructed the full $3 \times 3$ matrix of 3-point
correlation functions in their variational analysis, it is 13 versus
27 inversions for the 2-state versus the variational approach.
\item
They did not evaluate the change in the cost effectiveness of the two
methods as the quark mass is decreased and the lattice size $T$ is
increased correspondingly. With larger $N_{\rm meas}$, the relative
cost effectiveness of the 2-state fit method increases.
\end{itemize}

To summarize, we have compared the two methods using the optimal
smearing sizes. Our conclusion on cost effectiveness is based on the
best case scenario of a tuned value of $\sigma$ for both methods and
using three smearing sizes with a known result for $u_0$ in the
variational analysis. We have also assumed that the same choice of the
smearing parameters and $t_{\rm sep}$ are equally effective for all
matrix elements.  We find that both methods give results that are
consistent within errors. The variational method is more
cost effective if results at a single value of $t_{\rm sep}$ are
sufficient to obtain the $t_{\rm sep} \to \infty$ value and a good
estimate of $u_0$ is known beforehand. The 2-state fit with four
values of $t_{\rm sep}$ and double the statistics at the larger
$\tsep$ values has the advantage of the built in check of the
convergence to the $t_{\rm sep} \to \infty$ estimate that can be made
separately for each observable.  Lastly, the cost effectiveness of the
2-state fit method increases as the lattice size $T$ is increased and
the quark mass is lowered to its physical value because a larger
number of measurements, $N_{\rm meas}$, can be made simultaneously on
each configuration and $N_{\rm meas}$ sources at the sink timeslice
added in the coherent source method to produce the sequential
propagator.

\section{Conclusions}
\label{sec:conclusions}

We have presented a high statistics study of isovector charges of the
nucleon using (2+1)-flavor clover lattices generated using the RHMC
algorithm. The focus of this work is to investigate methods to improve
the statistical precision of the data and reduce the excited-state
contamination in matrix elements of quark bilinear operators within
nucleon states.  We show that both the variational method and the
2-state fit with data at multiple $\tsep$ are equally effective at
reducing excited-state contamination once the smearing parameters and
the values of $\tsep$ have been tuned.

With the current lattice parameters, our ability to conclude which
method gives a more reliable estimate of the $t_{\rm sep} \to \infty$
value and is more cost effective is limited by statistics since all
the estimates are consistent within $1\sigma$ error estimates.  To
demonstrate that the $t_{\rm sep} \to \infty$ estimate has been
obtained requires doing the variational calculation at two values of
$\tsep$ and in the 2-state fit using at least 3 values of $\tsep$ with
$\tsep \ge 1$~fm in both cases. The advantage of simulating multiple
values of $\tsep$ in either method is to be able to evaluate the
convergence to the $t_{\rm sep} \to \infty$ limit as a function of
$\tsep$. The cost of adding additional values of $\tsep$ is much less
in the 2-state fit method compared to a $3 \times 3$ variational
analysis.\looseness-1

For a fixed number of gauge configurations available and measurements
made, the errors in the variational method with a fixed $\tsep$($
\approx 1$~fm in our study) are consistent with those from the 2-state
fit to data with the same $\tsep$ but the excited-state contamination
is smaller, so it gives a better estimate of the $t_{\rm sep} \to
\infty$ limit.  The error in the 2-state fit with multiple $\tsep$
method are larger because data with/at larger $\tsep$ are needed to
reduce excited-state contamination and errors in the data for the
3-point functions grow rapidly with $\tsep$.

Assuming that the $t_{\rm sep} \to \infty$ estimate has been obtained
in all four runs R1--R4 analyzed in this study with $a=0.081$~fm,
$M_\pi=312$~MeV lattices of size $T=64$ and $N_{\rm meas}=3$, the $3
\times 3$ variational method is computationally more cost effective
than the 2-state fit to data at four values of $\tsep$ because the
errors are about $50\%$ smaller. The cost becomes the same if one
doubles the statistics in the 2-state method to make the errors
roughly equal and simulates a second $\tsep$ in the variational
calculation to confirm the convergence to the $t_{\rm sep} \to \infty$
limit.

The cost effectiveness of the 2-state method increases rapidly as the
light quark mass is reduced towards its physical value and the lattice
size $T$ is increased correspondingly because the number of
simultaneous measurements, $N_{\rm meas}$, that can be made on each
configuration and benefit from the coherent sequential source method
increases with $T$.  Since the cost of the lattice calculations at a
fixed value of the lattice spacing is dominated by the analysis of
ensembles close to the physical values of the quark mass, one should
carefully choose the method that is more cost effective in that limit.

Our overall conclusion is that both methods are effective in reducing
the excited-state contamination and have their relative strengths.
The choice depends on the number of gauge configurations available,
the value of the light quark mass, the lattice size, and the effort
needed to tune the smearing parameters, the eigenvector $u_0$ and the
values of $\tsep$ adequately prior to the calculation of the 3-point
functions.

\begin{acknowledgments}
R.G. thanks Jack Dragos for discussions on their variational analysis.
This research used resources of the Oak Ridge Leadership Computing
Facility at the Oak Ridge National Laboratory, which is supported by
the Office of Science of the U.S. Department of Energy under Contract
No. DE-AC05-00OR22725. The calculations used the Chroma software
suite~\cite{Edwards:2004sx}. The work of T.B., R.G. and B.Y. is
supported by the U.S. Department of Energy, Office of Science, Office
of High Energy Physics under contract number~DE-KA-1401020 and the
LANL LDRD program. The work of JG was supported by PRISMA Cluster of
Excellence at the University of Mainz.  The work of HWL is supported
in part by the M. Hildred Blewett Fellowship of the American Physical
Society.  J. N. is supported by the U.S. Department of Energy, Office
of Science, Office of Nuclear Physics under grant Contract Number
DE-SC0011090.  ME is supported by DOE grant number DE-FG02-96ER40965.
B.J., K.O., D.G.R., S.S. and F.W. are supported by the U.S. Department
of Energy, Office of Science, Office of Nuclear Physics under contract
DE-AC05-06OR23177.
\end{acknowledgments}

\clearpage
%
\bibliography{ref} 

\end{document}